\documentclass[submission, Phys]{SciPost}

\usepackage{xcolor}
\definecolor{RoyalBlue}{rgb}{0.255, 0.412, 0.882}
\definecolor{DeepSkyBlue}{rgb}{0.0, 0.749, 1.0}
\definecolor{Crimson}{rgb}{0.862, 0.078, 0.235}
\definecolor{ForestGreen}{rgb}{0.133, 0.545, 0.133}
\definecolor{OrangeRed}{rgb}{1.0, 0.271, 0.0}
\definecolor{Orchid}{rgb}{0.855, 0.439, 0.839}
\definecolor{Sienna}{rgb}{0.627, 0.322, 0.176}
\definecolor{Goldenrod}{rgb}{0.855, 0.647, 0.125}
\definecolor{CadetBlue}{rgb}{0.372, 0.619, 0.627}
\definecolor{CornflowerBlue}{rgb}{0.392, 0.584, 0.929}
\definecolor{RebeccaPurple}{rgb}{0.4, 0.2, 0.6}
\definecolor{Salmon}{rgb}{0.980, 0.502, 0.447}
\definecolor{HotPink}{rgb}{1.0, 0.412, 0.706}
\definecolor{Chocolate}{rgb}{0.824, 0.412, 0.118}
\definecolor{SteelBlue}{rgb}{0.275, 0.510, 0.706}
\definecolor{FireBrick}{rgb}{0.698, 0.133, 0.133}
\definecolor{bondiblue}{rgb}{0.0, 0.58, 0.71}
\definecolor{celestialblue}{rgb}{0.29, 0.59, 0.82}
\definecolor{coolblack}{rgb}{0.0, 0.18, 0.39}
\definecolor{frenchblue}{rgb}{0.0, 0.45, 0.73}
\definecolor{lapislazuli}{rgb}{0.15, 0.38, 0.61}
\definecolor{mediumpersianblue}{rgb}{0.0, 0.4, 0.65}
\definecolor{darkpowderblue}{rgb}{0.0, 0.2, 0.6}
\definecolor{darkcandyapplered}{rgb}{0.64, 0.0, 0.0}
\definecolor{darkscarlet}{rgb}{0.34, 0.01, 0.1}
\definecolor{falured}{rgb}{0.5, 0.09, 0.09}

\hypersetup{
    colorlinks,
    linkcolor={falured},
    citecolor={darkpowderblue},
    urlcolor={lapislazuli}
}

\usepackage[bitstream-charter]{mathdesign}

\usepackage{multirow}
\usepackage{hhline}
\usepackage{bigdelim}
\usepackage{graphicx}
\usepackage{color}
\usepackage{dcolumn}
\usepackage{physics}
\usepackage{mathtools}
\usepackage{bm}
\usepackage{array}
\usepackage{comment}
\usepackage{cancel}
\usepackage{colortbl}
\usepackage{amsmath}
\usepackage{amsthm}
\usepackage{here}

\newtheorem{thm}{Theorem}

\newtheorem{lem}{Lemma}

\newtheorem{cor}{Corollary}
\newtheorem*{coefficientrecurrencerestatement}{Theorem~\ref{thm:coefficient-recurrence-conservation}}

\newcommand{\paren}[1]{\left(#1\right)}
\newcommand{\bck}[1]{\left[#1\right]}
\newcommand{\bce}[1]{\left\{#1\right\}}

\newcommand{\floor}[1]{\left\lfloor#1\right\rfloor}

\newcommand{\imi}{\mathrm{i}}

\newcommand{\ebs}{\boldsymbol{e}}
\newcommand{\Mbs}{\boldsymbol{M}}
\newcommand{\sbs}{\va{\sigma}}
\newcommand{\lax}{\boldsymbol{L}}
\newcommand{\laxcal}{\boldsymbol{\mathcal{L}}}

\newcommand{\Jbs}{\boldsymbol{J}}
\newcommand{\Jtilbs}{\widetilde{\boldsymbol{J}}}
\newcommand{\Rbs}{\boldsymbol{R}}
\newcommand{\w}{\boldsymbol{W}}

\newcommand{\wcorr}{\overline{\w}}
\newcommand{\Wcorr}{\overline{W}}
\newcommand{\lr}[3]{\left#1 #3 \right#2}

\newcommand{\ntil}{\widetilde{n}}
\newcommand{\sfunc}[4]{S_{#1}^{\paren{#2, #3, #4}} }
\newcommand{\sfuncone}[2]{S_{#1}^{\paren{#2}} }
\newcommand{\shatfuncone}[2]{\widehat S_{#1}^{\paren{#2}} }
\newcommand{\rfuncpre}[5]{R_{#1, #2}^{\paren{#3, #4, #5}}}
\newcommand{\rfunc}[4]{R_{#1}^{\paren{#2, #3, #4}} }
\newcommand{\ffunc}[4]{F_{#1}^{\paren{#2, #3, #4}} }
\newcommand{\qfamily}[2]{Q_{#1}^{#2}}

\newcommand{\rfuncone}[2]{R_{#1}^{\paren{#2}} }
\newcommand{\ffuncone}[2]{F_{#1}^{\paren{#2}} }
\newcommand{\Gsfunc}[4]{G^{\qty(#1, #2, #3)}(#4) }
\newcommand{\Gsfuncone}[2]{G^{\qty(#1)}(#2) }
\newcommand{\obA}{\overline{\boldsymbol{A}}}

\newcommand*{\mask}[2]{%
    \mathord{\makebox[\widthof{\(#1\)}]{\(#2\)}}%
}

\providecommand{\I}{}
\renewcommand{\I}{\mask{X}{I}}

\newcommand{\ad}[2]{\mathrm{Ad}_{#1}{\qty[#2]}}

\usepackage{tikz}
\usetikzlibrary{positioning}
\usepackage{calc}

\newcommand{\actmark}[1]{%
  \fboxsep=1pt\fboxrule=0.8pt%
  \fcolorbox{orange}{white}{$#1$}%
}

\newcommand{\admark}[2]{%
  \tikz[baseline=(X.base)]{%
    \node[inner sep=0pt] (X) {\actmark{#2}};%
    \pgfmathsetmacro{\upperwidth}{width("\actmark{#2}")}%
    \pgfmathsetmacro{\lowerwidth}{width("$\overline{#1}$")}%
    \pgfmathsetmacro{\myscale}{min(1, 1.25*\upperwidth/\lowerwidth)}%
    \node[below=0.3ex of X, inner sep=0pt, scale=\myscale] {$\overline{#1}$};%
  }%
}

\newlength{\admarkboxheight}
\newlength{\admarkboxoffset}
\newlength{\admarkbaseline}
\newcommand{\admarknew}[1]{%
  \tikz[baseline=\admarkbaseline]{%
    \pgfmathsetmacro{\boxwidth}{width("\actmark{A_1}")}%
    \pgfmathsetmacro{\lowerwidth}{width("$\overline{#1}$")}%
    \pgfmathsetmacro{\myscale}{min(1, 1.2*\boxwidth/\lowerwidth)}%
    \node[draw, dotted, orange, thick, inner sep=1pt, minimum width=0.7*\boxwidth, minimum height=\admarkboxheight] (X) {};%
    \node[below=\admarkboxoffset of X, inner sep=0pt, scale=\myscale] {$\overline{#1}$};%
  }%
}

\usepackage{tikz}
\usepackage{xfp}

\usetikzlibrary{calc,intersections,math,matrix}
\usetikzlibrary{arrows.meta}
\usetikzlibrary{shapes.misc}

\pgfdeclarelayer{front}
\pgfdeclarelayer{middle}
\pgfdeclarelayer{back}
\pgfdeclarelayer{deep}
\pgfsetlayers{deep,back,main,middle,front}
\usetikzlibrary{decorations.pathreplacing,positioning, fit, patterns,calc}
\usetikzlibrary{spy}

\pgfqkeys{/qkfig}{
    cell size/.initial = 0.425em,
    circle radius/.initial = 0.17em,
    circle fill color/.initial = darkgray,
    circle fill opacity/.initial = 0.7,
    circle line width/.initial = 0.6pt,
    cross size/.initial = 0.12em,
    cross line width/.initial = 1.1pt,
    cross color/.initial = black,
    axis line width/.initial = 0.8pt,
    arrow scale/.initial = 1.0,
    region color/.initial = RoyalBlue,
    region opacity/.initial = 0.2,
    label scale/.initial = 1.0,
    label offset/.initial = 0.6,
}

\newcommand{\qkget}[1]{\pgfkeysvalueof{/qkfig/#1}}

\newcommand{\qkset}[1]{\pgfqkeys{/qkfig}{#1}}

\newcommand{\cd}[2]{(#1*\qkget{cell size}, #2*\qkget{cell size})}

\newcommand{\sh}[2]{\cd{\fpeval{#2+0.5}}{\fpeval{#1-1.5}}}

\newcommand{\shcirc}[2]{%
    \draw[
        fill = \qkget{circle fill color},
        fill opacity = \qkget{circle fill opacity},
        draw = black,
        line width = \qkget{circle line width}
    ] \sh{#1}{#2} circle [radius=\qkget{circle radius}];
}

\newcommand{\shtimes}[2]{%
    \draw[
        \qkget{cross color},
        line width = \qkget{cross line width}
    ] \sh{#1}{#2} ++(-\qkget{cross size}, -\qkget{cross size})
                  -- ++( 2*\qkget{cross size},  2*\qkget{cross size})
      \sh{#1}{#2} ++(-\qkget{cross size},  \qkget{cross size})
                  -- ++( 2*\qkget{cross size}, -2*\qkget{cross size});
}

\newcommand{\shup}[2]{%
    \node[scale=\qkget{arrow scale}] at \sh{#1}{#2} {$\uparrow$};
}
\newcommand{\shdown}[2]{%
    \node[scale=\qkget{arrow scale}] at \sh{#1}{#2} {$\downarrow$};
}
\newcommand{\shright}[2]{%
    \node[scale=\qkget{arrow scale}] at \sh{#1}{#2} {$\rightarrow$};
}
\newcommand{\shleft}[2]{%
    \node[scale=\qkget{arrow scale}] at \sh{#1}{#2} {$\leftarrow$};
}

\newcommand{\drawQkStructure}[2][]{%
    \qkset{#1}%
    \pgfmathtruncatemacro{\k}{#2}%
    \pgfmathtruncatemacro{\kHalf}{\k/2}%
    \pgfmathtruncatemacro{\kHalfMinus}{(\k-1)/2}%
    \pgfmathtruncatemacro{\dh}{mod(\k,2)==1 ? 1 : 0}%
    \draw[line width=\qkget{axis line width}]
        \cd{0}{\fpeval{2*\k+2}} -- \cd{0}{2};
    \draw[line width=\qkget{axis line width}]
        \cd{\fpeval{\k+1}}{\fpeval{2*\k+1}} -- \cd{-1}{\fpeval{2*\k+1}};
    \ifnum\dh=1
        \shtimes{4}{0}
        \shtimes{6}{0}
    \fi
    \node[scale=\qkget{label scale}] at \sh{2*\k-0.85}{\k/2+3} {Structure of $\qfamily{\k}{F}$};
    \node[rotate=90, scale=\qkget{label scale}] at \sh{2*(\k/2+1.5)}{-2} {Length $l$};
    \node[scale=\qkget{label scale}] at \sh{2*\k+4}{(\k-1)/2} {Hole $m$};
    \foreach \n in {0,...,\fpeval{\kHalf-1-\dh}}{%
        \pgfmathtruncatemacro{\deltahole}{2*(\kHalfMinus-\n)}%
        \pgfmathtruncatemacro{\baseRow}{2*(\k-2*\n)}%
        \fill[\qkget{region color}, opacity=\qkget{region opacity}]
            \sh{\baseRow}{-0.5}
            -- \sh{\baseRow+1}{-0.5}
            -- \sh{\baseRow-\deltahole}{\deltahole+0.5}
            -- \sh{\baseRow-\deltahole-1}{\deltahole+0.5}
            -- cycle;
        \pgfmathtruncatemacro{\labelRow}{2*(\k - \n - \kHalf + 1)}%
        \pgfmathtruncatemacro{\labelCol}{2*(\kHalf - \n - 1)}%
        \node[anchor=west, scale=\qkget{label scale}] at \sh{\labelRow-0.25}{\labelCol + \qkget{label offset}-0.25} {$n=\n$};
    }%
    \foreach \n in {0,...,\fpeval{\kHalf-1-\dh}}{%
        \foreach \m in {0,...,\fpeval{\kHalf-\n-1}}{%
            \shcirc{\fpeval{2*(\k-2*\n-\m)}}{2*\m}%
        }%
    }%
    \foreach \n in {0,...,\fpeval{\kHalf-1-\dh}}{%
        \foreach \m in {0,...,\fpeval{\kHalf-\n-1}}{%
            \shup{\fpeval{2*(\k-2*\n-\m)+1}}{2*\m}%
        }%
    }%
    \foreach \n in {0,...,\fpeval{\kHalf-1}}{%
        \pgfmathtruncatemacro{\mMax}{\kHalfMinus-\n-1}%
        \ifnum\mMax<0\else
            \foreach \m in {0,...,\mMax}{%
                \shdown{\fpeval{2*(\k-2*\n-\m)-1}}{2*\m}%
            }%
        \fi
    }%
    \foreach \n in {0,...,\fpeval{\kHalf-1}}{%
        \pgfmathtruncatemacro{\mMax}{\kHalfMinus-\n-1}%
        \ifnum\mMax<0\else
            \foreach \m in {0,...,\mMax}{%
                \shright{\fpeval{2*(\k-2*\n-\m)}}{2*\m+1}%
            }%
        \fi
    }%
    \foreach \n in {0,...,\fpeval{\kHalf-1}}{%
        \pgfmathtruncatemacro{\mMax}{\kHalf-\n-1}%
        \ifnum\mMax<1\else
            \foreach \m in {1,...,\mMax}{%
                \shleft{\fpeval{2*(\k-2*\n-\m)}}{2*\m-1}%
            }%
        \fi
    }%
    \foreach \n in {0,...,\fpeval{\kHalfMinus-\dh}}{%
        \foreach \m in {0,...,\fpeval{\kHalfMinus-\n}}{%
            \shtimes{\fpeval{2*(\k-2*\n-\m+1)}}{2*\m}%
        }%
    }%
    \foreach \n in {2,...,\fpeval{\k+1}}{%
        \pgfmathtruncatemacro{\lengthnow}{\n-1}
        \node[scale=\qkget{label scale}] at \sh{\fpeval{2*\n}}{-1} {$\lengthnow$};
    }%
    \foreach \n in {0,...,\fpeval{\kHalfMinus-\dh}}{%
        \node[scale=\qkget{label scale}] at \sh{\fpeval{2*(\k+1)+1}}{2*\n} {$\n$};
    }%
}

\newcommand{\drawBasicStructure}{%
    \fill[\qkget{region color}, opacity=\qkget{region opacity}]
        \sh{6}{1.5} -- \sh{7}{1.5} -- \sh{4}{4.5} -- \sh{3}{4.5} -- cycle;
    \node[scale=0.7*\qkget{label scale}] at \sh{3}{1} {$n$};
    \fill[\qkget{region color}, opacity=\qkget{region opacity}]
        \sh{4}{-0.5} -- \sh{5}{-0.5} -- \sh{2}{2.5} -- \sh{1}{2.5} -- cycle;
    \node[scale=0.6*\qkget{label scale}] at \sh{5}{3} {$n\!-\!1$};
    \draw[gray, thin, dashed] \sh{0}{0} -- \sh{8}{0};
    \draw[gray, thin, dashed] \sh{0}{2} -- \sh{8}{2};
    \draw[gray, thin, dashed] \sh{0}{4} -- \sh{8}{4};
    \draw[gray, thin, dashed] \sh{2}{-1} -- \sh{2}{5};
    \draw[gray, thin, dashed] \sh{4}{-1} -- \sh{4}{5};
    \draw[gray, thin, dashed] \sh{6}{-1} -- \sh{6}{5};
    \node[scale=\qkget{label scale}] at \sh{2}{-1.5} {$l-1$};
    \node[scale=\qkget{label scale}] at \sh{4}{-1.5} {$l$};
    \node[scale=\qkget{label scale}] at \sh{6}{-1.5} {$l+1$};
    \node[scale=\qkget{label scale}] at \sh{8.5}{0} {$m-1$};
    \node[scale=\qkget{label scale}] at \sh{8.5}{2} {$m$};
    \node[scale=\qkget{label scale}] at \sh{8.5}{4} {$m+1$};
    \node[scale=\qkget{label scale}, rotate=90, anchor=south] at \sh{4}{-2.75} {Length};
    \node[scale=\qkget{label scale}] at \sh{9.5}{2} {Hole};
    \shtimes{4}{2}
    \shcirc{2}{2}
    \node[scale=1.3*\qkget{arrow scale}, font=\bfseries] at \sh{3}{2} {$\uparrow$};
    \shcirc{6}{2}
    \node[scale=1.3*\qkget{arrow scale}, font=\bfseries] at \sh{5}{2} {$\downarrow$};
    \shcirc{4}{0}
    \node[scale=1.5*\qkget{arrow scale}, font=\bfseries] at (1.5*\qkget{cell size}, 2.5*\qkget{cell size}-0.35) {$\rightarrow$};
    \shcirc{4}{4}
    \node[scale=1.3*\qkget{arrow scale}, font=\bfseries] at (3.5*\qkget{cell size}, 2.5*\qkget{cell size}-0.35) {$\leftarrow$};
}

\DeclareSymbolFont{usualmathcal}{OMS}{cmsy}{m}{n}
\DeclareSymbolFontAlphabet{\mathcal}{usualmathcal}

\begin{document}

\begin{center}{\Large \textbf{
            Matrix product operator representations for the local conserved quantities of the spin-$1/2$ XYZ chain
        }}\end{center}

\begin{center}
    Kohei Fukai\textsuperscript{1,2$\star$} and
    Kyoichi Yamada\textsuperscript{3}
\end{center}

\begin{center}
    {\bf 1} RIKEN iTHEMS, Wako, Saitama 351-0198, Japan
    \\
    {\bf 2} Department of Physics, Graduate School of Science, \\ The University of Tokyo, \\7-3-1, Hongo, Bunkyo-ku, Tokyo, 113-0033, Japan
    \\
    {\bf 3} The Institute for Solid State Physics, The University of Tokyo, \\Kashiwa, Chiba 277-8581, Japan
    \\
    \vspace{3pt}
    ${}^\star$ {\small \sf kohei.fukai@riken.jp}
\end{center}

\begin{center}
    \today
\end{center}

\section*{Abstract}
 {\bf
  We present explicit matrix product operator (MPO) representations for the local conserved quantities of the spin-$1/2$ XYZ chain.
  Through these MPO representations, we simplify the coefficients appearing in the local conserved quantities originally derived by one of the authors~\cite{Nozawa2020}, and reveal their combinatorial meaning: the coefficients prove to be a polynomial generalization of the Catalan numbers, given by weighted sums over monotone lattice paths.
  We also prove that a single recurrence relation for the coefficients guarantees the conservation law.
  We further show that these MPO representations can also be recovered from a product of two eight-vertex transfer matrices built from Baxter's R-matrix, providing the first direct connection between the R-matrix formalism and closed-form expressions for all local conserved quantities.
  We also obtain a simple Lax operator for the XYZ chain that, unlike Baxter's R-matrix, involves no elliptic functions.
 }

\vspace{10pt}
\noindent\rule{\textwidth}{1pt}
\tableofcontents\thispagestyle{fancy}
\noindent\rule{\textwidth}{1pt}
\vspace{10pt}

\section{Introduction}
\label{sec:intro}

Quantum integrable systems admit exact solutions via the Bethe ansatz~\cite{Bethe1931,Baxter1982,sutherland2004beautiful} and have been studied extensively in both physics and mathematics.
A defining characteristic of these systems is the existence of an extensive number of mutually commuting local conserved quantities $\{Q_k\}_{k=2,3,4,\ldots}$, which underlies their exact solvability.
The local conserved quantities are also expected to be useful for computing correlation functions~\cite{fukai-TL-correlation-2024}, which remains a central challenge even in integrable systems.

Despite their importance, explicit expressions for all the local conserved quantities $Q_k$ in an integrable system are difficult to obtain, even when an R-matrix satisfying the Yang-Baxter equation is known~\cite{GRABOWSKI1995299}.
    In the quantum inverse scattering method (QISM)~\cite{korepin_bogoliubov_izergin_1993}, the local conserved quantities are generated from the logarithmic expansion of the transfer matrix.
    For higher-order $Q_k$, however, this expansion generates a rapidly growing number of nonlocal contributions that must cancel among themselves, making the extraction of $Q_k$ increasingly impractical; see Appendix~\ref{app:log-transfer-comparison} for details.
    For this reason, general closed-form expressions for the local conserved quantities have often been derived by more direct methods~\cite{grabowski-xxx-1994,GRABOWSKI1995299,Nozawa2020,Nienhuis2021,fukai-hubbard-charge-2023,fukai-completeness-hubbard-2024,fukai-doctoral-arxiv,Fukai-Pozsgay-Vona-GFFI-2026,Fukai-Pozsgay-Vona-GFFII-2026}, and they remain largely model-specific.

Explicit expressions for local conserved quantities have been obtained for the Heisenberg (XXX) chain and its $\mathrm{SU}(N)$ generalization~\cite{anshelevich1980first,grabowski-xxx-1994,GRABOWSKI1995299}, the Temperley-Lieb models including the XXZ chain~\cite{Nienhuis2021}, the XYZ chain~\cite{Nozawa2020}, the one-dimensional Hubbard model~\cite{fukai-hubbard-charge-2023,fukai-completeness-hubbard-2024,fukai-doctoral-arxiv}, and solvable spin chains defined by graph Clifford algebras associated with claw-free frustration graphs~\cite{Fukai-Pozsgay-Vona-GFFI-2026,Fukai-Pozsgay-Vona-GFFII-2026}.
Even in these cases, however, the resulting expressions are often complicated, and it is difficult to discern any underlying pattern at first glance.

Matrix product operator (MPO) representations offer an efficient approach to understanding the structure of local conserved quantities.
In our previous work~\cite{yamada2023matrix}, we constructed MPO representations of the local conserved quantities of the Heisenberg (XXX) chain and its $\mathrm{SU}(N)$ generalization.
In this construction, the local MPO tensor can be represented as an upper-triangular matrix whose entries are local operators.
The local conserved quantities are then recovered by expanding the resulting matrix product.
Unlike the logarithmic expansion of the transfer matrix, this expansion produces no nonlocal contributions that would have to be cancelled.
Moreover, the MPO form makes manifest a combinatorial structure that is obscured in the bare operator expressions: in the XXX and $\mathrm{SU}(N)$ cases, ordinary Catalan numbers appear in the local tensor, and the generalized Catalan numbers appearing in the explicit expressions for the local conserved quantities~\cite{grabowski-xxx-1994,GRABOWSKI1995299} are recovered only after the matrix product is evaluated.

    Given this progress on the XXX and $\mathrm{SU}(N)$ chains~\cite{yamada2023matrix}, it is natural to ask whether the same MPO approach extends to the XYZ chain and how the underlying Catalan-number combinatorics generalizes to the fully anisotropic case.
    The XYZ chain is considerably less tractable than the XXX and XXZ chains: it lacks the $U(1)$ symmetry, so there is no reference state on which to build the Bethe ansatz.
    This complexity has motivated several recent, more elementary approaches to the XYZ chain, including spin-helix eigenstates and a simplified route to its Bethe ansatz~\cite{Zhang-Klumper-Popkov-open-XYZ-2022,Zhang-Klumper-Popkov-XYZ-pedestrian-2024}.
    Another such approach is a transfer matrix which does not involve elliptic functions and has a four-dimensional auxiliary space~\cite{Fendley-xyz-2025}.
    This construction is closely related to the strong zero mode~\cite{Fendley-XYZ-strong-zero-2016,Vernier-Yeh-Piroli-Mitra-SZM-2024,Gehrmann-Essler-SZM-2026,Essler-Fendley-Vernier-SZM-spinS-2026}.
    However, an MPO representation of the local conserved quantities of the XYZ chain has remained unavailable, even though explicit expressions for all of them were obtained in Ref.~\cite{Nozawa2020}.
    Consequently, the combinatorial structure underlying these coefficients has remained less transparent than that of the Catalan numbers arising in the XXX case, since the coefficients here are polynomials in the coupling constants.

In this work, we present MPO representations for the local conserved quantities of the spin-$1/2$ XYZ chain~\cite{baxter-prl-1971, baxter-1973-i, baxter-1973-ii, baxter-1973-iii, Takhtadzhan_Faddeev_xyz_1979}, extending the construction of Ref.~\cite{yamada2023matrix}.
Through these MPO representations, we simplify the coefficients appearing in the local conserved quantities originally derived in~\cite{Nozawa2020}, and reveal that they constitute a polynomial generalization of the Catalan numbers~\cite{stanley-catalan-2015}.
Specifically, these coefficients generalize the counting of $p$-good lattice paths studied by Hilton and Pedersen~\cite{hilton-catalan-1991} to weighted path enumeration, with $p=4$ corresponding to the XYZ chain.
    We also prove that the conservation of $Q_k$ follows from a single recurrence relation for its coefficients.
    This leads to a general construction in which the freedom to add lower-order local conserved quantities remains unfixed, a particular choice of which yields a closed-form expression alternative to that of~\cite{Nozawa2020}.

By introducing a spectral parameter, we compress the MPOs for all local conserved quantities into a single MPO with a four-dimensional auxiliary space.
    We show that this compressed MPO also arises from a product of two eight-vertex transfer matrices built from Baxter's R-matrix.
    The local conserved quantities appear only after one spectral parameter is moved off the fusion point.
    To our knowledge, this is the first direct connection between the R-matrix formalism and closed-form expressions for all local conserved quantities, which had previously been obtained only by direct methods that bypass QISM~\cite{grabowski-xxx-1994,GRABOWSKI1995299,Nozawa2020,Nienhuis2021,fukai-hubbard-charge-2023}.
From this compressed MPO, we obtain a simple Lax operator for the XYZ chain: it has a three-dimensional auxiliary space and, unlike Baxter's R-matrix~\cite{Baxter1982}, does not involve elliptic functions.

This paper is organized as follows.
Section~\ref{sec:recap-xyz} reviews the local conserved quantities of the spin-$1/2$ XYZ chain, introducing a simplified representation for their coefficients and showing that their conservation follows from a single recurrence relation.
Section~\ref{sec:combinatoric} shows that these coefficients admit a combinatorial interpretation as weighted sums over monotone lattice paths.
Section~\ref{sec:mpo-xyz} presents the main result: the MPO representation for the local conserved quantities of the spin-$1/2$ XYZ chain.
Section~\ref{sec:new_mpo} introduces a spectral parameter to compress the MPOs and derives a Lax operator for the XYZ chain without elliptic functions.
The compressed MPO also arises from a product of two eight-vertex transfer matrices.
Section~\ref{sec:conclusion} contains our conclusions.
The appendices provide detailed proofs of the identities and theorems stated in the main text.
    In particular, Appendix~\ref{app:commutativity-Qk} proves Theorem~\ref{thm:coefficient-recurrence-conservation}, which reduces the conservation law $\qty[Q_k, H] = 0$ to a single recurrence relation for the coefficients.

\section{Local conserved quantities of the spin-$1/2$ XYZ chain}
\label{sec:recap-xyz}
In this section, we present explicit expressions for the local conserved quantities of the spin-$1/2$ XYZ chain, originally obtained in~\cite{Nozawa2020}.
We also present a simplified form of their coefficients.

The Hamiltonian of the spin-$1/2$ XYZ chain is
\begin{equation}
    \label{eq:XYZhamiltonian}
    H
    =
    \sum_{i = 1}^{L}
    \qty[J_XX_{i}X_{i+1}+J_YY_{i}Y_{i+1}+J_ZZ_{i}Z_{i+1}]
    \,.
\end{equation}
Here, $X_i, Y_i, Z_i$ are the Pauli matrices acting nontrivially on the $i$-th site, and $L$ is the system size.
We assume periodic boundary conditions: $X_{i+L}=X_{i}$, etc.
The Hamiltonian~\eqref{eq:XYZhamiltonian} is integrable, as established by Baxter~\cite{baxter-prl-1971,baxter-1973-i,baxter-1973-ii,baxter-1973-iii} and reformulated within QISM by Takhtajan and Faddeev~\cite{Takhtadzhan_Faddeev_xyz_1979}.

It possesses a macroscopic number of mutually commuting local conserved quantities $\bce{Q_k}_{k=2,3,4,\ldots}$, with $Q_2=H$.
We define the support of a Pauli string as the number of sites in the smallest contiguous interval outside which it acts trivially.
Each $Q_k$ is a $k$-support conserved quantity: every term has support at most $k$, and at least one term has support exactly $k$.

\subsection{Doubling-product notation}
To express the explicit formula for $Q_k$, we introduce the doubling-product notation~\cite{Shiraishi2019, Nozawa2020}:
\begin{equation}
    \label{eq:doubling-def}
    \overline{A_{1} A_{2} \cdots A_{l}}
    \coloneqq
    \sum_{i=1}^{L}
    (A_1)_i(A_1A_2)_{i+1}(A_2A_3)_{i+2}\cdots (A_{ l-1}A_l)_{i+l-1}(A_l)_{i+l}
    \,,
\end{equation}
where each $A_{\alpha}\in\{X, Y, Z\}$ is one of the Pauli matrices and $A_{\alpha}A_{\alpha+1}$ denotes the product of $A_\alpha$ and $A_{\alpha+1}$.
The subscript $(\cdot)_{i}$ indicates the operator acting on the $i$-th site.
We define the length of a doubling product as the number of letters $A_{\alpha}$ appearing in the sequence; thus the doubling product in Eq.~\eqref{eq:doubling-def} has length $l$.
The support of a doubling product is equal to the length plus one.
    Each local conserved quantity $Q_k$ is a linear combination of doubling products, with coefficients determined by the letter sequence; the doubling-product notation makes this dependence explicit, in contrast to an expansion in bare Pauli strings.

The expression for $Q_k$ below is organized by the length and also by the number of holes, defined as the number of indices $\alpha$ satisfying $A_{\alpha}=A_{\alpha+1}$.
We introduce an abbreviated notation for the doubling product, into which the phase factor $(-\imi)^{t-1}$ is absorbed:
\begin{equation}
    \label{doubling-abbreviated}
    \overline{
        A^{m_1+1}_{1} A^{m_2+1}_{2} \cdots A^{m_t+1}_{t}
    }
    \coloneqq
    \overline{
        \underbrace{A_1\ldots A_1}_{m_1+1}
        \underbrace{A_2\ldots A_2}_{m_2+1}
        \ldots
        \underbrace{A_t\ldots A_t}_{m_t+1}
    }
    \times (-\imi)^{t-1}
    \,,
\end{equation}
where the letter $A_\alpha$ is repeated $m_\alpha+1$ times, with $A_{\alpha}\neq A_{\alpha+1}$ and $m_\alpha\geq 0$.
The number of holes in the doubling product of Eq.~\eqref{doubling-abbreviated} is $m = \sum_{\alpha=1}^{t} m_\alpha$.
In the following, we use the abbreviated notation~\eqref{doubling-abbreviated} for doubling products.

The Hamiltonian~\eqref{eq:XYZhamiltonian} is then rewritten as
\begin{equation}
    H
    =
    J_X \overline{X} + J_Y \overline{Y} + J_Z \overline{Z}
    \,.
\end{equation}
    The next local conserved quantity $Q_3$ is given by
    \begin{equation}
        Q_3
        =
        J_X J_Y (\overline{XY} + \overline{YX})
        +
        J_Y J_Z (\overline{YZ} + \overline{ZY})
        +
        J_Z J_X (\overline{ZX} + \overline{XZ})
        \,.
    \end{equation}

\subsection{Local conserved quantities and simplification of the coefficients}
Here we present the explicit expression for the local conserved quantities $Q_k$, originally obtained in~\cite{Nozawa2020}.
    We found that its coefficients admit a further simplification:
\begin{equation}
    \label{eq:xyz-Qk}
    Q_{k}
    =
    \sum_{\substack{0 \leq n+m < \floor{k/2} \\ n,m \ge 0}}
    \sum_{\obA\in\mathcal{S}_k^{n,m}}
    \rfunc{n}{N_x^{\boldsymbol{A}}+m}{N_y^{\boldsymbol{A}}+m}{N_z^{\boldsymbol{A}}+m}
    J_{\boldsymbol{A}}
    \,
    \obA
    \,,
\end{equation}
where $\mathcal{S}_k^{n,m}$ is the set of all doubling products with length $k-2n-m-1$ and $m$ holes.
Here $\obA \coloneqq \overline{A^{m_1+1}_{1} A^{m_2+1}_{2} \cdots A^{m_t+1}_{t}}$, where the number of letters is $t \coloneqq k-2(n+m)-1$ and the number of holes is $m \coloneqq \sum_{\alpha=1}^{t} m_\alpha$.
The coupling factor is $J_{\boldsymbol{A}} \coloneqq (J_X J_Y J_Z)^m \prod_{\alpha=1}^{t} J_{A_\alpha}^{1-m_\alpha}$.
The integers $N_\rho^{\boldsymbol{A}}$, with $\rho=x,y,z$, denote the numbers of $X$, $Y$, and $Z$ among the letters $A_{1} A_{2} \cdots A_{t}$, respectively.
    We use the form in Eq.~\eqref{eq:xyz-Qk} because its dependence on the letter sequence and the number of holes is contained in the combinations $N_\rho^{\boldsymbol A}+m$.
    This makes the combinatorial interpretation developed in Section~\ref{sec:combinatoric} transparent.

The coefficient $\rfunc{n}{N_x}{N_y}{N_z}$ is given by
\begin{align}
    \label{eq:rfunc}
    \rfunc{n}{N_x}{N_y}{N_z}
     & \coloneqq
    4\sfunc{n}{N_x+n}{N_y+n}{N_z+n}
    -
    \sfunc{n}{N_x+n+1}{N_y+n}{N_z+n}
    \nonumber        \\
     & \hspace*{6em}
    -
    \sfunc{n}{N_x+n}{N_y+n+1}{N_z+n}
    -
    \sfunc{n}{N_x+n}{N_y+n}{N_z+n+1}
    \,,
\end{align}
where $\sfunc{n}{N_x}{N_y}{N_z}$ is defined as
\begin{equation}
    \label{eq:sfunc}
    \sfunc{n}{N_x}{N_y}{N_z}
    \coloneqq
    \sum_{\substack{n_x+n_y+n_z=n \\ n_x, n_y, n_z \geq 0}}
    \binom{n_x+N_x-1}{n_x}
    \binom{n_y+N_y-1}{n_y}
    \binom{n_z+N_z-1}{n_z}
    x^{n_x} y^{n_y} z^{n_z}
    \,.
\end{equation}
Here $\binom{n}{m} \equiv \frac{n!}{(n-m)! m!}$ is the binomial coefficient, and we abbreviate the squared couplings as
\begin{align}
    \label{eq:coupling-squared}
    x \coloneqq J_X^{2}\,,\qquad y \coloneqq J_Y^{2}\,,\qquad z \coloneqq J_Z^{2}\,.
\end{align}
Note that $\sfunc{0}{N_x}{N_y}{N_z} = 1$, $\sfunc{n}{0}{0}{0} = 0$ for $n>0$, and $\rfunc{0}{N_x}{N_y}{N_z} = 1$.
We also have $\rfunc{n}{0}{0}{0} = 0$ for $n>0$, which is proved in Appendix~\ref{app:sfunc-identities}.

The local conserved quantities satisfy the conservation law $\qty[Q_k, H] = 0$ and mutually commute, $\qty[Q_k, Q_l] = 0$ for $k, l = 2,3,4,\ldots$~\cite{Nozawa2020}.
    Section~\ref{subsec:recurrence-conserved-quantities} derives this conservation law from a recurrence relation for the coefficients.

\subsection{Isotropic limit of the coefficient}
The isotropic limit $J_X = J_Y = J_Z = 1$ of the coefficients $\rfunc{n}{N_x}{N_y}{N_z}$ does not coincide with those in the previous study for the XXX case~\cite{yamada2023matrix}, which involve the Catalan numbers~\cite{grabowski-xxx-1994}.
This is because the freedom to add lower-order local conserved quantities differs from that in~\cite{yamada2023matrix}.

The coefficient in the isotropic case is
\begin{equation}
    \label{eq:rfunc-isotropic}
    \rfunc{n}{N_x}{N_y}{N_z}
    \big|_{J_X=J_Y=J_Z=1}
    =
    \frac{m}{4n+m}
    \binom{4n+m}{n}
    \,,
\end{equation}
where $m = N_x + N_y + N_z$ and $(n,m) \neq (0,0)$.
We write $R^{n,m}$ for this isotropic value, with $R^{0,0} \equiv \rfunc{0}{0}{0}{0} = 1$.
In this limit, the recurrence relation~\eqref{eq:R-identity-diff}, derived in Section~\ref{sec:combinatoric}, reduces to $R^{n,m} = R^{n,m-1} + R^{n-1,m+3}$.
    For comparison, the ordinary Catalan numbers $C_n = \frac{1}{n+1}\binom{2n}{n}$ count monotone lattice paths that stay below a diagonal~\cite{stanley-catalan-2015}.
    The isotropic limit~\eqref{eq:rfunc-isotropic} coincides with the $p$-good lattice-path counts of Hilton and Pedersen~\cite{hilton-catalan-1991} for $p=4$; a $p$-good path is a monotone lattice path constrained by a line of slope $p-1$ and the case $p=2$ gives the ordinary Catalan numbers.

In Section~\ref{sec:combinatoric}, we generalize the result of Ref.~\cite{hilton-catalan-1991} to a weighted enumeration of $4$-good paths, where the weights encode the anisotropy of the couplings.

    \subsection{A general formulation of local conserved quantities}
    \label{subsec:recurrence-conserved-quantities}

    Equation~\eqref{eq:xyz-Qk} fixes the freedom to add lower-order local conserved quantities in one particular way.
    Replacing the coefficient $\rfunc{n}{N_x}{N_y}{N_z}$ by an arbitrary function $\ffunc{n}{N_x}{N_y}{N_z}$ of $n\in\mathbb Z$ and $(N_x,N_y,N_z)\in\mathbb Z^3$, we define
    \begin{align}
        \label{eq:Qk-F}
        \qfamily{k}{F}
         & \coloneqq
        \sum_{\substack{0 \leq n+m < \floor{k/2} \\ n,m \ge 0}}
        \sum_{\obA\in\mathcal{S}_k^{n,m}}
        \ffunc{n}{N_x^{\boldsymbol{A}}+m}{N_y^{\boldsymbol{A}}+m}{N_z^{\boldsymbol{A}}+m}
        J_{\boldsymbol{A}}
        \,
        \obA
        \,,
    \end{align}
    where $\mathcal{S}_k^{n,m}$, $J_{\boldsymbol{A}}$, and $N_\rho^{\boldsymbol{A}}$ are as in Eq.~\eqref{eq:xyz-Qk}.

    The following theorem gives a condition on $F$ under which $\qfamily{k}{F}$ is conserved.
    Here $\vec N=(N_x,N_y,N_z)$, $\ffuncone{n}{\vec N}\coloneqq\ffunc{n}{N_x}{N_y}{N_z}$, $\vec 1=(1,1,1)$, $\vec 0=(0,0,0)$, and $\vec e_A$ is the unit vector with $1$ in the component corresponding to $A\in\{X,Y,Z\}$.
    \begin{thm}
        \label{thm:coefficient-recurrence-conservation}
        Suppose $\ffuncone{n}{\vec N}=0$ for $n<0$,
        $\ffuncone{0}{\vec 0}\neq0$, and $F$ satisfies the recurrence relation
        \begin{equation}
            \label{eq:F-identity-diff}
            \ffuncone{n}{\vec N}
            =
            \ffuncone{n}{\vec N-\vec e_A}
            +
            J_A^2\ffuncone{n-1}{\vec N+\vec 1}
            \qquad
            \left(
            A\in\{X,Y,Z\}
            \right)
            \,.
        \end{equation}
        Then, for every $2\leq k\leq L$, $\qfamily{k}{F}$ is a
        $k$-support conserved quantity:
        \begin{equation}
            \label{eq:F-coefficient-conservation}
            \qty[\qfamily{k}{F},H]=0
            \,.
        \end{equation}
    \end{thm}
    The proof of Theorem~\ref{thm:coefficient-recurrence-conservation} is given in Appendix~\ref{app:commutativity-Qk}.
    The function $F=R$ satisfies Eq.~\eqref{eq:F-identity-diff}.
    For this choice, Eq.~\eqref{eq:Qk-F} reduces to Eq.~\eqref{eq:xyz-Qk}, so that $Q_k=\qfamily{k}{R}$, and Theorem~\ref{thm:coefficient-recurrence-conservation} proves $\qty[Q_k,H]=0$.

    The values $\ffuncone{a}{\vec 0}$ determine $F$, and each $\qfamily{k}{F}$ is a linear combination of $\qfamily{q}{R}$ with $q\le k$; Appendix~\ref{app:F-recurrence-boundary-solution-proof} gives the explicit relations.

    As a concrete alternative to the choice $F=R$, we construct local conserved quantities involving only the functions $S_n$.
    Define $\widehat S$ by
    \begin{equation}
        \label{eq:shifted-S-definition}
        \shatfuncone{n}{\vec N}
        \coloneqq
        \sfuncone{n}{\vec N+n\vec 1}
        =
        \sfunc{n}{N_x+n}{N_y+n}{N_z+n}
        \,.
    \end{equation}
    The recursion relations~\eqref{eq:sfunc_formula_recursion_x}--\eqref{eq:sfunc_formula_recursion_z} in Appendix~\ref{app:sfunc-identities} show that $\widehat S$ satisfies Eq.~\eqref{eq:F-identity-diff}.
    For $F=\widehat S$, Eq.~\eqref{eq:Qk-F} becomes
    \begin{equation}
        \label{eq:S-coefficient-Qhat}
        \qfamily{k}{\widehat S}
        =
        \sum_{\substack{0 \leq n+m < \floor{k/2} \\ n,m \ge 0}}
        \sum_{\obA\in\mathcal{S}_k^{n,m}}
        \sfunc{n}{N_x^{\boldsymbol{A}}+n+m}{N_y^{\boldsymbol{A}}+n+m}{N_z^{\boldsymbol{A}}+n+m}
        J_{\boldsymbol{A}}
        \,
        \obA
        \,.
    \end{equation}
    Thus $\qfamily{k}{\widehat S}$ involves only the functions $S_n$, and Theorem~\ref{thm:coefficient-recurrence-conservation} gives $\qty[\qfamily{k}{\widehat S},H]=0$.
    Equation~\eqref{eq:F-Q-triangular} gives the unitriangular relation
    \begin{equation}
        \label{eq:Qhat-triangular}
        \qfamily{k}{\widehat S}
        =
        \sum_{a=0}^{\floor{(k-2)/2}}
        \sfunc{a}{a}{a}{a}
        \qfamily{k-2a}{R}
        \,.
    \end{equation}
    In particular, $\qfamily{2}{\widehat S}=\qfamily{2}{R}=H$ and $\qfamily{4}{\widehat S}=\qfamily{4}{R}+(J_X^2+J_Y^2+J_Z^2)\qfamily{2}{R}$.
    Because Eq.~\eqref{eq:Qhat-triangular} is unitriangular, each $\qfamily{k}{\widehat S}$ is a linear combination of the $\qfamily{j}{R}$, and conversely.

\section{Combinatorics of the coefficients in the local conserved quantities}\label{sec:combinatoric}
In this section, we describe the combinatorial structure of the coefficients $\sfunc{n}{N_x}{N_y}{N_z}$ and $\rfunc{n}{N_x}{N_y}{N_z}$ defined in Eqs.~\eqref{eq:rfunc} and~\eqref{eq:sfunc}.
The combinatorial picture also yields two identities for $\rfunc{n}{N_x}{N_y}{N_z}$, which are used later in the proofs of the conservation law $\qty[Q_k, H] = 0$ and of the MPO representation in Section~\ref{sec:mpo-xyz}.

\begin{figure}[tbh]
    \centering
    \begin{tikzpicture}[scale=0.8]

        \begin{scope}[xshift=0cm]
            \def\ncols{6}
            \def\nrows{8}

            \definecolor{path1}{RGB}{220,80,60}

            \node[font=\small] at (3, \nrows+1.2) {(a) General case};

            \foreach \x in {0,...,\ncols} {
                    \draw[black!70] (\x, 0) -- (\x, \nrows);
                }

            \foreach \y in {0,...,\nrows} {
                    \draw[black!70, thick] (0, \y) -- (\ncols, \y);
                }

            \node[left, font=\small] at (-0.2, 0) {$\alpha_1$};
            \node[left, font=\small] at (-0.2, 1) {$\alpha_2$};
            \node[left, font=\small] at (-0.2, 2) {$\alpha_3$};
            \node[left, font=\small] at (-0.2, 3) {$\alpha_4$};
            \draw[dotted, thick] (-0.5, 3.75) -- (-0.5, 4.5){};
            \node[left, font=\small] at (-0.2, 5) {$\alpha_{j}$};
            \draw[dotted, thick] (-0.5, 5.75) -- (-0.5, 6.5){};
            \node[left, font=\small] at (-0.2, 7) {$\alpha_{N-1}$};
            \node[left, font=\small] at (-0.2, 8) {$\alpha_N$};

            \draw[decorate, decoration={brace, amplitude=10pt}]
            (-1.35, -0.3) -- (-1.35, \nrows+0.3);
            \node[left, font=\footnotesize, align=left] at (-1.75, 4) {%
                $N_x$ rows of $x$\\[2pt]
                $N_y$ rows of $y$\\[2pt]
                $N_z$ rows of $z$
            };

            \draw[decorate, decoration={brace, amplitude=8pt, mirror}]
            (0, -0.45) -- (\ncols, -0.45);
            \node[below, font=\normalsize] at (3, -0.8) {$n$ East steps};

            \draw[path1, ultra thick, line cap=round, line join=round]
            (0,0) -- (1,0) -- (1,2) -- (3,2) -- (3,5) -- (4,5) -- (4,7) -- (5,7) -- (5,8) -- (6,8);

            \fill[black] (0,0) circle (4pt);
            \fill[black] (\ncols,\nrows) circle (4pt);
            \node[below left, font=\small] at (0,-0.3) {$(0,1)$};
            \node[above, font=\small] at (\ncols,\nrows) {$(n, N)$};
        \end{scope}

        \draw[dashed] (7, -1.5) -- (7, 10);

        \begin{scope}[xshift=10cm, yshift=6cm]
            \def\ncols{2}
            \def\nrows{2}

            \definecolor{pathRed}{RGB}{220,80,60}

            \node[font=\small] at (1, \nrows+1.2) {(b) Example: $\sfunc{2}{2}{1}{0}$};

            \foreach \x in {0,...,\ncols} {
                    \draw[black!70] (\x, 0) -- (\x, \nrows);
                }

            \foreach \y in {0,...,\nrows} {
                    \draw[black!70, thick] (0, \y) -- (\ncols, \y);
                }

            \node[left, font=\small] at (-0.1, 0) {$x$};
            \node[left, font=\small] at (-0.1, 1) {$x$};
            \node[left, font=\small] at (-0.1, 2) {$y$};

            \draw[pathRed, very thick, line cap=round, line join=round]
            (0,0) -- (0,1) -- (1,1) -- (1,2) -- (2,2);

            \fill[black] (0,0) circle (3pt);
            \fill[black] (\ncols,\nrows) circle (3pt);

            \node[font=\small, align=left] at (1, -1.0) {%
                \textcolor{pathRed}{\raisebox{1.75pt}{\rule{0.5cm}{1.25pt}}} : example path (weight $xy$)
            };

            \node[font=\small, align=center] at (1, -4) {%
                \begin{tabular}{c|c}
                    weight & \# paths \\ \hline
                    $x^2$  & 3        \\
                    $xy$   & 2        \\
                    $y^2$  & 1
                \end{tabular}
            };

        \end{scope}

    \end{tikzpicture}
    \caption{
        Combinatorial representation of $\sfunc{n}{N_x}{N_y}{N_z}$ as a weighted sum over monotone lattice paths.
        (a) General case:
        The grid has width $n$ and $N = N_x + N_y + N_z$ rows.
        Each row $j$ is assigned a flavor $\alpha_j \in \{x, y, z\}$, with $N_\alpha$ rows of flavor $\alpha$ for each $\alpha \in \{x, y, z\}$.
        Paths run from $(0,1)$ to $(n, N)$, with each East step along row $j$ contributing a factor $\alpha_j$ to the path weight.
        The red line illustrates one path, with weight $\alpha_1 \alpha_3^2 \alpha_j \alpha_{N-1} \alpha_N$.
        (b) Example: $n=2$, $(N_x, N_y, N_z) = (2,1,0)$.
        The table enumerates all path weights and their multiplicities.
        Summing all path contributions gives $\sfunc{2}{2}{1}{0} = 3x^2 + 2xy + y^2$.
    }
    \label{fig:Sfunc-combinatoric}
\end{figure}

\subsection{Combinatorial representation of $\sfunc{n}{N_x}{N_y}{N_z}$}\label{sec:combinatoric-sfunc}
We first show that $\sfunc{n}{N_x}{N_y}{N_z}$, defined in Eq.~\eqref{eq:sfunc}, can be represented as a weighted sum over monotone lattice paths.

Consider a rectangular grid of width $n$ with $N \coloneqq N_x + N_y + N_z$ rows, as illustrated in Fig.~\ref{fig:Sfunc-combinatoric}(a).
Each row is assigned a \emph{flavor} $\alpha \in \{x, y, z\}$, such that exactly $N_\alpha$ rows have flavor $\alpha$.
    The weighted sum defined below depends only on $(N_x, N_y, N_z)$, not on the specific arrangement of flavors among the rows.
A \emph{monotone lattice path} is a sequence of unit steps from $(0,1)$ to $(n, N)$, where each step moves either to the right (East) or upward (North).
A concrete example with $n = 2$ and $(N_x, N_y, N_z) = (2,1,0)$ is shown in Fig.~\ref{fig:Sfunc-combinatoric}(b).

An East step along a row of flavor $\alpha \in \{x, y, z\}$ contributes a factor of $\alpha$ to the path weight.
The total weight of a path $\gamma$ is the product over all its East steps:
\begin{equation}
    w(\gamma) = x^{n_x(\gamma)} y^{n_y(\gamma)} z^{n_z(\gamma)},
\end{equation}
where $n_\alpha(\gamma)$ denotes the number of East steps taken along rows of flavor $\alpha$.

The coefficient $\sfunc{n}{N_x}{N_y}{N_z}$ equals the sum of weights over all monotone lattice paths from $(0,1)$ to $(n, N)$:
\begin{equation}
    \sfunc{n}{N_x}{N_y}{N_z} = \sum_{\gamma : \text{path}} w(\gamma).
\end{equation}
To verify this identity, note that paths with exactly $n_\alpha$ East steps along rows of flavor $\alpha$ contribute weight $x^{n_x} y^{n_y} z^{n_z}$.
The number of such paths is $\binom{n_x + N_x - 1}{N_x - 1} \binom{n_y + N_y - 1}{N_y - 1} \binom{n_z + N_z - 1}{N_z - 1}$.
Summing over all $(n_x, n_y, n_z)$ with $n_x + n_y + n_z = n$ recovers Eq.~\eqref{eq:sfunc}.

\begin{figure}[tbh]
    \centering
    \begin{tikzpicture}[scale=0.75]

        \begin{scope}[xshift=0cm]
            \def\ncols{6}      %
            \def\nrowsbelow{6} %
            \def\nbundles{6}   %

            \pgfmathsetmacro{\totalrows}{\nrowsbelow + \nbundles}

            \definecolor{pathRed}{RGB}{220,80,60}
            \definecolor{forbiddenRegion}{RGB}{245,240,250}
            \definecolor{bundleColor}{RGB}{0,0,50}
            \definecolor{flavorX}{RGB}{200,60,60}
            \definecolor{flavorY}{RGB}{50,150,50}
            \definecolor{flavorZ}{RGB}{50,50,200}

            \fill[forbiddenRegion]
            (0, \nrowsbelow) -- (\ncols, \totalrows) -- (\ncols, \totalrows+0.1)
            -- (0, \totalrows+0.1) -- cycle;

            \foreach \x in {0,...,\ncols} {
                    \draw[black!40] (\x, 0) -- (\x, \nrowsbelow);
                }
            \foreach \y in {0,...,\nrowsbelow} {
                    \draw[black!70, thick] (0, \y) -- (\ncols, \y);
                }

            \foreach \i in {1,...,\nbundles} {
                    \pgfmathsetmacro{\ypos}{\nrowsbelow + \i}
                    \draw[bundleColor, line width=0.8pt, double, double distance=0.75pt]
                    (0, \ypos) -- (\ncols, \ypos);
                }

            \foreach \x in {0,...,\ncols} {
                    \draw[black!25] (\x, \nrowsbelow) -- (\x, \totalrows);
                }

            \draw[blue!70!black, line width=1.8pt] (0, \nrowsbelow) -- (\ncols, \totalrows);
            \node[blue!70!black, font=\footnotesize, rotate=45, fill=white, inner sep=2pt, rounded corners=1pt] at (3.7, 10.5) {Forbidden region};

            \draw[decorate, decoration={brace, amplitude=10pt, mirror}]
            (-0.3, \nrowsbelow+0.2) -- (-0.3, -0.2);
            \node[left, font=\footnotesize, align=left] at (-0.8, 3) {%
                $N_x$ rows of $x$\\[2pt]
                $N_y$ rows of $y$\\[2pt]
                $N_z$ rows of $z$
            };

            \draw[decorate, decoration={brace, amplitude=10pt, mirror}]
            (-0.3, \totalrows+0.25) -- (-0.3, \nrowsbelow+1-0.25);
            \node[left, font=\footnotesize, align=center] at (-0.8, 9.5) {$n$ bundles};

            \draw[decorate, decoration={brace, amplitude=8pt, mirror}]
            (0, -0.55) -- (\ncols, -0.55);
            \node[below, font=\normalsize] at (3, -1.0) {$n$ East steps};

            \draw[pathRed, ultra thick, line cap=round, line join=round]
            (0,0) -- (1,0) -- (1,2) -- (3,2) -- (3,5) -- (4,5) -- (4,6)
            -- (4,7) -- (5,7) -- (5,9) -- (6,9) -- (6,12);

            \fill[black] (0,0) circle (4pt);
            \fill[black] (\ncols,\totalrows) circle (4pt);
            \node[below left, font=\small] at (0,-0.2) {$(0,1)$};
            \node[above = 0.35cm, font=\small] at (\ncols,\totalrows) {$(n, N\!+\!3n)$};

            \node[right, font=\small] at (\ncols+0.1, \nrowsbelow) {$(n, N)$};
            \fill[black] (\ncols, \nrowsbelow) circle (4pt);

            \def\smallR{0.4}
            \def\smallX{2.1}
            \def\smallY{7.9}
            \draw[thick, black] (\smallX, \smallY) circle (\smallR cm);

        \end{scope}

        \def\largeX{8.5}
        \def\largeY{8}
        \draw[thick, black] (\largeX, \largeY) circle (1.31 cm);

        \draw[black] (2, 8.3) -- (8.25, 9.3);
        \draw[black] (2, 7.5) -- (8.25, 6.7);

        \begin{scope}[xshift=\largeX cm, yshift=\largeY cm, scale=0.4]
            \definecolor{flavorX}{RGB}{0,0,60}
            \definecolor{flavorY}{RGB}{0,0,50}
            \definecolor{flavorZ}{RGB}{0,0,50}
            \definecolor{forbiddenRegion}{RGB}{245,240,250}

            \fill[white] (0,0) circle (3.2cm);

            \clip (0,0) circle (3.2cm);

            \fill[forbiddenRegion] (-4.5, -1.6) -- (-4.5, 4) -- (4, 4) -- (1.5, 4.4) -- cycle;

            \draw[flavorX, line width=1pt] (-3.5, 1.5) -- (3.5, 1.5);
            \draw[flavorY, line width=1pt] (-3.5, 0) -- (3.5, 0);
            \draw[flavorZ, line width=1pt] (-3.5, -1.5) -- (3.5, -1.5);

            \draw[black!30, line width=0.8pt] (-1.5, -3) -- (-1.5, 3);

            \node[font=\small, flavorX, fill=white, inner sep=1pt] at (1.75, 1.5) {$x$};
            \node[font=\small, flavorY, fill=white, inner sep=1pt] at (1.75, 0) {$y$};
            \node[font=\small, flavorZ, fill=white, inner sep=1pt] at (1.75, -1.5) {$z$};

            \draw[blue!70!black, line width=2pt] (-4.5, -1.5) -- (1.5, 4.5);

        \end{scope}
        \draw[decorate, decoration={brace, amplitude=5pt}]
        (10, 8.8) -- (10, 7.2);
        \node[right, font=\footnotesize, align=left] at (10.25, 8) {1 bundle};

    \end{tikzpicture}
    \caption{
        Combinatorial representation of $\rfunc{n}{N_x}{N_y}{N_z}$ as a weighted sum over good paths.
        The lower region is a rectangular grid identical to Fig.~\ref{fig:Sfunc-combinatoric}, with $N = N_x + N_y + N_z$ flavored rows.
        The upper region contains $n$ bundles, each bundle consisting of three rows $\{x, y, z\}$ shown as double lines.
        An East step taken within a bundle runs along exactly one of its three rows and contributes that row's flavor to the weight, just as in the lower region.
        The diagonal line connects $(0, N)$ and $(n, N+3n)$ with slope $3$; good paths must stay weakly below this line, and the forbidden region above it is shaded.
        The red path illustrates an example of a good path.
        The magnified view shows the internal structure of one bundle.
    }
    \label{fig:Rfunc-combinatoric}
\end{figure}

\begin{figure}
    \centering
    \begin{tikzpicture}

        \begin{scope}[scale=0.7]

            \def\nexample{2}
            \def\Nexample{3}
            \pgfmathsetmacro{\totalex}{8}

            \definecolor{forbiddenRegion}{RGB}{245,240,250}

            \node[font=\small] at (1, \totalex+1.5) {Example: $\rfunc{2}{2}{1}{0}$};

            \fill[forbiddenRegion] (0, 2) -- (0, \totalex) -- (\nexample, \totalex) -- (\nexample, \totalex) -- cycle;

            \foreach \x in {0,...,\nexample} {
                    \draw[black!40] (\x, 0) -- (\x, \Nexample);
                }
            \foreach \y in {0,...,\Nexample} {
                    \draw[black, thick] (0, \y) -- (\nexample, \y);
                }

            \foreach \y in {3,...,8} {
                    \draw[black, thick] (0, \y) -- (\nexample, \y);
                }

            \foreach \x in {0,...,\nexample} {
                    \draw[black!30] (\x, \Nexample) -- (\x, \totalex);
                }

            \draw[blue!70!black, thick] (0, 2) -- (\nexample, \totalex);

            \node[left, font=\small] at (-0.1, 0) {$x$};
            \node[left, font=\small] at (-0.1, 1) {$x$};
            \node[left, font=\small] at (-0.1, 2) {$y$};
            \node[left, font=\small] at (-0.1, 3) {$z$};
            \node[left, font=\small] at (-0.1, 4) {$y$};
            \node[left, font=\small] at (-0.1, 5) {$x$};
            \node[left, font=\small] at (-0.1, 6) {$z$};
            \node[left, font=\small] at (-0.1, 7) {$y$};
            \node[left, font=\small] at (-0.1, 8) {$x$};

            \draw[decorate, decoration={brace, amplitude=5pt, mirror}]
            (-0.65, 5.25) -- (-0.65, 2.75);
            \node[left, font=\scriptsize] at (-0.8, 4) {bundle 1};

            \draw[decorate, decoration={brace, amplitude=5pt, mirror}]
            (-0.65, 8.25) -- (-0.65, 5.75);
            \node[left, font=\scriptsize] at (-0.8, 7) {bundle 2};

            \draw[red, very thick, line cap=round, line join=round]
            (0,0) -- (0,2) -- (1,2) -- (1,5) -- (2,5) -- (2,8);

            \fill[black] (0,0) circle (3pt);
            \fill[black] (\nexample,\totalex) circle (3pt);
            \fill[black] (\nexample, 2) circle (3pt);
            \node[below left, font=\footnotesize] at (0,0) {$(0,1)$};
            \node[above right, font=\footnotesize] at (\nexample,\totalex) {$(2,9)$};
            \node[right, font=\footnotesize] at (\nexample+0.1, 2) {$(2,3)$};

            \node[font=\small, align=left] at (8, 8) {%
                \textcolor{red}{\raisebox{1.75pt}{\rule{0.5cm}{1.25pt}}} : example path (weight $xy$)
            };

            \node[font=\small, align=center] at (8, 4) {%
                \renewcommand{\arraystretch}{1.4}%
                \begin{tabular}{c|c}
                    weight & \# paths \\ \hline
                    $x^2$  & 5        \\
                    $xy$   & 5        \\
                    $xz$   & 2        \\
                    $y^2$  & 2        \\
                    $yz$   & 1
                \end{tabular}
            };

        \end{scope}
    \end{tikzpicture}
    \caption{
        Example of $\rfunc{2}{2}{1}{0}$ with $n=2$, $N_x = 2$, $N_y = 1$, and $N_z = 0$.
        The diagonal line connects $(0, N) = (0, 3)$ and $(n, N+3n) = (2, 9)$.
        Good paths must stay weakly below the diagonal, and the forbidden region is shaded.
        The table enumerates all path weights and their multiplicities.
        Summing all path contributions gives $\rfunc{2}{2}{1}{0} = 5x^2 + 5xy + 2y^2 + 2xz + yz$.
        The diagonal constraint prevents paths from taking an East step along any row of bundle 2.
    }
    \label{fig:Rfunc-example}
\end{figure}

\subsection{Combinatorial representation of $\rfunc{n}{N_x}{N_y}{N_z}$}\label{sec:combinatoric-rfunc}
We next show that $\rfunc{n}{N_x}{N_y}{N_z}$, defined in Eq.~\eqref{eq:rfunc}, can be expressed as a weighted generalization of $p$-good path counting with $p = 4$~\cite{hilton-catalan-1991}.

The lattice structure for $\rfunc{n}{N_x}{N_y}{N_z}$ is illustrated in Fig.~\ref{fig:Rfunc-combinatoric}.
It consists of two regions: a lower rectangular region and an upper diagonal region.
The lower region is identical to the lattice for $\sfunc{n}{N_x}{N_y}{N_z}$ described in the previous subsection, containing $N = N_x + N_y + N_z$ rows with the corresponding flavors.
The upper region contains $n$ bundles, where each bundle consists of three rows with flavors $x$, $y$, and $z$.
Neither the arrangement of flavors in the lower region nor the ordering of the three rows within each bundle affects the weighted sum defined below.

A \emph{good path} is a monotone lattice path from $(0, 1)$ to $(n, N + 3n)$, consisting of $N + 3n - 1$ North steps and $n$ East steps.
In addition, a good path must stay weakly below the diagonal line connecting $(0, N)$ and $(n, N + 3n)$.

The weight $w(\gamma)$ is the product of the row flavors traversed by its $n$ East steps, hence a monomial of degree $n$ in $x$, $y$, and $z$.

The coefficient $\rfunc{n}{N_x}{N_y}{N_z}$ is given by
\begin{equation}
    \label{eq:rfunc-combinatoric}
    \rfunc{n}{N_x}{N_y}{N_z} = \sum_{\gamma:\,\text{good path}} w(\gamma).
\end{equation}
Here the sum runs over all good paths $\gamma$ from $(0,1)$ to $(n, N+3n)$.
An example is shown in Fig.~\ref{fig:Rfunc-example} for the case $n = 2$ and $(N_x, N_y, N_z) = (2,1,0)$.

The equivalence between the combinatorial representation~\eqref{eq:rfunc-combinatoric} and the original definition~\eqref{eq:rfunc} is proved in Appendix~\ref{app:rfunc-combinatoric}.

\subsection{Identities for the coefficient $\rfunc{n}{N_x}{N_y}{N_z}$}

The coefficient $\rfunc{n}{N_x}{N_y}{N_z}$ satisfies two identities analogous to those of the Catalan numbers.
    The first is the recurrence relation used in Theorem~\ref{thm:coefficient-recurrence-conservation}; the second is a convolution formula used in the proof of Theorem~\ref{thm:Qk-MPO}.

Setting $\rfunc{n}{N_x}{N_y}{N_z}=0$ for $n<0$, the recurrence relation is
\begin{align}
    \label{eq:R-identity-diff}
    \rfunc{n}{N_x}{N_y}{N_z}
     & =
    \rfunc{n}{N_x-1}{N_y}{N_z}
    +
    x \, \rfunc{n-1}{N_x+1}{N_y+1}{N_z+1}
    \,.
\end{align}
Analogous identities hold with $(N_y, y)$ or $(N_z, z)$ in place of $(N_x, x)$.
The identity follows from a first-step decomposition: since the arrangement of flavors is arbitrary, take the bottom row of the lower region to have flavor $x$.
The first term corresponds to paths whose first step is North, skipping this row.
The second term corresponds to paths whose first step is East, gaining weight $x$; the $+1$ in each upper index arises from absorbing one bundle into the lower region.
    Appendix~\ref{app:R-identity-diff-proof} gives an alternative proof of Eq.~\eqref{eq:R-identity-diff}.

The second identity is the convolution formula:
\begin{equation}
    \label{eq:R-identity}
    \sum_{\ntil = 0}^{n}
    \rfunc{\ntil}{N_x}{N_y}{N_z}
    \rfunc{n-\ntil}{N^\prime_x}{N^\prime_y}{N^\prime_z}
    =
    \rfunc{n}{N_x+N^\prime_x}{N_y+N^\prime_y}{N_z+N^\prime_z}
    \,.
\end{equation}
The proof of Eq.~\eqref{eq:R-identity} is given in Appendix~\ref{app:R-identity-proof}.

\section{MPO representation for the local conserved quantities}
\label{sec:mpo-xyz}
In this section, we present the matrix product operator (MPO) representation for the local conserved quantities of the spin-$1/2$ XYZ chain, extending the construction of Ref.~\cite{yamada2023matrix}.
We show that the local conserved quantities can be represented by a periodic-boundary MPO constructed from almost upper triangular matrices $\Gamma_k^i$ involving dual numbers.

\newcommand{\bigzero}{\mbox{\normalfont\Large\bfseries O}}

\subsection{Building blocks of the MPO}
\label{sec:Building-blocks}
We introduce a vector $\sbs_i$ of size $3$ and additional $3\times3$ matrices $\Mbs_i$ and $\ebs$ as building blocks of the MPO for the local conserved quantities of the spin-$1/2$ XYZ chain.
    These building blocks reproduce the doubling products in Eq.~\eqref{eq:xyz-Qk} and their coefficients $\rfunc{n}{N_x}{N_y}{N_z}$.

The vector $\sbs_i$ is defined by
\begin{equation}
    \label{eq:building_blocl_sigma}
    \sbs_i
    \coloneqq
    \begin{pmatrix}
        X_i & Y_i & Z_i
    \end{pmatrix}
    \,,
\end{equation}
where the elements are the Pauli matrices acting on the $i$-th site.
The matrix $\Mbs_i$ is defined by
\begin{equation}
    \label{eq:building_blocl_M}
    \Mbs_i
    \coloneqq
    -i
    \begin{pmatrix}
        O      & X_iY_i & X_iZ_i \\
        Y_iX_i & O      & Y_iZ_i \\
        Z_iX_i & Z_iY_i & O
    \end{pmatrix}
    =
    \begin{pmatrix}
        O    & Z_i  & -Y_i \\
        -Z_i & O    & X_i  \\
        Y_i  & -X_i & O
    \end{pmatrix}
    \,,
\end{equation}
where the off-diagonal elements are Pauli matrices acting on the $i$-th site.
The matrix $\Mbs_i$ also serves as a building block in the isotropic case~\cite{yamada2023matrix}; it has the form of an $\mathfrak{so}(3)$ generator whose entries are Pauli matrices.
The matrix $\ebs$ is a $3\times 3$ matrix with identity operators on the diagonal:
\begin{equation}
    \label{eq:building_blocl_e}
    \ebs
    \coloneqq
    \begin{pmatrix}
        I & O & O \\
        O & I & O \\
        O & O & I
    \end{pmatrix}
    \,.
\end{equation}
We also define diagonal $3 \times 3$ matrices with scalar elements, which carry the coupling factor $J_{\boldsymbol{A}}$:
\begin{equation}
    \label{eq:building_blocl_J}
    \boldsymbol{J}
    \coloneqq
    \begin{pmatrix}
        J_X & 0   & 0   \\
        0   & J_Y & 0   \\
        0   & 0   & J_Z
    \end{pmatrix}
    ,
    \quad
    \Jtilbs
    \coloneqq
    \begin{pmatrix}
        J_YJ_Z & 0      & 0      \\
        0      & J_XJ_Z & 0      \\
        0      & 0      & J_XJ_Y
    \end{pmatrix}
    \,.
\end{equation}
We further define the diagonal $3 \times 3$ matrix
\begin{equation}
    \label{eq:building_blocl_Rn}
    \Rbs^n
    \coloneqq
    \begin{pmatrix}
        \rfunc{n}{1}{0}{0} & 0                  & 0                  \\
        0                  & \rfunc{n}{0}{1}{0} & 0                  \\
        0                  & 0                  & \rfunc{n}{0}{0}{1}
    \end{pmatrix}
    \,.
\end{equation}

Combining the building blocks defined in Eqs.~\eqref{eq:building_blocl_sigma}--\eqref{eq:building_blocl_Rn}, we define
\begin{equation}
    \sbs_i^n
    \coloneqq
    \sbs_i \boldsymbol{J} \Rbs^n
    ,
    \quad
    \Mbs_i^n
    \coloneqq
    \Mbs_i \boldsymbol{J} \Rbs^n
    ,
    \quad
    \ebs^n
    \coloneqq
    \ebs
    \Jtilbs \rfunc{n}{1}{1}{1}
    \,.
\end{equation}
In the following, we treat these building blocks as symbols rather than explicit matrices.

\subsection{MPO for the Hamiltonian}
\label{subsec:mpo_hamiltonian}

We introduce a periodic MPO formulation for the Hamiltonian of the spin-$1/2$ XYZ chain in Eq.~\eqref{eq:XYZhamiltonian} using the dual number $\epsilon$, which satisfies $\epsilon^2 = 0$~\cite{Study1903}.
The Hamiltonian is represented by a periodic MPO with a four-dimensional auxiliary space, whose local tensor is
\begin{equation}
    \label{eq:mpo-Q2}
    \Gamma^{i}_{2}
    \coloneqq
    \begin{pmatrix}
        I            & J_X  X_i & J_Y Y_i & J_Z Z_i \\
        \epsilon X_i & O        & O       & O       \\
        \epsilon Y_i & O        & O       & O       \\
        \epsilon Z_i & O        & O       & O       \\
    \end{pmatrix}
    =
    \begin{pmatrix}
        I                   & \sbs_{i}^{0} \\
        \epsilon\sbs_i^\top & O            \\
    \end{pmatrix}
    \,,
\end{equation}
where $\sbs_i^0 = (J_X X_i, J_Y Y_i, J_Z Z_i)$ and $O$ on the right-hand side denotes the $3 \times 3$ zero matrix.
The Hamiltonian is reproduced by taking the trace of the periodic MPO over the auxiliary space:
\begin{equation}
    \label{eq:mpo-Q2-44}
    \Tr_{a}
    \bck{{\Gamma}^{1}_{2} {\Gamma}^{2}_{2} \cdots {\Gamma}^{L}_{2}}
    =
    I+\epsilon H
    \,.
\end{equation}
    The dual number $\epsilon$ prevents the appearance of unwanted terms in the expansion of the MPO.

The local tensor with a dual number in Eq.~\eqref{eq:mpo-Q2} is simpler than the open-boundary MPO used in previous studies~\cite{SCHOLLWOCK201196,yamada2023matrix}.
For comparison, we present the local tensor $\Gamma_2^{o,i}$ for open boundaries, which has bond dimension $5$~\cite{yamada2023matrix}:
\begin{equation}
    \label{eq:hamiltonian_mpo_open}
    \Gamma_2^{o,i}
    \coloneqq
    \begin{pmatrix}
        I & X_i J_X & Y_i J_Y & Z_i J_Z & O   \\
        O & O       & O       & O       & X_i \\
        O & O       & O       & O       & Y_i \\
        O & O       & O       & O       & Z_i \\
        O & O       & O       & O       & I
    \end{pmatrix}
    \,.
\end{equation}
The matrix product sandwiched between the boundary vectors $\bra{L}=(1,0,0,0,0)$ and $\ket{R}=(0,0,0,0,1)^\top$ reproduces the open-chain Hamiltonian $H_{\mathrm{open}}\coloneqq \sum_{j=1}^{L-1} h_{j, j+1}^{\mathrm{XYZ}}$:
\begin{equation}
    \label{eq:bulk_mpo}
    H_{\mathrm{open}} = \mel{L}{\Gamma_2^{o,1} \Gamma_2^{o,2} \cdots \Gamma_2^{o,L}}{R}
    \,.
\end{equation}
To obtain the full Hamiltonian under periodic boundary conditions, one must add the boundary term $h_{L, 1}^{\mathrm{XYZ}}$ connecting the last and first sites: $H = H_{\mathrm{open}} + h_{L, 1}^{\mathrm{XYZ}}$.
In the periodic formulation, the trace automatically generates this boundary term.

\subsection{Explicit expressions of the MPO for local conserved quantities}
\label{subsec:explicit_mpo}

We construct the local tensor $\Gamma_k^i$ for the MPO of the higher-order local conserved quantities $Q_k$ ($k>2$) from the building blocks defined in Section~\ref{sec:Building-blocks}.

Generalizing $\Gamma_2^i$ in Eq.~\eqref{eq:mpo-Q2}, the local tensor $\Gamma_k^i$ is a $k \times k$ block matrix with bond dimension $3k-2$.
Its nonzero blocks are given by
\begin{equation}
    \label{eq:mpo_rep_result}
    \begin{aligned}
        \qty(\Gamma_k^i)_{1,1}
         & = I,
        \qquad
        \qty(\Gamma_k^i)_{k,1}
        = \epsilon \sbs_i^\top,
        \\
        \qty(\Gamma_k^i)_{1,2n+2}
         & = \sbs_i^n,
        \qquad
        0 \leq n < \floor{\frac{k}{2}},
        \\
        \qty(\Gamma_k^i)_{a,a+2n+1}
         & = \Mbs_i^n,
        \qquad
        n \geq 0,\quad 2 \leq a < k-2n,
        \\
        \qty(\Gamma_k^i)_{a,a+2n+2}
         & = \ebs^n,
        \qquad
        n \geq 0,\quad 2 \leq a < k-2n-1
        \,.
    \end{aligned}
\end{equation}
All block entries not listed in \eqref{eq:mpo_rep_result} vanish.
Zero blocks have the dimensions dictated by their positions: first-row zero blocks are $1 \times 3$, first-column zero blocks are $3 \times 1$, and all other zero blocks are $3 \times 3$.

\begin{thm}\label{thm:Qk-MPO}
    For $k \le L$, the local conserved quantity $Q_k$ is given by
    \begin{equation}
        \label{eq:MPOall}
        \Tr_{a}
        \bck{
            \Gamma^{1}_k \Gamma^{2}_{k} \cdots \Gamma^{L}_{k}
        }
        =
        I + \epsilon Q_k
        \,.
    \end{equation}
\end{thm}

The proof of Theorem~\ref{thm:Qk-MPO} is given in Appendix~\ref{app:Qk-MPO}.

For $k=3$ and $k=4$, one obtains
\begin{align}
    \label{eq:mpo-Q3}
    \Gamma^{i}_{3}
     & =
    \begin{pmatrix}
        I                    & \sbs_i^0 & O        \\
        O                    & O        & \Mbs_i^0 \\
        \epsilon \sbs_i^\top & O        & O        \\
    \end{pmatrix}
    \,,
    \hspace{2em}
    \Gamma^{i}_{4}
    =
    \begin{pmatrix}
        I                    & \sbs_i^0 & O        & \sbs_i^1 \\
        O                    & O        & \Mbs_i^0 & \ebs^0   \\
        O                    & O        & O        & \Mbs_i^0 \\
        \epsilon \sbs_i^\top & O        & O        & O        \\
    \end{pmatrix}
    \,.
\end{align}
The larger examples $\Gamma_5^i,\ldots,\Gamma_8^i$ are collected in Appendix~\ref{app:gamma-examples}.

\section{Lax operator without elliptic functions}
\label{sec:new_mpo}
In this section, we compress the local tensors for all local conserved quantities $Q_k$ into a single local tensor with a four-dimensional auxiliary space.
The local tensor in Eq.~\eqref{eq:mpo_rep_result} introduced in the previous section has many vanishing lower-triangular blocks; we eliminate this redundancy by introducing a spectral parameter.
Using the resulting local tensor as a guide, we obtain a Lax operator without elliptic functions.

    In the quantum inverse scattering method (QISM), the auxiliary trace of an ordered product of Lax operators defines a transfer matrix, and the RLL relation implies that transfer matrices at different spectral parameters commute~\cite{Takhtadzhan_Faddeev_xyz_1979,korepin_bogoliubov_izergin_1993}.
    For the XYZ chain the standard choice is Baxter's R-matrix~\cite{Baxter1982}, whose entries are elliptic functions, whereas the Lax operator constructed below has entries algebraic in the spectral parameter and the couplings.

\subsection{Compression of the MPO}
Here we treat the MPOs for all local conserved quantities $Q_k$ simultaneously by introducing a spectral parameter $s$.
The local tensors $\Gamma_k^i$ of Theorem~\ref{thm:Qk-MPO} repeat the same building blocks for different $k$, most of which vanish.
Assigning powers of $s$ to the repeated building blocks therefore packages the $\Gamma_k^i$ for all $k$ into a single tensor, from which $Q_k$ is read off as the term proportional to $\epsilon s^{k-1}$.

We define the following $4 \times 4$ local tensor depending on $s$:
\begin{equation}
    \label{eq:w-mpo}
    \w_{i}(s; \epsilon)
    \coloneqq
    \begin{pmatrix}
        I                      & \sbs_{i}^{\infty}(s) \\
        \epsilon \sbs_{i}^\top & \lax_{i}(s)
    \end{pmatrix}
    =
    \begin{pmatrix}
        \I           & s f_X(s) J_X X_i     & s f_Y(s) J_Y Y_i     & s f_Z(s) J_Z Z_i    \\
        \epsilon X_i & \omega(s) J_Y J_Z  I & sf_Y(s) J_Y Z_i      & -  sf_Z(s) J_Z Y_i  \\
        \epsilon Y_i & -  sf_X(s) J_X Z_i   & \omega(s) J_X J_Z  I & sf_Z(s) J_Z X_i     \\
        \epsilon Z_i & sf_X(s) J_X Y_i      & -  sf_Y(s) J_Y X_i   & \omega(s) J_X J_Y I \\
    \end{pmatrix}
    \,,
\end{equation}
where we define
\begin{align}
    \sbs_{i}^{\infty}(s)
     & \coloneqq
    \sum_{n = 0}^{\infty} \sbs_{i}^{n} s^{2n+1}
    =
    \sbs_{i} \boldsymbol{J} s \Rbs^{\infty}(s)
    \,,
    \\
    \lax_{i}(s)
     & \coloneqq
    \sum_{n = 0}^{\infty} \bck{ \Mbs_{i}^{n} s^{2n+1} + \ebs^{n} s^{2n+2} }
    =
    \Mbs_{i} \boldsymbol{J} s \Rbs^{\infty}(s) + \ebs \Jtilbs \omega(s)
    \,.
\end{align}
Here, $\Rbs^{\infty}(s)$ is defined by
\begin{align}
    \Rbs^{\infty}(s)
     & \coloneqq
    \sum_{n = 0}^{\infty} \Rbs^{n}s^{2n}
    =
    \begin{pmatrix}
        f_X(s) & 0      & 0      \\
        0      & f_Y(s) & 0      \\
        0      & 0      & f_Z(s)
    \end{pmatrix}
    \,,
\end{align}
with the generating functions $f_X(s)$, $f_Y(s)$, and $f_Z(s)$ given by
\begin{align}
    f_X(s) \coloneqq \sum_{n = 0}^{\infty} \rfunc{n}{1}{0}{0} s^{2n}
    \,, \quad
    f_Y(s) \coloneqq \sum_{n = 0}^{\infty} \rfunc{n}{0}{1}{0} s^{2n}
    \,, \quad
    f_Z(s) \coloneqq \sum_{n = 0}^{\infty} \rfunc{n}{0}{0}{1} s^{2n}
    \,,
\end{align}
and the generating function $\omega(s)$ is defined by
\begin{equation}\label{eq:fx-omega}
    \omega(s) \coloneqq \sum_{n = 0}^{\infty} \rfunc{n}{1}{1}{1} s^{2n+2}
    \,.
\end{equation}

The generating functions $f_X(s)$, $f_Y(s)$, $f_Z(s)$, and $\omega(s)$ are not independent.
Using Eq.~\eqref{eq:R-identity-diff}, we obtain
\begin{align}
    \label{eq:fx-ito-fe}
    f_{\alpha}(s)
    =
    \frac{1}{1 - \omega(s) J_{\alpha}^2}
    \qquad(\alpha \in \{X,Y,Z\})
    \,,
\end{align}
and using Eq.~\eqref{eq:R-identity}, we obtain
\begin{align}\label{eq:omega-fxyz}
    \omega(s)
    =
    s^2 f_X(s) f_Y(s) f_Z(s)
    \,.
\end{align}
The proofs are given in Appendices~\ref{app:fx-omega} and~\ref{app:omega-fxyz}, respectively.
From Eqs.~\eqref{eq:fx-ito-fe} and~\eqref{eq:omega-fxyz}, we obtain the following equation for $\omega(s)$:
\begin{equation}
    \label{eq:eq-for-omega}
    \omega(s) (1 - J_{X}^2 \omega(s)) (1 - J_{Y}^2 \omega(s)) (1 - J_{Z}^2 \omega(s)) = s^2
    \,.
\end{equation}
This quartic equation determines $\omega(s)$ as an explicit algebraic function of $s$, which we do not present.

Using the local tensor $\w_i(s; \epsilon)$, we define the MPO $W(s; \epsilon)$ by
\begin{equation}
    \label{eq:W-mpo}
    W(s; \epsilon)
    =
    \Tr_{a}
    \bck{
        \w_{1}(s; \epsilon) \w_{2}(s; \epsilon) \cdots \w_{L}(s; \epsilon)
    }
    \,.
\end{equation}
    For a finite chain, $W(s; \epsilon)$ splits into a conserved part containing the local conserved quantities $Q_k$ and finite-size wrap-around terms:
    \begin{align}
        \label{eq:W-mpo_expand}
        W(s; \epsilon)
         & =
        \underbrace{I + V(s)
            + \epsilon \sum_{k = 2}^{L} s^{k - 1} Q_{k}}_{\substack{\text{Conserved part}}}
        + \epsilon
        \underbrace{\sum_{k = L + 1}^{\infty} s^{k - 1} \widetilde{Q}_{k}^{(L)}}_{\substack{\text{Non-conserved part} \\\text{for generic XYZ}}}
        \,,
    \end{align}
where $V(s)$ is given by
\begin{equation}
    \label{eq:Lax-mpo}
    V(s)
    =
    \Tr_{a} \bck{ \lax_{1}(s) \lax_{2}(s) \cdots \lax_{L}(s)  }
    \,.
\end{equation}
Here, the auxiliary traces appearing in~\eqref{eq:W-mpo} and~\eqref{eq:Lax-mpo} are taken over four- and three-dimensional spaces, respectively.

    Theorem~\ref{thm:fused-expansion} below implies the conservation of $V(s)$, while Theorem~\ref{thm:Qk-MPO} guarantees the appearance of $Q_k$ for $2\le k\le L$.
    The remaining, generically non-conserved part consists of finite-size wrap-around terms.
    For $k>L$, the formula~\eqref{eq:xyz-Qk} no longer defines a local conserved quantity at system size $L$, because some doubling products have support on more than $L$ sites; hence $\widetilde Q_k^{(L)}$ is not conserved for generic XYZ couplings.

    \subsection{Product of two eight-vertex transfer matrices}
    \label{subsec:baxter-fusion}
    A product of two eight-vertex transfer matrices, expanded to first order off the fusion point of Baxter's R-matrix, repairs this obstruction~\cite{Kulish-Reshetikhin-Sklyanin-1981,Takebe-1995,Konno2006}.
    Fusion at that point projects the four-dimensional auxiliary space of the product onto its triplet.

    Baxter's eight-vertex R-matrix, which underlies the solution of the XYZ chain~\cite{baxter-prl-1971,baxter-1973-i,baxter-1973-ii,baxter-1973-iii,Takhtadzhan_Faddeev_xyz_1979,Baxter1982}, is given by
    \begin{equation}
        \label{eq:baxter-R-main}
        R^{8V}_{ai}(z)
        =
        \begin{pmatrix}
            a(z) & 0    & 0    & d(z) \\
            0    & b(z) & c(z) & 0    \\
            0    & c(z) & b(z) & 0    \\
            d(z) & 0    & 0    & a(z)
        \end{pmatrix}
        =
        \sum_{\mu=0,X,Y,Z}
        \varphi_\mu(z)\sigma_a^\mu\sigma_i^\mu
        \,.
    \end{equation}
    Here, $\sigma^{X,Y,Z}$ are the Pauli matrices and $\sigma^0=I$, with the subscripts $a$ and $i$ indicating action on auxiliary space $a$ and physical site $i$, respectively.
    The matrix entries satisfy
    $a=\varphi_0+\varphi_Z$,
    $b=\varphi_0-\varphi_Z$,
    $c=\varphi_X+\varphi_Y$, and
    $d=\varphi_X-\varphi_Y$.
    The weights are given by~\cite{Takebe-1995}
    \begin{equation}
        \label{eq:baxter-theta-weights-main}
        \begin{aligned}
            \varphi_0(z)
             & =
            \frac{\vartheta_1(z+\eta_{8V}\mid\tau)}
            {\vartheta_1(\eta_{8V}\mid\tau)}
            \,,
             &
            \varphi_X(z)
             & =
            \frac{\vartheta_2(z+\eta_{8V}\mid\tau)}
            {\vartheta_2(\eta_{8V}\mid\tau)}
            \,,
            \\
            \varphi_Y(z)
             & =
            \frac{\vartheta_3(z+\eta_{8V}\mid\tau)}
            {\vartheta_3(\eta_{8V}\mid\tau)}
            \,,
             &
            \varphi_Z(z)
             & =
            \frac{\vartheta_4(z+\eta_{8V}\mid\tau)}
            {\vartheta_4(\eta_{8V}\mid\tau)}
            \,.
        \end{aligned}
    \end{equation}
    Here $\vartheta_1,\ldots,\vartheta_4$ are the Jacobi theta functions in the convention of Ref.~\cite[Chapter~21]{Whittaker-Watson-1927}, with nome $q=e^{\imi\pi\tau}$.
    The argument $z$ is the spectral parameter, $\eta_{8V}$ is the crossing parameter, and $\tau$ is the modular parameter.
    The eight-vertex transfer matrix is given by~\cite{Takhtadzhan_Faddeev_xyz_1979,Baxter1982}:
    \begin{equation}
        \label{eq:baxter-transfer-main}
        t^{8V}(z)
        \coloneqq
        \Tr_a\qty(
        R^{8V}_{aL}(z)
        R^{8V}_{a,L-1}(z)
        \cdots
        R^{8V}_{a1}(z)
        )
        \,.
    \end{equation}

    For fixed $(\tau,\eta_{8V})$ and every $z$, the weights and the couplings satisfy~\cite{Baxter1982}
    \begin{equation}
        \label{eq:baxter-invariants-main}
        \varphi_0(z)^2
        +\varphi_\alpha(z)^2
        -\varphi_\beta(z)^2
        -\varphi_\gamma(z)^2
        =
        2\nu(z)J_\alpha
        \,,
        \qquad
        \{\alpha,\beta,\gamma\}=\{X,Y,Z\}
        \,,
    \end{equation}
    with a factor $\nu(z)$ independent of $\alpha$.
    For given coupling ratios, $(\tau,\eta_{8V})$ is chosen to satisfy these relations.

    \begin{thm}
        \label{thm:fused-expansion}
        There exist functions $z(s)$, $\kappa(s)$, and $\Lambda_0(s,\delta)$ such that
        \begin{equation}
            \label{eq:fused-expansion-factorization}
            T^{8V}_{\mathrm{fused}}(s,\delta)
            \coloneqq
            \Lambda_0(s,\delta)^{-L}\,
            t^{8V}\qty(z(s)-2\eta_{8V}+s\,\kappa(s)\,\delta)\,
            t^{8V}\qty(z(s))
            \,,
        \end{equation}
        satisfies the first-order identity
        \begin{equation}
            \label{eq:fused-expansion-Wcorr}
            T^{8V}_{\mathrm{fused}}(s,\delta)
            =
            \Wcorr(s; \delta)
            +
            O(\delta^2)
            \,.
        \end{equation}
        Here $\Wcorr(s; \delta)$ is defined by the following MPO:
        \begin{equation}
            \label{eq:W-corr-mpo}
            \wcorr_i(s; \delta)
            \coloneqq
            \begin{pmatrix}
                I                 & \sbs_i^\infty(s)              \\
                \delta\sbs_i^\top & \lax_i(s)+\delta\,\Delta_i(s)
            \end{pmatrix}
            \,,
            \qquad
            \Wcorr(s; \delta)
            \coloneqq
            \Tr_a
            \qty(
            \wcorr_1(s; \delta)
            \cdots
            \wcorr_L(s; \delta)
            )
            \,.
        \end{equation}
        The local tensor $\wcorr_i(s;\delta)$ differs from $\w_i(s;\delta)$ only by the triplet-block term $\delta\Delta_i(s)$, where $\Delta_i(0)=0$.
        Expanding $\Wcorr(s; \delta)$ to first order and then substituting $\delta=\epsilon$ gives
        \begin{align}
            \label{eq:W-corr-mpo-expand}
            \Wcorr(s; \epsilon)
             & =
            I+V(s)
            +\epsilon
            \sum_{k=2}^{L}s^{k-1}Q_k
            +\epsilon
            \sum_{k=L+1}^{\infty}
            s^{k-1}
            \qty(
            \widetilde Q_k^{(L)}
            +
            \mathcal C_k^{(L)}
            )
            \,,
        \end{align}
        where $\mathcal C_k^{(L)}$ is the correction term from single insertions of $\Delta_i(s)$, given by Eq.~\eqref{eq:baxter-fusion-Cp-definition-rev} in Appendix~\ref{app:fused-expansion}.
        In particular, the term proportional to $\delta\,s^{k-1}$ in $T^{8V}_{\mathrm{fused}}(s,\delta)$ is the local conserved quantity $Q_k$ for $2\le k\le L$.
    \end{thm}
    The proof, in Appendix~\ref{app:fused-expansion}, constructs $z(s)$, $\kappa(s)$, and $\Lambda_0(s,\delta)$ and derives the triplet-block correction $\Delta_i(s)$.
    For $k>L$, the wrap-around term $\widetilde Q_k^{(L)}$ is generically not conserved; in $\Wcorr(s;\epsilon)$ it is replaced by the conserved combination $\widetilde Q_k^{(L)}+\mathcal C_k^{(L)}$.
    \begin{cor}
        \label{cor:Wcorr-baxter-H}
        Let $\epsilon$ be a dual number with $\epsilon^2=0$.
        Then
        \begin{equation}
            \label{eq:Wcorr-baxter-H-commute}
            \bck{\Wcorr(s; \epsilon),t^{8V}(v)}=0
            \,.
        \end{equation}
        In particular, $\bck{\Wcorr(s; \epsilon),H}=0$, $\bck{V(s),t^{8V}(v)}=0$, and $\bck{Q_k,t^{8V}(v)}=0$ for $2\le k\le L$.
    \end{cor}
    Since $T^{8V}_{\mathrm{fused}}(s,\delta)$ is a product of two eight-vertex transfer matrices, it commutes with every $t^{8V}(v)$.
    For $\epsilon^2=0$, Eq.~\eqref{eq:fused-expansion-Wcorr} gives $T^{8V}_{\mathrm{fused}}(s,\epsilon)=\Wcorr(s;\epsilon)$.
    Hence $\Wcorr(s;\epsilon)$ commutes with every $t^{8V}(v)$, proving Corollary~\ref{cor:Wcorr-baxter-H}.

In the isotropic (XXX) case $J_X = J_Y = J_Z$, by contrast, $W(s; \epsilon)$ itself becomes a conserved quantity, so the part labelled as non-conserved in Eq.~\eqref{eq:W-mpo_expand} also becomes conserved in this limit.
    Appendix~\ref{app:W-XXX-commutativity} proves this conservation and the mutual commutativity of the isotropic $W(s;\epsilon)$ using a four-dimensional Lax operator obtained from the product of two fundamental XXX Lax operators.
    This construction retains the Catalan-number structure of the XXX MPOs in Ref.~\cite{yamada2023matrix}.
This Lax operator is a generalization of the integrable MPO with a four-dimensional auxiliary space ($D=4$) constructed by Katsura~\cite{katsura2015} and the isotropic limit of the Lax operator obtained by Fendley et al.~\cite{Fendley-xyz-2025}.

\subsection{Lax operator in symmetric form}
Applying a gauge transformation to $\lax_i(s)$ and introducing the spectral parameter $u \coloneqq \sqrt{\omega}$, we obtain the Lax operator in a symmetric form:
\begin{align}\label{eq:laxcal}
    \laxcal_i(u)
    =
    \begin{pmatrix}
        \sqrt{\dfrac{u J_{Y}J_{Z}}{J_X}} I & g(u J_Z) Z_i                       & - g(u J_Y) Y_i                     \\[1em]
        - g(u J_Z) Z_i                     & \sqrt{\dfrac{u J_{Z}J_{X}}{J_Y}} I & g(u J_X) X_i                       \\[1em]
        g(u J_Y) Y_i                       & - g(u J_X)  X_i                    & \sqrt{\dfrac{u J_{X}J_{Y}}{J_Z}} I
    \end{pmatrix}
    \,,
\end{align}
where $g(x) \coloneqq \sqrt{x^{-1}-x}$.
The details of the derivation are given in Appendix~\ref{app:simple-lax-derivation}.

As with $\lax_i(s)$, we can construct the transfer matrix from the Lax operator $\laxcal_i(u)$:
\begin{equation}
    \label{eq:Laxcal-mpo}
    T(u)
    =
    \Tr_{a} \bck{ \laxcal_{1}(u) \laxcal_{2}(u) \cdots \laxcal_{L}(u)}
    \,.
\end{equation}
The relation between the transfer matrices $T(u)$ and $V(s)$ is
\begin{align}
    \label{eq:V-T-relation}
    V(s)
     & =
    \qty{u^3 J_X J_Y J_Z}^{\frac{L}{2}}
    T(u)
    \,,
\end{align}
where $u$ and $s$ are related by Eq.~\eqref{eq:eq-for-omega} with $\omega(s) = u^2$.
    Corollary~\ref{cor:commutativity-Lax-H} follows immediately from Theorem~\ref{thm:fused-expansion} by setting $\delta=0$.
    \begin{cor}
        \label{cor:commutativity-Lax-H}
        $T(u)$ commutes with the Hamiltonian of the spin-$1/2$ XYZ chain~\eqref{eq:XYZhamiltonian}:
        \begin{equation}
            \label{eq:commutativity-Lax-H}
            \bck{ T(u), H } = 0
            \,.
        \end{equation}
    \end{cor}
    An alternative telescoping proof of Corollary~\ref{cor:commutativity-Lax-H} is given in Appendix~\ref{app:commutativity-Lax-H}.

    \begin{cor}
        \label{cor:commutativity-Lax}
        The transfer matrices $T(u)$ are mutually commuting:
        \begin{equation}
            \label{eq:commutativity-Lax}
            \bck{ T(u), T(v) } = 0
            \,.
        \end{equation}
    \end{cor}
    Eight-vertex transfer matrices at different spectral parameters commute, so the same factorization gives $\bck{V(s),V(s')}=0$, and Eq.~\eqref{eq:V-T-relation} gives Eq.~\eqref{eq:commutativity-Lax}.

    \begin{cor}
        \label{cor:3by3-baxter}
        The transfer matrices $T(u)$ and $t^{8V}(v)$ commute:
        \begin{equation}
            \label{eq:commutativity-Lax-baxter}
            \bck{T(u), t^{8V}(v)} = 0
            \,.
        \end{equation}
    \end{cor}
    Corollary~\ref{cor:Wcorr-baxter-H} gives $\bck{V(s),t^{8V}(v)}=0$, and Eq.~\eqref{eq:V-T-relation} gives Eq.~\eqref{eq:commutativity-Lax-baxter}.

    \begin{cor}
        \label{cor:Lax-Qk-commute}
        $T(u)$ commutes with the local conserved quantities of the XYZ chain:
        \begin{equation}
            \label{eq:commutativity-Lax-Qk}
            \bck{T(u), Q_k} = 0
            \quad (k\in\{2,3,\ldots,L\})
            \,.
        \end{equation}
    \end{cor}
    Since $T^{8V}_{\mathrm{fused}}(s,\delta)$ is a product of two eight-vertex transfer matrices, Corollary~\ref{cor:3by3-baxter} implies $\bck{T(u),T^{8V}_{\mathrm{fused}}(s,\delta)}=0$.
    By Theorem~\ref{thm:fused-expansion}, extracting the term proportional to $\delta\,s^{k-1}$ from this commutation relation yields Eq.~\eqref{eq:commutativity-Lax-Qk} for $2\le k\le L$.

The transfer matrix $T(u)$ can be written as
\begin{equation}
    \label{eq:T-u-pauli-string}
    T(u)
    =
    \sum_{\{a_j\}, \{A_j\}}
    \mathcal{L}_{A_L,A_1}^{a_1} \mathcal{L}_{A_1,A_2}^{a_2}
    \cdots
    \mathcal{L}_{A_{L-2},A_{L-1}}^{a_{L-1}} \mathcal{L}_{A_{L-1},A_{L}}^{a_{L}}
    \sigma_{1}^{a_1} \sigma_{2}^{a_2}
    \cdots
    \sigma_{L}^{a_L}
    \,,
\end{equation}
where the lower indices are $A_j \in \{X, Y, Z\}$ and the upper indices are $a_j \in \{0, X, Y, Z\}$.
The nonzero values of $\mathcal{L}_{k,l}^{a}$ can be read from the expression for the Lax operator~\eqref{eq:laxcal}:
\begin{align}
    \label{eq:T-u-Lax-entries}
    \mathcal{L}_{k,k}^{0} = \frac{\sqrt{u J_X J_Y J_Z}}{J_k}
    \,,\quad
    \mathcal{L}_{k,l}^{m} = \varepsilon_{k,l,m} \, g(u J_m)
    \,,
\end{align}
where $k,l,m \in \{X, Y, Z\}$ and $\varepsilon_{k,l,m}$ is the totally antisymmetric tensor.
    The simple form of Eq.~\eqref{eq:laxcal}, which involves no elliptic functions, yields a closed formula for the conserved quantities generated by $T(u)$, as derived in Appendix~\ref{app:T-u-expansion}.

    Another Lax operator without elliptic functions for the XYZ chain was independently found in~\cite{Fendley-xyz-2025}.
    Recently, a two-parameter family of conserved MPOs for the XYZ chain has been constructed~\cite{Yashin-two-parameter-MPO-2026}; it contains, as special cases, the present MPO~\eqref{eq:Laxcal-mpo} and the conserved MPO constructed in Ref.~\cite{Fendley-xyz-2025}.

In the XXX limit $J_X = J_Y = J_Z = 1$, the Lax operator reduces to
\begin{align}\label{eq:laxcal-xxx}
    \laxcal_i^{\mathrm{XXX}}(v)
    =
    I + v \, \boldsymbol{S}_a \cdot \boldsymbol{\sigma}_i
    \,,
\end{align}
where $\boldsymbol{S}_a = (S^X_a, S^Y_a, S^Z_a)$ denotes the standard spin-$1$ generators of $\mathfrak{su}(2)$ acting on the three-dimensional auxiliary space, and $\boldsymbol{\sigma}_i = (\sigma^X_i, \sigma^Y_i, \sigma^Z_i)$ are the Pauli matrices acting on the physical site $i$.
The spectral parameter $v$ is related to $u$ through $v = \imi g(u)/\sqrt{u}$.
In the XXX limit, $\laxcal_i^{\mathrm{XXX}}(v) = u^{-1/2} \laxcal_i(u)$.
This form is the Lax operator for the spin-$1/2$ XXX chain with a spin-$1$ auxiliary space~\cite{quasi-local-XXX-prl-2015}.

\section{Conclusion}
\label{sec:conclusion}

In this paper, we have extended the matrix product operator (MPO) representation previously developed for the Heisenberg (XXX) chain~\cite{yamada2023matrix} to the spin-$1/2$ XYZ chain.
Through these MPOs, we have substantially simplified the coefficients of the local conserved quantities obtained in~\cite{Nozawa2020}.
The simplified coefficient $\rfunc{n}{N_x}{N_y}{N_z}$ is a polynomial generalization of the Catalan numbers of Hilton and Pedersen~\cite{hilton-catalan-1991}, given by weighted sums over monotone lattice paths whose weights encode the anisotropy of the XYZ chain.
    We also proved that the conservation law follows from a single recurrence relation for the coefficients, and gave a general construction of local conserved quantities that leaves unfixed the freedom to add lower-order ones.

Grabowski and Mathieu~\cite{GrabowskiCatalantreepattern} extracted from the XXX chain a set of algebraic conditions sufficient for the existence of infinitely many local conserved quantities, which they called the Catalan-tree pattern, and models satisfying the same conditions---such as $\mathrm{SU}(N)$ chains and the octonionic chain---share the same combinatorial structure in their local conserved quantities.
    More recently, the Catalan-tree pattern was also found in the local conserved quantities of the periodic Fendley model~\cite{Fendley-disguise-2019, Fukai-Pozsgay-Vona-GFFI-2026, Fukai-Pozsgay-Vona-GFFII-2026}.
The weighted polynomial generalization of the Catalan numbers found in this work may provide an anisotropic extension of the Catalan-tree pattern.

By introducing a spectral parameter, we compress the MPO representations of the local conserved quantities $Q_k$ into a single MPO with a four-dimensional auxiliary space.
    Theorem~\ref{thm:fused-expansion} shows that this compressed MPO arises from a product of two eight-vertex transfer matrices built from Baxter's R-matrix.
    To our knowledge, this is the first direct connection between the R-matrix formalism and closed-form expressions for all local conserved quantities, which had previously been obtained only by direct, model-specific methods outside QISM~\cite{grabowski-xxx-1994,GRABOWSKI1995299,Nozawa2020,Nienhuis2021,fukai-hubbard-charge-2023}.
    From this compressed MPO with a spectral parameter, we derive a Lax operator $\laxcal_i(u)$ with a three-dimensional auxiliary space and no elliptic functions, which can be identified with the spin-$1$ Lax operator associated with the Sklyanin algebra~\cite{Kulish-Reshetikhin-Sklyanin-1981,Sklyanin1982,Sklyanin1983,Takebe-1995,Takebe-1996,Hasegawa1997,Konno2006}.
    Corollaries~\ref{cor:commutativity-Lax-H} and~\ref{cor:commutativity-Lax} derive from Theorem~\ref{thm:fused-expansion} that its transfer matrices commute with the Hamiltonian and with one another.
For the XXZ chain, Lax operators with higher-spin auxiliary spaces have been used to construct quasilocal conserved charges~\cite{quasi-local-review-2016}; the XXZ limit of our Lax operator may also be related to these constructions.

    The same viewpoint may also be useful beyond the XYZ chain.
    It suggests looking for suitable fusion limits of known R-matrices that lead to compact MPO representations of local conserved quantities, rather than deriving general explicit expressions only through direct, model-specific computations~\cite{anshelevich1980first,grabowski-xxx-1994,GRABOWSKI1995299,Nozawa2020,Nienhuis2021,fukai-hubbard-charge-2023,fukai-completeness-hubbard-2024,fukai-doctoral-arxiv}.

The one-dimensional Hubbard model is a natural next target for this approach: general explicit expressions for its local conserved quantities have been obtained~\cite{fukai-hubbard-charge-2023,fukai-completeness-hubbard-2024,fukai-doctoral-arxiv}.
Although recurrence relations for the coefficients appearing in these quantities are known, they are considerably more involved than in the XYZ case, and closed-form expressions have not yet been found.
An MPO construction may reveal the combinatorial interpretation of these coefficients and yield a closed-form expression for them.

More broadly, it would be interesting to investigate whether combinatorial structures of this kind underlie local conserved quantities across integrable models and whether they are directly connected to integrability itself.

\section*{Acknowledgements}
The authors thank Yuji Nozawa and Hosho Katsura for insightful comments and fruitful discussions.
K.F.\ was supported by JSPS KAKENHI Grant-in-Aid for Research Activity Start-up (Grant No.~JP25K23354), JSPS KAKENHI Grant-in-Aid for Early-Career Scientists (Grant No.~JP26K17045), JSPS KAKENHI Grant-in-Aid for JSPS Fellows (Grant No.~JP26KJ0404), the JSPS Bilateral Program with MTA (Project No.~120263802), and MEXT KAKENHI Grant-in-Aid for Transformative Research Areas A ``Extreme Universe'' (KAKENHI Grant No.~JP21H05191).

\begin{appendix}

    \section{Review of previous study~\cite{Nozawa2020}}
    \label{app:review_xyz}

    This appendix reviews the results and notation of Ref.~\cite{Nozawa2020} and recasts the original coefficient in the sequence-independent form~\eqref{eq:rfunc_pre_rewrite}, so that the equivalence with our simplified expression~\eqref{eq:xyz-Qk} reduces to the single identity~\eqref{eq:to-prove-app1}.
    The local conserved quantities obtained in~\cite{Nozawa2020} are given by
    \begin{equation}
        \label{eq:xyz-Qk-pre}
        Q_{k}
        \coloneqq
        \sum_{\substack{0 \leq n+m < \floor{k/2} \\ n, m \ge 0}}
        \sum_{\obA\in\mathcal{S}_k^{n,m}}
        R^{n, m}\left(A_{1} A_{2} \cdots A_{t}\right)
        J_{\boldsymbol{A}}
        \,
        \obA
        \,,
    \end{equation}
    where $t$ and the symbols $\mathcal{S}_k^{n,m}$, $J_{\boldsymbol{A}}$, and $\obA$ are defined below Eq.~\eqref{eq:xyz-Qk}.

    The main statement of~\cite{Nozawa2020} is that the $Q_k$ are local conserved quantities of the spin-$1/2$ XYZ chain, $\qty[Q_k, H] = 0$, and that they mutually commute, $\qty[Q_k, Q_l] = 0$.

    The coefficient $R^{n, m}\left(A_{1} A_{2} \cdots A_{t}\right)$ is defined by
    \begin{equation}
        \label{eq:rfunc-pre}
        R^{n, m}\left(A_{1} A_{2} \cdots A_{t}\right)
        \coloneqq
        \sum_{\tilde{n}=0}^{n} f(n-\tilde{n}, m+\tilde{n}) S_{\tilde{n}}\left(A_{1} A_{2} \cdots A_{t}\right)
        ,
    \end{equation}
    where the sequence dependence enters only through $S_{\tilde{n}}\paren{A_{1} A_{2} \cdots A_{t}}$, defined in Eq.~\eqref{eq:sfunc-pre} below, and $f(n,m)$ is a symmetric polynomial in $J_X$, $J_Y$, and $J_Z$ given by
    \begin{align}
        f(n, m)
         & \coloneqq
        \frac{m}{n+m} \sum_{p=1}^{n}\binom{n+m}{p} \sum_{\substack{j_1+j_2+\cdots+j_p=n \\ j_1, j_2, \ldots, j_p \geq 1}} a_{j_1} a_{j_2} \cdots a_{j_p}  \quad (n>0)
        ,
    \end{align}
    and $f(0, m) \equiv 1$.

    The building block $a_{n}$ is a symmetric polynomial in $J_X$, $J_Y$, and $J_Z$, defined by
    \begin{align}
        \label{eq:an-def}
        a_{n}
         & \coloneqq
        \frac{
            x\paren{y^{n+2}-z^{n+2}}
            +
            y\paren{z^{n+2}-x^{n+2}}
            +
            z\paren{x^{n+2}-y^{n+2}}
        }{
            \left(x-y\right)\left(y-z\right)\left(z-x\right)
        }
        ,
    \end{align}
    with $x$, $y$, and $z$ as in Eq.~\eqref{eq:coupling-squared}.
    The denominator in Eq.~\eqref{eq:an-def} always divides the numerator, so each $a_n$ is a polynomial; the first two are $a_{1}= x+y+z$ and $a_{2}=x^2+y^2+z^2+xy+yz+zx$.

    The function $S_{n}\left(A_{1} A_{2} \cdots A_{t}\right)$ is given by
    \begin{align}
        \label{eq:sfunc-pre}
        S_{n}\paren{A_{1} A_{2} \cdots A_{t}}
         & \coloneqq
        \sum_{1\leq j_1\leq j_2\leq\ldots\leq j_n\leq t}
        J^2_{A_{j_1}}J^2_{A_{j_2}}\cdots J^2_{A_{j_n}}
        \quad (n > 0)
        \,,
    \end{align}
    and $S_{0}\paren{A_{1} A_{2} \cdots A_{t}} \coloneqq 1$.
    $S_{n}\left(A_{1} A_{2} \cdots A_{t}\right)$ is independent of the ordering of the $A_i$ and depends only on the numbers of $X$, $Y$, and $Z$ in the sequence $A_{1} A_{2} \cdots A_{t}$.
    From Eq.~\eqref{eq:sfunc-pre}, we can therefore identify it with the weighted sum over monotone lattice paths of Section~\ref{sec:combinatoric-sfunc}:
    \begin{align}
        \label{eq:sfunc_equivalence}
        S_{n}\paren{A_{1} A_{2} \cdots A_{t}}
        =
        \sfunc{n}{N_x}{N_y}{N_z}
        \,,
    \end{align}
    where $N_x$, $N_y$ and $N_z$ are the numbers of $X$, $Y$ and $Z$ in the sequence $A_{1} A_{2} \cdots A_{t}$, respectively.

    Therefore the coefficient in Eq.~\eqref{eq:rfunc-pre} depends on the sequence $A_{1} A_{2} \cdots A_{t}$ only through $N_x$, $N_y$, and $N_z$, and we write
    \begin{align}
        \label{eq:rfunc_pre_rewrite}
        \rfuncpre{n}{m}{N_x}{N_y}{N_z}
        \coloneqq
        R^{n, m}\left(A_{1} A_{2} \cdots A_{t}\right)
        \,,
    \end{align}
    with $N_x$, $N_y$, and $N_z$ as in Eq.~\eqref{eq:sfunc_equivalence}.
    With this notation, the statement that Eq.~\eqref{eq:xyz-Qk-pre} simplifies to Eq.~\eqref{eq:xyz-Qk} is equivalent to the identity
    \begin{align}
        \label{eq:to-prove-app1}
        \rfuncpre{n}{m}{N_x}{N_y}{N_z}
        =
        \rfunc{n}{N_x + m}{N_y+m}{N_z+m}
        \,,
    \end{align}
    which is proven in Appendix~\ref{app:proof-identities}.

        \section{Practical cost of logarithmic-derivative extraction}
        \label{app:log-transfer-comparison}

        The logarithmic derivative of the eight-vertex transfer matrix generates the local conserved quantities, but extracting them explicitly at higher orders requires cancellations of nonlocal terms.
        Let $\widetilde t^{8V}(z) \coloneqq t^{8V}(0)^{-1}t^{8V}(z)$ be the transfer matrix of Eq.~\eqref{eq:baxter-transfer-main}, with $\widetilde t^{8V}(0)=I$, and write $\widetilde t^{8V}(z) = \sum_{k \ge 0} t_k z^k$ with $t_0 = I$.
        Its logarithmic derivative is
        \begin{equation}
            \mathcal{Q}(z)
            =
            \bigl(\widetilde t^{8V}(z)\bigr)^{-1}\frac{d}{dz}\widetilde t^{8V}(z)
            =
            \sum_{k \ge 2} \mathcal{Q}_k z^{k-2}
            \,.
        \end{equation}
        On a sufficiently long chain, $\mathcal{Q}_k$ has support at most $k$ and hence~\cite{Nozawa2020}
        \begin{equation}
            \label{eq:log-coeff-triangular-basis}
            \mathcal{Q}_k
            =
            \sum_{q=1}^{k} c_{k,q} Q_q
            \,,
            \qquad
            c_{k,k}\ne0
            \,,
        \end{equation}
        where $Q_1\equiv I$ and $c_{k,k}\ne0$ follows from the nonzero range-$k$ term in $\mathcal{Q}_k$~\cite{GRABOWSKI1995299,Nozawa2020}.
        Hence $Q_k$ is a combination of $\mathcal{Q}_2,\ldots,\mathcal{Q}_k$.

        The identity $\frac{d}{dz}\widetilde t^{8V}(z)=\widetilde t^{8V}(z)\mathcal{Q}(z)$ gives
        \begin{equation}
            \label{eq:log-derivative-recursion-app}
            \mathcal{Q}_k
            =
            (k-1)t_{k-1}
            -
            \sum_{r=1}^{k-2} t_r \mathcal{Q}_{k-r}
            \,,
            \qquad
            k\ge2
            \,.
        \end{equation}
        Thus $\mathcal{Q}_2=t_1$, whereas for $k\ge3$ the products $t_r\mathcal{Q}_{k-r}$ contain contributions supported on separated intervals that cancel only in the full expression.

        Obtaining $\mathcal{Q}_k$ from the logarithmic derivative requires $t_1,\ldots,t_{k-1}$, each evaluated as an operator on the full chain at a cost growing exponentially with $L$.
        Applying Eq.~\eqref{eq:log-derivative-recursion-app} successively then requires expanding $t_{k-1}$ and all products $t_r\mathcal{Q}_{k-r}$ and cancelling their nonlocal contributions exactly.
        Explicit expressions for higher-order local conserved quantities are therefore impractical to extract this way.

        The MPO of Eq.~\eqref{eq:MPOall} requires no such cancellation.
        Expanding its auxiliary trace directly gives the local conserved quantity $Q_k$, while $\epsilon^2=0$ excludes products of extensive operators.
        The expansion yields Eq.~\eqref{eq:xyz-Qk}, the closed form of $Q_k$ for arbitrary $k$.

    \section{Identities of $\sfunc{n}{N_x}{N_y}{N_z}$}
    \label{app:sfunc-identities}
    This appendix proves the identities of $\sfunc{n}{N_x}{N_y}{N_z}$ used in Appendix~\ref{app:proof-identities}, including the relation $\rfunc{n}{0}{0}{0} = 0$ stated below Eq.~\eqref{eq:rfunc}.

    \subsection{Definitions and notation}
    We define the generating function of $\sfunc{n}{N_x}{N_y}{N_z}$ by
    \begin{align}
        \Gsfunc{N_x}{N_y}{N_z}{s}
         & \coloneqq
        \sum_{n=0}^{\infty}
        \sfunc{n}{N_x}{N_y}{N_z}
        s^{n}
        \,.
    \end{align}
    Using the definition of $\sfunc{n}{N_x}{N_y}{N_z}$ in Eq.~\eqref{eq:sfunc}, we have
    \begin{align}
        \Gsfunc{N_x}{N_y}{N_z}{s}
         & =
        \sum_{n=0}^{\infty}
        \sum_{\substack{n_x+n_y+n_z = n \\ n_x, n_y, n_z \ge 0}}
        \binom{N_x+n_x-1}{n_x}
        \binom{N_y+n_y-1}{n_y}
        \binom{N_z+n_z-1}{n_z}
        x^{n_x} y^{n_y} z^{n_z}
        s^{n}
        \nonumber                       \\
         & =
        \prod_{\alpha \in \{x,y,z\}}
        \sum_{n_\alpha=0}^{\infty}
        \binom{N_\alpha+n_\alpha-1}{n_\alpha}
        (s\alpha)^{n_\alpha}
        =
        \frac{1}{(1-sx)^{N_x}(1-sy)^{N_y}(1-sz)^{N_z}}
        \,,
        \label{eq:generating_function_of_Sn}
    \end{align}
    where in the last equality we used the binomial series $(1-t)^{-N} = \sum_{n=0}^{\infty} \binom{N+n-1}{n} t^n$.
        Reading $\binom{N+n-1}{n}$ as a generalized binomial coefficient extends $S_n$ and the generating function~\eqref{eq:generating_function_of_Sn} to arbitrary integer upper indices.

    When the three upper indices coincide, we abbreviate:
    \begin{align}
        \sfuncone{n}{N}  & \coloneqq \sfunc{n}{N}{N}{N} \,, \label{eq:sfuncone-def}   \\
        \rfuncone{n}{N}  & \coloneqq \rfunc{n}{N}{N}{N} \,, \label{eq:rfuncone-def}   \\
        \Gsfuncone{N}{s} & \coloneqq \Gsfunc{N}{N}{N}{s} \,. \label{eq:Gsfuncone-def}
    \end{align}

    \subsection{List of identities}

    Of the relations below, the recursion relations, the vanishing identity, and Eq.~\eqref{eq:f_r_s_relation} are used in Appendix~\ref{app:proof-identities}; the rest are lemmas for their proofs.

    \paragraph{Convolution identity.}
    \begin{align}
        \sfunc{n}{N_x+N^\prime_x}{N_y+N^\prime_y}{N_z+N^\prime_z}
        =
        \sum_{\ntil=0}^{n}
        \sfunc{\ntil}{N_x}{N_y}{N_z}
        \sfunc{n-\ntil}{N^\prime_x}{N^\prime_y}{N^\prime_z}
        \,.
        \label{eq:sfunc_formula_divide}
    \end{align}

    \paragraph{Recursion relations.}
    We use the convention $\sfunc{n}{N_x}{N_y}{N_z}=0$ for $n<0$.
    On the nonnegative lattice, the $x$-, $y$-, and $z$-recursions below are applied for $N_x>0$, $N_y>0$, and $N_z>0$, respectively.
    \begin{align}
        \sfunc{n}{N_x}{N_y}{N_z}
         & =
        \sfunc{n}{N_x-1}{N_y}{N_z}
        +
        x \sfunc{n-1}{N_x}{N_y}{N_z}
        \label{eq:sfunc_formula_recursion_x}
        \\
         & =
        \sfunc{n}{N_x}{N_y-1}{N_z}
        +
        y \sfunc{n-1}{N_x}{N_y}{N_z}
        \label{eq:sfunc_formula_recursion_y}
        \\
         & =
        \sfunc{n}{N_x}{N_y}{N_z-1}
        +
        z \sfunc{n-1}{N_x}{N_y}{N_z}
        \,.
        \label{eq:sfunc_formula_recursion_z}
    \end{align}

    \paragraph{Four-term relation (equal indices).}
    For $N > 0$,
    \begin{align}
        n \sfuncone{n}{N}
        =
        N \qty(
        \sfunc{n}{N + 1}{N}{N} + \sfunc{n}{N}{N+1}{N} + \sfunc{n}{N}{N}{N+1} - 3\sfuncone{n}{N}
        )
        \,.
        \label{eq:sfunc_four_term_relation}
    \end{align}

    \paragraph{Vanishing identity.}
    For $n > 0$,
    \begin{align}
        0
         & =
        4\sfuncone{n}{n} - \sfunc{n}{n + 1}{n}{n} - \sfunc{n}{n}{n+1}{n} - \sfunc{n}{n}{n}{n+1}
        \,.
        \label{eq:sfunc-vanish-identity}
    \end{align}

    \paragraph{Relation between $a_n$ and $\sfuncone{n}{1}$.}
    \begin{align}
        a_n
         & =
        \sfuncone{n}{1}
        \,,
        \label{eq:an_sn}
    \end{align}
    where $a_n$ is the symmetric polynomial defined in Eq.~\eqref{eq:an-def}.

    \paragraph{Relations between $f(n,m)$, $\rfuncone{n}{m}$, and $\sfuncone{n}{N}$.}
    \begin{align}
        f(n,m)
         & =
        \rfuncone{n}{m}
        =
        \frac{m}{n+m} \sfuncone{n}{n+m}
        \,,
        \label{eq:f_r_s_relation}
    \end{align}
    where $f(n,m)$ is the coefficient function defined in Appendix~\ref{app:review_xyz}.

    \subsection{Proof of the identities~\eqref{eq:sfunc_formula_divide}--\eqref{eq:f_r_s_relation}}

    \subsubsection*{Proof of the convolution identity~\eqref{eq:sfunc_formula_divide}}

    Equation~\eqref{eq:generating_function_of_Sn} gives
    \begin{align}
        \Gsfunc{N_x+N^\prime_x}{N_y+N^\prime_y}{N_z+N^\prime_z}{s}
        =
        \Gsfunc{N_x}{N_y}{N_z}{s}
        \Gsfunc{N^\prime_x}{N^\prime_y}{N^\prime_z}{s}
        \,.
    \end{align}
    Extracting the coefficient of $s^n$ from both sides yields Eq.~\eqref{eq:sfunc_formula_divide}.
    \hfill $\square$

    \subsubsection*{Proof of the recursion relations~\eqref{eq:sfunc_formula_recursion_x}--\eqref{eq:sfunc_formula_recursion_z}}

    We prove Eq.~\eqref{eq:sfunc_formula_recursion_x}; the cases for $y$ and $z$ follow by the same argument.
    \begin{align}
        \Gsfunc{N_x}{N_y}{N_z}{s}
         & =
        \Gsfunc{N_x-1}{N_y}{N_z}{s}
        (1-sx)^{-1}
        \nonumber \\
         & =
        \Gsfunc{N_x-1}{N_y}{N_z}{s}
        +
        \qty[ (1-sx)^{-1}-1 ]
        \Gsfunc{N_x-1}{N_y}{N_z}{s}
        \nonumber \\
         & =
        \Gsfunc{N_x-1}{N_y}{N_z}{s}
        +
        sx
        \Gsfunc{N_x}{N_y}{N_z}{s}
        \,.
    \end{align}
    Extracting the coefficient of $s^n$ from both sides yields Eq.~\eqref{eq:sfunc_formula_recursion_x}.
    \hfill $\square$
    \subsubsection*{Proof of the four-term relation~\eqref{eq:sfunc_four_term_relation}}

    Rearranging the final line of the previous proof shows that $sx\,\Gsfunc{N_x}{N_y}{N_z}{s}+\Gsfunc{N_x-1}{N_y}{N_z}{s}$ equals $\Gsfunc{N_x}{N_y}{N_z}{s}$.
    The analogous identities hold for $y$ and $z$.
    Using these relations, we have
    \begin{align}
        s
        \pdv{s}
        \Gsfunc{N_x}{N_y}{N_z}{s}
         & =
        N_x x \Gsfunc{N_x+1}{N_y}{N_z}{s}
        +
        N_y y \Gsfunc{N_x}{N_y+1}{N_z}{s}
        +
        N_z z \Gsfunc{N_x}{N_y}{N_z+1}{s}
        \nonumber \\
         & =
        N_x \qty(\Gsfunc{N_x+1}{N_y}{N_z}{s} - \Gsfunc{N_x}{N_y}{N_z}{s})
        +
        N_y \qty(\Gsfunc{N_x}{N_y+1}{N_z}{s} - \Gsfunc{N_x}{N_y}{N_z}{s})
        \nonumber \\
         & \qquad
        +
        N_z \qty(\Gsfunc{N_x}{N_y}{N_z+1}{s} - \Gsfunc{N_x}{N_y}{N_z}{s})
        \,.
    \end{align}
    For equal indices $N_x = N_y = N_z = N$, this becomes
    \begin{align}
        s
        \pdv{s}
        \Gsfuncone{N}{s}
        =
        N \qty(
        \Gsfunc{N + 1}{N}{N}{s} + \Gsfunc{N}{N + 1}{N}{s} + \Gsfunc{N}{N}{N + 1}{s} - 3\Gsfuncone{N}{s}
        )
        \,.
    \end{align}
    Extracting the coefficient of $s^n$ yields Eq.~\eqref{eq:sfunc_four_term_relation}.
    \hfill $\square$
    \subsubsection*{Proof of the vanishing identity~\eqref{eq:sfunc-vanish-identity}}

    Rearranging the four-term relation~\eqref{eq:sfunc_four_term_relation}, we obtain
    \begin{align}
        \qty(1-\frac{n}{N})\sfuncone{n}{N}
        =
        4\sfuncone{n}{N} - \sfunc{n}{N + 1}{N}{N} - \sfunc{n}{N}{N+1}{N} - \sfunc{n}{N}{N}{N+1}
        \,.
        \label{eq:sfunc_simplify}
    \end{align}
    Setting $n = N$ makes the left-hand side vanish, yielding Eq.~\eqref{eq:sfunc-vanish-identity}.
    By the definition~\eqref{eq:rfunc}, the right-hand side of Eq.~\eqref{eq:sfunc-vanish-identity} equals $\rfunc{n}{0}{0}{0}$, so $\rfunc{n}{0}{0}{0}=0$ for $n>0$.
    \hfill $\square$

    \subsubsection*{Proof of Eq.~\eqref{eq:an_sn}}

    From Eq.~\eqref{eq:an-def}, the generating function of $a_n$ is
    \begin{align}
        \sum_{n = 0}^{\infty} a_n s^n
         & =
        \frac{
            x\paren{\frac{y^2}{1-sy}-\frac{z^2}{1-sz}}
            +
            y\paren{\frac{z^2}{1-sz}-\frac{x^2}{1-sx}}
            +
            z\paren{\frac{x^2}{1-sx}-\frac{y^2}{1-sy}}
        }{
            \left(x-y\right)\left(y-z\right)\left(z-x\right)
        }
        \nonumber \\
         & =
        \frac{1}
        {
            (1-sx)(1-sy)(1-sz)
        }
        =
        \Gsfuncone{1}{s}
        \,,
    \end{align}
    where we used Eq.~\eqref{eq:generating_function_of_Sn} in the last equality.
    Extracting the coefficient of $s^n$ from both sides yields Eq.~\eqref{eq:an_sn}.
    \hfill $\square$

    \subsubsection*{Proof of Eq.~\eqref{eq:f_r_s_relation}}

    We first show that $\rfuncone{n}{m} = \frac{m}{n+m} \sfuncone{n}{n+m}$.
    Using Eqs.~\eqref{eq:rfunc} and~\eqref{eq:rfuncone-def} for the first equality and Eq.~\eqref{eq:sfunc_simplify} for the second, we have
    \begin{align}
        \rfuncone{n}{m}
         & =
        4 \sfuncone{n}{n+m}
        -
        \sfunc{n}{n+m+1}{n+m}{n+m}
        -
        \sfunc{n}{n+m}{n+m+1}{n+m}
        -
        \sfunc{n}{n+m}{n+m}{n+m+1}
        \nonumber \\
         & =
        \qty(1-\frac{n}{n+m}) \sfuncone{n}{n+m}
        =
        \frac{m}{n+m} \sfuncone{n}{n+m}
        \,.
    \end{align}

    Next, we show that $f(n,m) = \rfuncone{n}{m}$:
    \begin{align}
        f(n,m)
         & =
        \frac{m}{n+m} \sum_{p=1}^{n}\binom{n+m}{p} \sum_{\substack{j_1+j_2+\cdots+j_p=n \\ j_1, j_2, \ldots, j_p \geq 1}} a_{j_1} a_{j_2} \cdots a_{j_p}
        \nonumber                                                                       \\
         & =
        \frac{m}{n+m} \sum_{p=1}^{n}\binom{n+m}{p}
        \lr{.}{|}{
            \paren{ \Gsfuncone{1}{s}-1 }^p
        }_{s^n}
        \nonumber                                                                       \\
         & =
        \frac{m}{n+m}
        \lr{.}{|}{
            \qty(\Gsfuncone{1}{s})^{n+m}
        }_{s^n}
        =
        \frac{m}{n+m}
        \sfuncone{n}{n+m}
        =
        \rfuncone{n}{m}
        \,,
    \end{align}
    where in the first equality we used the definition of $f(n,m)$ in Appendix~\ref{app:review_xyz}, in the second equality we used Eq.~\eqref{eq:an_sn}, in the third equality we used the binomial theorem and $\lr{.}{|}{\paren{ \Gsfuncone{1}{s}-1 }^p}_{s^n} = 0$ for $p > n$, and in the last equality we used the relation just proven above.
    Here $\lr{.}{|}{g(s)}_{s^n}$ denotes the coefficient of $s^n$ in the power series $g(s)$.
    \hfill $\square$

    \section{Proof of identities}

    \label{app:proof-identities}

    Appendix~\ref{app:xyz-Qk} shows that the simplified expression~\eqref{eq:xyz-Qk} of the local conserved quantities coincides with the original result of Ref.~\cite{Nozawa2020}.
    Appendix~\ref{app:rfunc-combinatoric} proves the combinatorial representation~\eqref{eq:rfunc-combinatoric}.
    The remaining subsections prove Eqs.~\eqref{eq:R-identity-diff}, \eqref{eq:R-identity}, \eqref{eq:fx-ito-fe}, and \eqref{eq:omega-fxyz}.

    \begin{figure}[tbh]
        \centering
        \begin{tikzpicture}[scale=0.45]

            \def\n{11}            %
            \def\N{4}             %
            \def\m{2}             %
            \def\ntilcol{5}       %
            \pgfmathsetmacro{\nminusntil}{\n - \ntilcol}
            \pgfmathsetmacro{\diagStartY}{\N + 3*\m}                %
            \pgfmathsetmacro{\totalH}{\N + \n + 3*\m}               %
            \pgfmathsetmacro{\blueTopLeftY}{\diagStartY + \ntilcol} %

            \definecolor{redRegion}{RGB}{255,230,230}
            \definecolor{blueRegion}{RGB}{230,230,255}
            \definecolor{redPath}{RGB}{220,80,60}
            \definecolor{bluePath}{RGB}{60,60,200}

            \coordinate (O) at (0, 0);
            \coordinate (A) at (\ntilcol, \N);                          %
            \coordinate (A0) at (\ntilcol, 0);
            \coordinate (B) at (\ntilcol, \N + 1);                      %
            \coordinate (F) at (\n, \totalH);                           %

            \coordinate (DiagStart) at (0, \diagStartY);
            \coordinate (DiagEnd) at (\n, \totalH);

            \coordinate (BlueTopLeft) at (\ntilcol, \blueTopLeftY);
            \coordinate (BlueBottomRight) at (\n, \N + 1);

            \fill[redRegion] (O) -- (0, \N) -- (A) -- (A0) -- cycle;

            \fill[blueRegion] (B) -- (BlueTopLeft) -- (F) -- (BlueBottomRight) -- cycle;

            \draw[black, thick] (O) rectangle (\n, \totalH);
            \draw[blue!70!black, line width=1pt] (DiagStart) -- (DiagEnd);
            \draw[dashed, black!50] (0, \N) -- (\n, \N);

            \draw[redPath, ultra thick, line cap=round, line join=round]
            (O) -- ++(0,1) -- ++(1,0) -- ++(0,2) -- ++(2,0) -- ++(0,1) -- ++(1,0) -- (A);

            \draw[bluePath, ultra thick, line cap=round, line join=round]
            (B) -- ++(0,2) -- ++(2,0) -- ++(0,3) -- ++(2,0) -- ++(0,4) -- ++(2,0) -- (F);

            \fill[black] (O) circle (4pt);
            \fill[black] (F) circle (4pt);
            \fill[black] (A) circle (4pt);
            \fill[black] (B) circle (4pt);
            \draw[black, thick] (A) -- (B);
            \draw[black, ->] (\ntilcol + 1, \N + 0.5) -- (\ntilcol + 0.3, \N + 0.5);
            \node[right, font=\scriptsize] at (\ntilcol + 1, \N + 0.5) {1 North step};

            \node[below left, font=\small] at (O) {$(0,1)$};
            \node[above right, font=\small] at (F) {$(n, N+3n+3m)$};

            \draw[decorate, decoration={brace, amplitude=8pt}]
            (-0.4, 0) -- (-0.4, \N);
            \node[left, font=\scriptsize] at (-1.0, \N*0.5) {$N$ rows};

            \draw[decorate, decoration={brace, amplitude=6pt}]
            (-0.4, \N+0.5) -- (-0.4, \diagStartY);
            \node[left, font=\scriptsize] at (-1.0, \N + 1.5*\m) {$m$ bundles};

            \draw[decorate, decoration={brace, amplitude=8pt}]
            (-0.4, \diagStartY) -- (-0.4, \totalH);
            \node[left, font=\scriptsize] at (-1.0, \diagStartY + \n*0.5) {$n$ bundles};

            \draw[decorate, decoration={brace, amplitude=6pt, mirror}]
            (0, -0.5) -- (\ntilcol - 0.1, -0.5);
            \node[below, font=\scriptsize] at (\ntilcol*0.5, -1.0) {$\ntil$};

            \draw[decorate, decoration={brace, amplitude=6pt, mirror}]
            (\ntilcol + 0.1, -0.5) -- (\n, -0.5);
            \node[below, font=\scriptsize] at (\ntilcol + \nminusntil*0.5, -1.0) {$n{-}\ntil$};

            \draw[decorate, decoration={brace, amplitude=8pt}]
            (\ntilcol - 0.5, \N + 1-0.5) -- (\ntilcol - 0.5, \blueTopLeftY);
            \node[left, font=\scriptsize] at (\ntilcol - 1, {(\N + 1 - 0.5 + \blueTopLeftY)*0.5}) {$m{+}\ntil$ bundles};

            \node[redPath, font=\small] at (\ntilcol*0.5+0.5, \N*0.5-0.3) {$\sfunc{\ntil}{N_x}{N_y}{N_z}$};
            \node[bluePath, font=\small] at (\ntilcol + \nminusntil*0.5 + 1.5, \N + 3.5) {$\rfuncone{n{-}\ntil}{m{+}\ntil}$};

        \end{tikzpicture}
        \caption{
            Combinatorial proof of the identity~\eqref{eq:sfunc-R1-identity}.
            The lattice represents $\rfunc{n}{N_x+m}{N_y+m}{N_z+m}$.
            We arrange the $N + 3m$ flavored rows of the lower region as $N = N_x + N_y + N_z$ rows followed by $m$ bundles, and place $n$ bundles in the upper region.
            Good paths must stay weakly below the diagonal line, which starts above the $N$ rows and the $m$ bundles.
            Any path can be decomposed at column $\ntil$ where it crosses the horizontal line $y = N$: the lower red region contributes $\sfunc{\ntil}{N_x}{N_y}{N_z}$, while the upper blue region contributes $\rfuncone{n-\ntil}{m+\ntil}$.
            Summing over $\ntil$ establishes the identity.
        }
        \label{fig:S-R1-identity}
    \end{figure}
    \subsection{Proof of Eq.~\eqref{eq:xyz-Qk}}
    \label{app:xyz-Qk}

    Appendix~\ref{app:review_xyz} reduces this claim to Eq.~\eqref{eq:to-prove-app1}, namely $\rfuncpre{n}{m}{N_x}{N_y}{N_z} = \rfunc{n}{N_x+m}{N_y+m}{N_z+m}$.
    From the definition~\eqref{eq:rfunc_pre_rewrite} and Eq.~\eqref{eq:f_r_s_relation}, we have
    \begin{equation*}
        \rfuncpre{n}{m}{N_x}{N_y}{N_z}
        =
        \sum_{\ntil=0}^{n} f(n-\ntil, m+\ntil) \sfunc{\ntil}{N_x}{N_y}{N_z}
        =
        \sum_{\ntil=0}^{n} \rfuncone{n-\ntil}{m+\ntil} \sfunc{\ntil}{N_x}{N_y}{N_z}
        \,.
    \end{equation*}
    Therefore, Eq.~\eqref{eq:to-prove-app1} is equivalent to the identity
    \begin{align}
        \label{eq:sfunc-R1-identity}
        \sum_{\ntil=0}^{n} \rfuncone{n-\ntil}{m+\ntil} \sfunc{\ntil}{N_x}{N_y}{N_z}
        =
        \rfunc{n}{N_x+m}{N_y+m}{N_z+m}
        \,.
    \end{align}

    Let $N = N_x + N_y + N_z$.
    The right-hand side counts good paths of $n$ East steps on a lattice with $N + 3m$ flavored rows and $n$ bundles, where the lower $N$ rows consist of $N_x$ rows of flavor $x$, $N_y$ rows of flavor $y$, and $N_z$ rows of flavor $z$, followed by $m$ bundles (each containing one row of each flavor).
    In units of single rows, the diagonal line connects $(0, N + 3m)$ to $(n, N + 3n + 3m)$ with slope $3$; in Fig.~\ref{fig:S-R1-identity}, each three-row bundle is drawn as a single row.

    As shown in Fig.~\ref{fig:S-R1-identity}, any good path from $(0,1)$ to $(n, N + 3n + 3m)$ can be uniquely decomposed at column $\ntil$ where it first crosses the horizontal line $y = N$.
    This decomposition separates the path into two portions:
    \begin{itemize}
        \item The lower portion (red region) consists of $\ntil$ steps in the region $y \le N$, counted by $\sfunc{\ntil}{N_x}{N_y}{N_z}$.

        \item The upper portion (blue region) consists of $n - \ntil$ East steps in the region $y > N$ and stays weakly below the diagonal.
              Its left edge spans $m + \ntil$ bundles, so this portion is counted by $\rfuncone{n - \ntil}{m + \ntil}$.
    \end{itemize}
        Splitting the path at the North step from $(\ntil,N)$ to $(\ntil,N+1)$, the total weight factorizes into the weights of the lower and upper portions.

    \newcommand{\pathSRHS}{%
        \begin{tikzpicture}[baseline=-0.5ex, scale=0.20]
            \draw[thick] (0,0) rectangle (8,14);
            \draw[dashed, black!50] (0,3) -- (8,3);
            \draw[blue!70!black, line width=1pt] (0,6) -- (8,14);
            \draw[decorate, decoration={brace, amplitude=2pt}]
            (-0.3, 0) -- (-0.3, 3);
            \node[left, font=\tiny] at (-0.5, 1.5) {$N$};
            \draw[decorate, decoration={brace, amplitude=2pt}]
            (-0.3, 3) -- (-0.3, 6);
            \node[left, font=\tiny] at (-0.5, 4.5) {$3m$};
        \end{tikzpicture}%
    }

    \newcommand{\pathSDecomp}{%
        \begin{tikzpicture}[baseline=-0.5ex, scale=0.20]
            \definecolor{redRegion}{RGB}{255,230,230}
            \definecolor{blueRegion}{RGB}{230,230,255}
            \fill[redRegion] (0,0) -- (0,3) -- (3,3) -- (3,0) -- cycle;
            \fill[blueRegion] (3,3.5) -- (3,9) -- (8,14) -- (8,3.5) -- cycle;
            \draw[thick] (0,0) rectangle (8,14);
            \draw[dashed, black!50] (0,3) -- (8,3);
            \draw[blue!70!black, line width=1pt] (0,6) -- (8,14);
            \fill[black] (3,3) circle (3pt);
            \fill[black] (3,3.5) circle (3pt);
            \draw[black, thick] (3,3) -- (3,3.5);
            \draw[decorate, decoration={brace, amplitude=2pt, mirror}]
            (0, -0.3) -- (3, -0.3);
            \node[below, font=\tiny] at (1.5, -0.5) {$\ntil$};
            \draw[decorate, decoration={brace, amplitude=2pt, mirror}]
            (3.1, -0.3) -- (8, -0.3);
            \node[below, font=\tiny] at (5.5, -0.5) {$n{-}\ntil$};
        \end{tikzpicture}%
    }

    \newcommand{\pathSRed}{%
        \begin{tikzpicture}[baseline=-0.5ex, scale=0.2]
            \definecolor{redRegion}{RGB}{255,230,230}
            \fill[redRegion] (0,0) -- (0,3) -- (3,3) -- (3,0) -- cycle;
            \draw[thick] (0,0) rectangle (3,3);
            \draw[decorate, decoration={brace, amplitude=2pt}]
            (-0.3, 0) -- (-0.3, 3);
            \node[left, font=\tiny] at (-0.5, 1.5) {$N$};
            \draw[decorate, decoration={brace, amplitude=2pt, mirror}]
            (0, -0.3) -- (3, -0.3);
            \node[below, font=\tiny] at (1.5, -0.5) {$\ntil$};
        \end{tikzpicture}%
    }

    \newcommand{\pathSBlue}{%
        \begin{tikzpicture}[baseline=-0.5ex, scale=0.16]
            \definecolor{blueRegion}{RGB}{230,230,255}
            \fill[blueRegion] (0,0) -- (0,5) -- (5,10) -- (5,0) -- cycle;
            \draw[thick] (0,0) rectangle (5,10);
            \draw[blue!70!black, line width=0.8pt] (0,5) -- (5,10);
            \draw[decorate, decoration={brace, amplitude=2pt}]
            (-0.3, 0) -- (-0.3, 5);
            \node[left, font=\tiny] at (-0.7, 2.5) {$m{+}\ntil$};
            \draw[decorate, decoration={brace, amplitude=2pt, mirror}]
            (0, -0.3) -- (5, -0.3);
            \node[below, font=\tiny] at (2.5, -0.5) {$n{-}\ntil$};
        \end{tikzpicture}%
    }

    Summing over $\ntil$ gives the identity~\eqref{eq:sfunc-R1-identity}.
    Schematically, the decomposition reads
    \begin{align}
        \rfunc{n}{N_x+m}{N_y+m}{N_z+m}
         & =
        \raisebox{-2.5em}{\pathSRHS}
        \ = \
        \sum_{\ntil=0}^{n}
        \raisebox{-2.5em}{\pathSDecomp}
        \notag \\[0.5em]
         & = \
        \sum_{\ntil=0}^{n}
        \raisebox{-0.8em}{\pathSRed}
        \times
        \raisebox{-1.5em}{\pathSBlue}
        \ = \
        \sum_{\ntil=0}^{n}
        \sfunc{\ntil}{N_x}{N_y}{N_z}
        \cdot
        \rfuncone{n-\ntil}{m+\ntil}
        \,.
        \notag
    \end{align}
    \hfill $\square$

    \subsection{Proof of Eq.~\eqref{eq:rfunc-combinatoric}}\label{app:rfunc-combinatoric}
    \begin{figure}[tbh]
        \centering
        \begin{tikzpicture}[scale=0.75]

            \def\ncols{8}
            \def\nrowsbelow{4}
            \def\nbundles{8}
            \def\ntil{4}
            \pgfmathsetmacro{\totalrows}{\nrowsbelow + \nbundles}
            \pgfmathsetmacro{\contactY}{\nrowsbelow + \ntil}
            \pgfmathsetmacro{\blueBottomY}{\contactY + 1}

            \definecolor{goodRegion}{RGB}{255,230,230}
            \definecolor{remainRegion}{RGB}{230,230,255}
            \definecolor{goodPathColor}{RGB}{220,80,60}
            \definecolor{remainPathColor}{RGB}{60,60,200}

            \fill[goodRegion] (0, 0) -- (0, \nrowsbelow) -- (\ntil, \contactY) -- (\ntil, 0) -- cycle;

            \fill[remainRegion] (\ntil, \blueBottomY) -- (\ntil, \totalrows) -- (\ncols, \totalrows) -- (\ncols, \blueBottomY) -- cycle;

            \draw[black, thick] (0, 0) -- (0, \totalrows) -- (\ncols, \totalrows) -- (\ncols, 0) -- cycle;

            \draw[blue!70!black, line width=2pt] (0, \nrowsbelow) -- (\ncols, \totalrows);

            \draw[goodPathColor, ultra thick, line cap=round, line join=round]
            (0,0) -- (0,2) -- (1,2) -- (1,4) -- (2,4) -- (2,6) -- (3,6) -- (4,6) -- (4,8);

            \draw[black, ultra thick, line cap=round, line join=round]
            (4,8) -- (4,9);

            \draw[remainPathColor, ultra thick, line cap=round, line join=round]
            (\ntil,9) -- (\ntil,10) -- (\ntil+3,10) -- (\ntil+3,12) -- (\ncols,\nrowsbelow+\nbundles);

            \fill[black] (\ntil, 8) circle (4pt);
            \node[right, font=\small] at (4.1, 7.9) {cross};

            \fill[black] (\ntil, 9) circle (4pt);

            \fill[black] (0,0) circle (4pt);
            \fill[black] (\ncols,\totalrows) circle (4pt);
            \node[below left, font=\small] at (0,0) {$(0,1)$};
            \node[above right, font=\small] at (\ncols,\totalrows) {$(n, N+3n)$};

            \draw[dashed, black!50] (\ntil, 0) -- (\ntil, \totalrows);

            \draw[decorate, decoration={brace, amplitude=8pt, mirror}]
            (-0.3, \nrowsbelow) -- (-0.3, 0);
            \node[left, font=\footnotesize, align=right] at (-0.8, 2) {$N$ rows};

            \draw[decorate, decoration={brace, amplitude=8pt, mirror}]
            (-0.3, \totalrows) -- (-0.3, \nrowsbelow);
            \node[left, font=\footnotesize, align=right] at (-0.8, 7) {$n$ bundles};

            \draw[decorate, decoration={brace, amplitude=6pt, mirror}]
            (0, -0.6) -- (\ntil-0.1, -0.6);
            \node[below, font=\small] at (2, -1.1) {$\tilde{n}$ steps};

            \draw[decorate, decoration={brace, amplitude=6pt, mirror}]
            (\ntil+0.1, -0.6) -- (\ncols, -0.6);
            \node[below, font=\small] at (6.25, -1.1) {$n - \tilde{n}$ steps};

            \node[goodPathColor, font=\small, align=center] at (2.5, 1.5) {$\rfunc{\tilde{n}}{N_x}{N_y}{N_z}$};

            \node[remainPathColor, font=\small, align=center] at (5, 11) {$\sfuncone{n-\tilde{n}}{n-\tilde{n}}$};

            \draw[black, ->, thick] (3, 8.5) -- (4, 8.5);
            \node[left, font=\footnotesize] at (3, 8.5) {1 North step};

        \end{tikzpicture}
        \caption{
            Decomposition of paths for proving Eqs.~\eqref{eq:sfunc-decomp-1} and \eqref{eq:sfunc-decomp-2}.
            The lower region contains $N = N_x + N_y + N_z$ flavored rows, as in Fig.~\ref{fig:Sfunc-combinatoric}, and the upper region contains $n$ bundles, each consisting of one row of each flavor $x$, $y$, $z$, as in Fig.~\ref{fig:Rfunc-combinatoric}.
            The diagonal line connects $(0, N)$ and $(n, N+3n)$ with slope $3$.
            Any path from $(0,1)$ to $(n, N+3n)$ can be uniquely decomposed at the column $\tilde{n}$ of the first North step that takes the path strictly above the diagonal.
            The portion before the crossing (red region) is a good path, contributing to $\rfunc{\tilde{n}}{N_x}{N_y}{N_z}$, while the portion after crossing (blue region) contributes to $\sfuncone{n-\tilde{n}}{n-\tilde{n}}$.
            Summing over $\tilde{n} = 0, 1, \ldots, n$ yields Eq.~\eqref{eq:sfunc-decomp-1}.
            For Eq.~\eqref{eq:sfunc-decomp-2}, one adds an $x$-flavored row on top, so that paths end at $(n, N+3n+1)$; the diagonal remains unchanged, and the portion after crossing contributes to $\sfunc{n-\tilde{n}}{n-\tilde{n}+1}{n-\tilde{n}}{n-\tilde{n}}$.
        }
        \label{fig:sfunc-decomposition}
    \end{figure}
    The proof adapts the argument of Ref.~\cite{hilton-catalan-1991} to weighted lattice-path counting.

    By the definition~\eqref{eq:rfunc} of $\rfunc{n}{N_x}{N_y}{N_z}$, this amounts to proving
    \begin{align}
        \label{eq:rfunc-combinatoric-re}
         & 4\sfunc{n}{N_x+n}{N_y+n}{N_z+n}
        -
        \sfunc{n}{N_x+n+1}{N_y+n}{N_z+n}
        \nonumber                          \\
         & \hspace*{6em}
        -
        \sfunc{n}{N_x+n}{N_y+n+1}{N_z+n}
        -
        \sfunc{n}{N_x+n}{N_y+n}{N_z+n+1} = \sum_{\gamma: \text{good path}} w(\gamma),
    \end{align}
    where the sum on the right-hand side runs over all good paths $\gamma$ from $(0,1)$ to $(n, N+3n)$ with $N \equiv N_x + N_y + N_z$, as described in Section~\ref{sec:combinatoric-rfunc}.

    In the remainder of this subsection, $\rfunc{\tilde{n}}{N_x}{N_y}{N_z}$ denotes the weighted sum over good paths on the right-hand side of Eq.~\eqref{eq:rfunc-combinatoric}.

    We first consider the decomposition of $\sfunc{n}{N_x+n}{N_y+n}{N_z+n}$.
    As illustrated in Fig.~\ref{fig:sfunc-decomposition}, any path from $(0,1)$ to $(n, N+3n)$ can be uniquely decomposed according to where it first touches the diagonal line connecting $(0, N)$ and $(n, N+3n)$.
    Let $\tilde{n}$ denote the column index of this first contact point.
    The portion of the path before this point constitutes a good path contributing to $\rfunc{\tilde{n}}{N_x}{N_y}{N_z}$, while the remaining portion, obtained by discarding the crossing step, contributes to $\sfuncone{n-\tilde{n}}{n-\tilde{n}}$.
    This yields the decomposition:
    \begin{align}
        \label{eq:sfunc-decomp-1}
        \sfunc{n}{N_x+n}{N_y+n}{N_z+n}
        =
        \sum_{\tilde{n} = 0}^{n}
        \rfunc{\tilde{n}}{N_x}{N_y}{N_z}
        \sfuncone{n-\tilde{n}}{n-\tilde{n}}.
    \end{align}

    The same decomposition argument applies to $\sfunc{n}{N_x+n+1}{N_y+n}{N_z+n}$.
    In this case, we consider paths from $(0,1)$ to $(n, N+3n+1)$, which corresponds to adding one $x$-flavored row on top of the $n$ bundles.
    Since the diagonal line remains unchanged, the good path portion before the first crossing still contributes to $\rfunc{\tilde{n}}{N_x}{N_y}{N_z}$.
    The remaining portion after crossing now traverses $n - \tilde{n}$ bundles plus the extra $x$-flavored row, contributing to $\sfunc{n-\tilde{n}}{n-\tilde{n}+1}{n-\tilde{n}}{n-\tilde{n}}$.
    This yields:
    \begin{align}
        \label{eq:sfunc-decomp-2}
        \sfunc{n}{N_x+n+1}{N_y+n}{N_z+n}
        =
        \sum_{\tilde{n} = 0}^{n}
        \rfunc{\tilde{n}}{N_x}{N_y}{N_z}
        \sfunc{n-\tilde{n}}{n-\tilde{n}+1}{n-\tilde{n}}{n-\tilde{n}},
    \end{align}
    and analogous expressions hold for $\sfunc{n}{N_x+n}{N_y+n+1}{N_z+n}$ and $\sfunc{n}{N_x+n}{N_y+n}{N_z+n+1}$, where the extra row carries flavor $y$ or $z$, respectively.

    Substituting Eq.~\eqref{eq:sfunc-decomp-1}, Eq.~\eqref{eq:sfunc-decomp-2}, and its $y$- and $z$-flavor analogues into the left-hand side of Eq.~\eqref{eq:rfunc-combinatoric-re}, we obtain
    \begin{align}
          & 4\sfunc{n}{N_x+n}{N_y+n}{N_z+n}
        -
        \sfunc{n}{N_x+n+1}{N_y+n}{N_z+n}
        \nonumber                           \\
          & \hspace*{6em}
        -
        \sfunc{n}{N_x+n}{N_y+n+1}{N_z+n}
        -
        \sfunc{n}{N_x+n}{N_y+n}{N_z+n+1}
        \nonumber                           \\
        = &
        \sum_{\tilde{n} = 0}^{n}
        \rfunc{\tilde{n}}{N_x}{N_y}{N_z}
        \Bigl[
            4\sfuncone{n-\tilde{n}}{n-\tilde{n}}
            -
            \sfunc{n-\tilde{n}}{n-\tilde{n} + 1}{n-\tilde{n}}{n-\tilde{n}}
        \nonumber                           \\
          & \hspace*{8em}
            -
            \sfunc{n-\tilde{n}}{n-\tilde{n}}{n-\tilde{n}+1}{n-\tilde{n}}
            -
            \sfunc{n-\tilde{n}}{n-\tilde{n}}{n-\tilde{n}}{n-\tilde{n}+1}
            \Bigr]
        \nonumber                           \\
        = & \;
        \rfunc{n}{N_x}{N_y}{N_z},
    \end{align}
    where we used the identity~\eqref{eq:sfunc-vanish-identity} for the terms with $\tilde{n} \neq n$, which all vanish.
    The term with $\tilde{n} = n$ survives because $4\sfuncone{0}{0} - \sfunc{0}{1}{0}{0} - \sfunc{0}{0}{1}{0} - \sfunc{0}{0}{0}{1} = 4 - 1 - 1 - 1 = 1$.

    \hfill $\square$

    \subsection{Proof of Eq.~\eqref{eq:R-identity-diff}}
    \label{app:R-identity-diff-proof}
        For $n<0$, Eq.~\eqref{eq:R-identity-diff} is immediate from $\rfunc{n}{N_x}{N_y}{N_z}=0$.
        For $n\ge0$, the proof below uses only the recursion relations of Appendix~\ref{app:sfunc-identities}, which hold for all integer upper indices, and therefore proves Eq.~\eqref{eq:R-identity-diff} for every $\vec N\in\mathbb Z^3$.
        We prove the identity for $(N_x, x)$; the $(N_y, y)$ and $(N_z, z)$ cases follow in the same way.
        Starting from the definition~\eqref{eq:rfunc}, we compute
        \begingroup
        \allowdisplaybreaks[4]
        \begin{align}
              &
            \rfunc{n}{N_x-1}{N_y}{N_z}
            +
            x \rfunc{n-1}{N_x+1}{N_y+1}{N_z+1}
            \nonumber         \\
            = &
            4\sfunc{n}{N_x-1+n}{N_y+n}{N_z+n}
            -
            \sfunc{n}{N_x+n}{N_y+n}{N_z+n}
            \nonumber         \\
              & \hspace*{6em}
            -
            \sfunc{n}{N_x-1+n}{N_y+n+1}{N_z+n}
            -
            \sfunc{n}{N_x-1+n}{N_y+n}{N_z+n+1}
            \nonumber         \\
              &
            \hspace*{3em}
            +
            x\bigl[4\sfunc{n-1}{N_x+n}{N_y+n}{N_z+n}
                -
                \sfunc{n-1}{N_x+n+1}{N_y+n}{N_z+n}
            \nonumber         \\
              & \hspace*{9em}
                -
                \sfunc{n-1}{N_x+n}{N_y+n+1}{N_z+n}
                -
                \sfunc{n-1}{N_x+n}{N_y+n}{N_z+n+1}
                \bigr]
            \nonumber         \\
            = &
            4\qty( \sfunc{n}{N_x-1+n}{N_y+n}{N_z+n} + x \sfunc{n-1}{N_x+n}{N_y+n}{N_z+n})
            \nonumber         \\
              & \hspace*{1em}
            -
            \qty( \sfunc{n}{N_x+n}{N_y+n}{N_z+n} + x \sfunc{n-1}{N_x+n+1}{N_y+n}{N_z+n} )
            \nonumber         \\
              & \hspace*{1em}
            -
            \qty( \sfunc{n}{N_x-1+n}{N_y+n+1}{N_z+n} + x \sfunc{n-1}{N_x+n}{N_y+n+1}{N_z+n})
            \nonumber         \\
              & \hspace*{1em}
            -
            \qty( \sfunc{n}{N_x-1+n}{N_y+n}{N_z+n+1} + x \sfunc{n-1}{N_x+n}{N_y+n}{N_z+n+1} )
            \nonumber         \\
            = &
            4 \sfunc{n}{N_x+n}{N_y+n}{N_z+n}
            -
            \sfunc{n}{N_x+n + 1 }{N_y+n}{N_z+n}
            \nonumber         \\
              & \hspace*{3em}
            -
            \sfunc{n}{N_x+n}{N_y+n + 1}{N_z+n}
            -
            \sfunc{n}{N_x+n}{N_y+n}{N_z+n+ 1}
            \nonumber         \\
            = &
            \rfunc{n}{N_x}{N_y}{N_z}
            \,,
        \end{align}
        \endgroup
        where the first and last equalities use the definition~\eqref{eq:rfunc}, the second equality regroups the terms into four pairs, and the third equality applies the recursion relation~\eqref{eq:sfunc_formula_recursion_x} to each pair.
        \hfill $\square$

    \begin{figure}[tb]
        \centering
        \begin{tikzpicture}[scale=0.43]

            \def\n{12}            %
            \def\N{3}             %
            \def\Np{4}            %
            \def\nx{3}            %
            \def\ntilcol{6}       %
            \pgfmathsetmacro{\nrest}{\n - \nx - \ntilcol}           %
            \pgfmathsetmacro{\NplusNp}{\N + \Np}                    %
            \pgfmathsetmacro{\totalH}{\N + \Np + \n}                %

            \definecolor{redRegion}{RGB}{255,215,215}
            \definecolor{redRegionLight}{RGB}{255,242,242}  %
            \definecolor{blueRegion}{RGB}{230,230,255}
            \definecolor{redPath}{RGB}{220,80,60}
            \definecolor{bluePath}{RGB}{60,60,200}

            \begin{scope}
                \node[font=\bfseries] at (1, \totalH + 1.2) {(a)};

                \coordinate (O) at (0, 0);
                \coordinate (A) at (\nx, \N);                                       %
                \coordinate (A0) at (\nx, 0);                                       %

                \coordinate (B) at (\nx, \N + 1);                                   %
                \coordinate (C) at (\nx, \NplusNp);                                 %
                \coordinate (D) at (\nx + \ntilcol, \NplusNp + \ntilcol);           %
                \coordinate (D0) at (\nx + \ntilcol, \N + 1);                     %

                \coordinate (E) at (\nx + \ntilcol, \NplusNp + \ntilcol + 1);       %
                \coordinate (E0) at (\nx + \ntilcol, \NplusNp + \ntilcol + \nrest); %
                \coordinate (F) at (\n, \totalH);                                   %
                \coordinate (Fbot) at (\n, \NplusNp + \ntilcol + 1);                %

                \coordinate (DiagStart) at (0, \NplusNp);
                \coordinate (DiagEnd) at (\n, \totalH);

                \fill[redRegion] (O) -- (0, \N) -- (A) -- (A0) -- cycle;

                \fill[blueRegion] (B) -- (C) -- (D) -- (D0) -- cycle;

                \fill[redRegion] (E) -- (Fbot) -- (F) -- (E0) -- cycle;

                \draw[black, thick] (O) rectangle (\n, \totalH);
                \draw[blue!70!black, line width=1pt] (DiagStart) -- (DiagEnd);
                \draw[dashed, black!50] (0, \N) -- (\n, \N);
                \draw[dashed, black!50] (0, \NplusNp) -- (\n, \NplusNp);

                \draw[redPath, ultra thick, line cap=round, line join=round]
                (O) -- ++(0,1) -- ++(1,0) -- ++(0,1) -- ++(2,0) -- (A);

                \draw[bluePath, ultra thick, line cap=round, line join=round]
                (B) -- ++(1,0) -- ++(0,3) -- ++(2,0) -- ++(0,3) -- ++(3,0) -- (D);

                \draw[redPath, ultra thick, line cap=round, line join=round]
                (E) -- ++(0,1) -- ++(1,0) -- ++(0,2) -- ++(1,0) -- ++(0,1) -- ++(1,0) -- (F);

                \fill[black] (O) circle (4pt);
                \fill[black] (F) circle (4pt);
                \fill[black] (A) circle (4pt);
                \fill[black] (B) circle (4pt);
                \draw[black, thick] (A) -- (B);
                \fill[black] (D) circle (4pt);
                \fill[black] (E) circle (4pt);
                \draw[black, thick] (D) -- (E);

                \node[below left, font=\small] at (O) {$(0,1)$};
                \node[above left, font=\small] at (F) {$(n, N+N'+3n)$};

                \draw[decorate, decoration={brace, amplitude=8pt, mirror}]
                (-0.3, \NplusNp) -- (-0.3, 0);
                \node[left, font=\scriptsize] at (-0.8, \NplusNp/2) {$N+N'$ rows};

                \draw[decorate, decoration={brace, amplitude=6pt}]
                (\nx, \N) -- (\nx, 0);
                \node[right, font=\scriptsize] at (\nx+0.5, \N/2) {$N$ rows};

                \draw[decorate, decoration={brace, amplitude=4pt, mirror}]
                (\nx, \NplusNp) -- (\nx, \N+1);
                \node[left, font=\scriptsize] at (\nx-0.2, \N + 0.5 + \Np/2) {$N'$ rows};

                \draw[decorate, decoration={brace, amplitude=8pt, mirror}]
                (-0.3, \totalH) -- (-0.3, \NplusNp);
                \node[left, font=\footnotesize] at (-0.8, \NplusNp + \n/2) {$n$ bundles};

                \draw[decorate, decoration={brace, amplitude=6pt, mirror}]
                (0, -0.3) -- (\n, -0.3);
                \node[below, font=\small] at (\n/2, -0.8) {$n$ steps};

                \draw[decorate, decoration={brace, amplitude=6pt, mirror}]
                (\nx+0.1, \N+0.8) -- (\nx+\ntilcol, \N+0.8);
                \node[below, font=\scriptsize] at (\nx + \ntilcol/2, \N+0.2) {$\ntil$ steps};

                \draw[decorate, decoration={brace, amplitude=5pt}]
                (0, \N) -- (\nx, \N);
                \node[above, font=\scriptsize] at (\nx/2, \N+0.2) {$q$ steps};

                \draw[decorate, decoration={brace, amplitude=5pt, mirror}]
                (\nx+\ntilcol, \N+\Np) -- (\nx+\ntilcol, \N+\Np+\ntilcol);
                \node[right,  font=\fontsize{7}{6}\selectfont] at (\nx+\ntilcol+0.2, \N+\Np+\ntilcol*0.5) {$\ntil$ bundles};

                \node[bluePath, font=\small] at (\nx + \ntilcol/2+1, \N + \Np/2 + 0.5)
                {$\rfunc{\ntil}{N'_x}{N'_y}{N'_z}$};

            \end{scope}

            \begin{scope}[xshift=14.5cm]
                \node[font=\bfseries] at (1, \totalH + 1.2) {(b)};

                \pgfmathsetmacro{\redWidth}{\nx + \nrest}                           %
                \pgfmathsetmacro{\blueStartX}{\nx + \nrest}                         %
                \pgfmathsetmacro{\blueStartY}{\N + \nx + \nrest + 1}                %

                \coordinate (O) at (0, 0);
                \coordinate (F) at (\n, \totalH);

                \coordinate (R1) at (\redWidth, \N);                                %
                \coordinate (R2) at (\redWidth, \N + \nx + \nrest);                 %
                \coordinate (R0) at (\redWidth, 0);

                \coordinate (B1) at (\blueStartX, \blueStartY);                     %
                \coordinate (B2) at (\blueStartX, \NplusNp + \nx + \nrest);         %
                \coordinate (B3) at (\n, \totalH);                                  %
                \coordinate (B4) at (\n, \blueStartY);                              %

                \fill[redRegionLight] (O) -- (0, \N) -- (R2) -- (R0) -- cycle;

                \fill[redRegion] (O) -- (0, \N) -- (\nx, \N) -- (\nx, 0) -- cycle;

                \fill[redRegion]
                (\nx, \N+1) -- (\nx, \N + \nx) -- (R2) -- (\redWidth, \N+1) -- cycle;

                \fill[blueRegion] (B1) -- (B2) -- (B3) -- (B4) -- cycle;

                \draw[black, thick] (O) rectangle (\n, \totalH);
                \draw[blue!70!black, line width=1pt] (0, \NplusNp) -- (\n, \totalH);

                \draw[redPath, ultra thick, line cap=round, line join=round]
                (O) -- ++(0,1) -- ++(1,0) -- ++(0,1) -- ++(2,0) -- (\nx, \N);

                \draw[redPath, ultra thick, line cap=round, line join=round]
                (\nx, \N+1)  -- ++(0,1) -- ++(1,0) -- ++(0,2) -- ++(1,0) -- ++(0,1) -- ++(1,0) -- (R2);

                \draw[bluePath, ultra thick, line cap=round, line join=round]
                (B1) -- ++(1,0) -- ++(0,3) -- ++(2,0) -- ++(0,3) -- ++(3,0) -- (F);

                \node[above left, font=\small] at (F) {$(n, N+N'+3n)$};

                \draw[decorate, decoration={brace, amplitude=6pt, mirror}]
                (0, -0.3) -- (\redWidth, -0.3);
                \node[below, font=\small] at (\redWidth/2, -0.8) {$n-\ntil$ steps};

                \draw[decorate, decoration={brace, amplitude=6pt, mirror}]
                (\redWidth + 0.2, \blueStartY - 0.3) -- (\n, \blueStartY - 0.3);
                \node[below, font=\footnotesize] at (\redWidth + \ntilcol/2, \blueStartY - 0.8) {$\ntil$ steps};

                \draw[decorate, decoration={brace, amplitude=8pt, mirror}]
                (\n-\ntilcol, \N+1) -- (\n-\ntilcol, \N+\n-\ntilcol);
                \node[right, font=\scriptsize] at (\n-\ntilcol+0.4, \N+\n*0.5-\ntilcol*0.5) {$n-\ntil$ bundles};

                \node[redPath, font=\footnotesize] at (4.25, 1) {$\rfunc{n-\ntil}{N_x}{N_y}{N_z}$};
                \node[bluePath, font=\footnotesize] at (10, 11.5) {$\rfunc{\ntil}{N'_x}{N'_y}{N'_z}$};

                \draw[red, dashed, ultra thick] (O) -- (0, \N) -- (R2) -- (R0) -- cycle;

                \fill[black] (O) circle (4pt);
                \fill[black] (F) circle (4pt);
                \fill[black] (\nx, \N) circle (4pt);
                \fill[black] (\nx, \N+1) circle (4pt);
                \draw[black, thick] (\nx, \N) -- (\nx, \N+1);
                \fill[black] (R2) circle (4pt);
                \fill[black] (B1) circle (4pt);
                \draw[black, thick] (R2) -- (B1);

            \end{scope}

        \end{tikzpicture}
        \caption{
            Combinatorial proof of the R-function identity~\eqref{eq:R-identity}.
            The lattice has $N + N'$ flavored rows, where the lower $N$ rows consist of $N_x$ rows of flavor $x$, $N_y$ rows of flavor $y$, and $N_z$ rows of flavor $z$, while the upper $N'$ rows consist of $N'_x$ rows of flavor $x$, $N'_y$ rows of flavor $y$, and $N'_z$ rows of flavor $z$.
            (a) For a fixed $\ntil$, a path from $(0,1)$ to $(n, N+N'+3n)$ is decomposed into three portions.
            The first portion (lower red) contains $q$ East steps until the path crosses the line $y = N$.
            The second portion (blue) is a good path of $\ntil$ East steps on a lattice with $N'$ flavored rows, starting from the crossing point.
            The third portion (upper red) contains the remaining $n - q - \ntil$ East steps.
            (b) The same path is rearranged by moving the blue portion to the upper-right corner.
            The combined red region (dashed outline) forms a good path of $n-\ntil$ steps on a lattice with $N$ rows, contributing to $\rfunc{n-\ntil}{N_x}{N_y}{N_z}$, while the blue region contributes to $\rfunc{\ntil}{N'_x}{N'_y}{N'_z}$.
        }
        \label{fig:R-identity}
    \end{figure}

    \subsection{Proof of Eq.~\eqref{eq:R-identity}}
    \label{app:R-identity-proof}
    We prove the convolution identity~\eqref{eq:R-identity} by constructing a bijection between lattice paths, as illustrated in Fig.~\ref{fig:R-identity}.

    Let $N = N_x + N_y + N_z$ and $N' = N'_x + N'_y + N'_z$.
    The right-hand side counts good paths of $n$ steps on a lattice with $N + N'$ flavored rows and $n$ bundles.
    As shown in Fig.~\ref{fig:R-identity}(a), we arrange the lattice so that the lower $N$ rows consist of $N_x$, $N_y$, and $N_z$ rows of flavors $x$, $y$, and $z$, respectively, and the upper $N'$ rows likewise consist of $N'_x$, $N'_y$, and $N'_z$ rows.

    For a fixed $\ntil$, any good path can be uniquely decomposed into three portions as follows.
    The first portion contains $q$ East steps until the path crosses the line $y = N$.
    The second portion, starting from the crossing point, is a good path of $\ntil$ East steps on a sublattice with $N'$ flavored rows and $\ntil$ bundles.
    The third portion contains the remaining $n - q - \ntil$ East steps.

    As illustrated in Fig.~\ref{fig:R-identity}(b), we rearrange the path by moving the blue portion (of $\ntil$ steps) to the upper-right corner.
    This rearrangement is a weight-preserving bijection, since every step keeps the flavor of its row.
    After rearrangement, the combined red region forms a single good path of $n - \ntil$ steps on a lattice with $N$ flavored rows, which is counted by $\rfunc{n-\ntil}{N_x}{N_y}{N_z}$.
    The blue region becomes an independent good path of $\ntil$ steps on a lattice with $N'$ flavored rows, counted by $\rfunc{\ntil}{N'_x}{N'_y}{N'_z}$.

    \newcommand{\pathRHS}{%
        \begin{tikzpicture}[baseline=-0.5ex, scale=0.22]
            \draw[thick] (0,0) rectangle (8,12);
            \draw[dashed, black!50] (0,2) -- (8,2);
            \draw[dashed, black!50] (0,4) -- (8,4);
            \draw[blue!70!black, line width=1pt] (0,4) -- (8,12);
            \draw[decorate, decoration={brace, amplitude=3pt}]
            (-0.4, 0) -- (-0.4, 2);
            \node[left, font=\tiny] at (-0.6, 1) {$N$};
            \draw[decorate, decoration={brace, amplitude=3pt}]
            (-0.4, 2) -- (-0.4, 4);
            \node[left, font=\tiny] at (-0.6, 3) {$N'$};
        \end{tikzpicture}%
    }

    \newcommand{\pathDecompA}{%
        \begin{tikzpicture}[baseline=-0.5ex, scale=0.22]
            \definecolor{redRegion}{RGB}{255,230,230}
            \definecolor{blueRegion}{RGB}{230,230,255}
            \fill[redRegion] (0,0) -- (0,2) -- (2,2) -- (2,0) -- cycle;
            \fill[blueRegion] (2,2.5) -- (2,4) -- (6,8) -- (6,2.5) -- cycle;
            \fill[redRegion] (6,8.5) -- (6,10) -- (8,12) -- (8,8.5) -- cycle;
            \draw[thick] (0,0) rectangle (8,12);
            \draw[dashed, black!50] (0,2) -- (8,2);
            \draw[dashed, black!50] (0,4) -- (8,4);
            \draw[blue!70!black, line width=1pt] (0,4) -- (8,12);
            \fill[black] (2,2) circle (3pt);
            \fill[black] (2,2.5) circle (3pt);
            \draw[black, thick] (2,2) -- (2,2.5);
            \fill[black] (6,8) circle (3pt);
            \fill[black] (6,8.5) circle (3pt);
            \draw[black, thick] (6,8) -- (6,8.5);
            \draw[decorate, decoration={brace, amplitude=3pt, mirror}]
            (0, -0.4) -- (2, -0.4);
            \node[below, font=\tiny] at (1, -0.6) {$x$};
            \draw[decorate, decoration={brace, amplitude=3pt, mirror}]
            (2.1, 1.75) -- (6, 1.75);
            \node[below, font=\tiny] at (4, 1.7) {$\ntil$};
        \end{tikzpicture}%
    }

    \newcommand{\pathDecompB}{%
        \begin{tikzpicture}[baseline=-0.5ex, scale=0.22]
            \definecolor{redRegion}{RGB}{255,230,230}
            \definecolor{blueRegion}{RGB}{230,230,255}
            \fill[redRegion] (0,0) -- (0,2) -- (2,2) -- (2,0) -- cycle;
            \fill[redRegion] (2,2.5) -- (2,4) -- (4,6) -- (4,2.5) -- cycle;
            \fill[blueRegion] (4,6.5) -- (4,8) -- (8,12) -- (8,6.5) -- cycle;
            \draw[red, dashed, thick] (0,0) -- (0,2) -- (4,6) -- (4,0) -- cycle;
            \draw[thick] (0,0) rectangle (8,12);
            \draw[blue!70!black, line width=1pt] (0,4) -- (8,12);
            \fill[black] (2,2) circle (3pt);
            \fill[black] (2,2.5) circle (3pt);
            \draw[black, thick] (2,2) -- (2,2.5);
            \fill[black] (4,6) circle (3pt);
            \fill[black] (4,6.5) circle (3pt);
            \draw[black, thick] (4,6) -- (4,6.5);
            \draw[decorate, decoration={brace, amplitude=2pt}]
            (0, 0.1) -- (2, 0.1);
            \node[above, font=\tiny] at (1, 0) {$x$};
            \draw[decorate, decoration={brace, amplitude=3pt, mirror}]
            (0, -0.4) -- (4, -0.4);
            \node[below, font=\tiny] at (2, -0.6) {$n{-}\ntil$};
            \draw[decorate, decoration={brace, amplitude=3pt, mirror}]
            (4.1, 6.5) -- (8, 6.5);
            \node[below, font=\tiny] at (6, 6.3) {$\ntil$};
        \end{tikzpicture}%
    }

    \newcommand{\pathDecompC}{%
        \begin{tikzpicture}[baseline=-0.5ex, scale=0.22]
            \definecolor{redRegion}{RGB}{255,230,230}
            \definecolor{blueRegion}{RGB}{230,230,255}
            \fill[redRegion] (0,0) -- (0,2) -- (4,6) -- (4,0) -- cycle;
            \fill[blueRegion] (4,6.5) -- (4,8) -- (8,12) -- (8,6.5) -- cycle;
            \draw[thick] (0,0) rectangle (8,12);
            \draw[blue!70!black, line width=1pt] (0,4) -- (8,12);
            \fill[black] (4,6) circle (3pt);
            \fill[black] (4,6.5) circle (3pt);
            \draw[black, thick] (4,6) -- (4,6.5);
            \draw[decorate, decoration={brace, amplitude=3pt, mirror}]
            (0, -0.4) -- (4, -0.4);
            \node[below, font=\tiny] at (2, -0.6) {$n{-}\ntil$};
            \draw[decorate, decoration={brace, amplitude=3pt, mirror}]
            (4.1, 6.5) -- (8, 6.5);
            \node[below, font=\tiny] at (6, 6.3) {$\ntil$};
        \end{tikzpicture}%
    }

    \newcommand{\pathRedOnly}{%
        \begin{tikzpicture}[baseline=-0.5ex, scale=0.18]
            \definecolor{redRegion}{RGB}{255,230,230}
            \fill[redRegion] (0,0) -- (0,2) -- (4,6) -- (4,0) -- cycle;
            \draw[thick] (0,0) rectangle (4,6);
            \draw[blue!70!black, line width=0.8pt] (0,2) -- (4,6);
            \draw[decorate, decoration={brace, amplitude=2pt, mirror}]
            (0, -0.3) -- (4, -0.3);
            \node[below, font=\tiny] at (2, -0.5) {$n{-}\ntil$};
            \draw[decorate, decoration={brace, amplitude=2pt}]
            (-0.3, 0) -- (-0.3, 2);
            \node[left, font=\tiny] at (-0.5, 1) {$N$};
        \end{tikzpicture}%
    }

    \newcommand{\pathBlueOnly}{%
        \begin{tikzpicture}[baseline=-0.5ex, scale=0.18]
            \definecolor{blueRegion}{RGB}{230,230,255}
            \fill[blueRegion] (0,0) -- (0,2) -- (4,6) -- (4,0) -- cycle;
            \draw[thick] (0,0) rectangle (4,6);
            \draw[blue!70!black, line width=0.8pt] (0,2) -- (4,6);
            \draw[decorate, decoration={brace, amplitude=2pt, mirror}]
            (0, -0.3) -- (4, -0.3);
            \node[below, font=\tiny] at (2, -0.5) {$\ntil$};
            \draw[decorate, decoration={brace, amplitude=2pt}]
            (-0.3, 0) -- (-0.3, 2);
            \node[left, font=\tiny] at (-0.5, 1) {$N'$};
        \end{tikzpicture}%
    }

    Summing over all possible values of $\ntil$, we obtain the identity.
    Schematically, the bijection reads:
    \begin{align}
        \rfunc{n}{N_x+N^\prime_x}{N_y+N^\prime_y}{N_z+N^\prime_z}
         & =
        \raisebox{-2.5em}{\pathRHS}
        \ = \
        \sum_{\ntil=0}^{n} \sum_{q=0}^{n-\ntil}
        \raisebox{-2.5em}{\pathDecompA}
        \ = \
        \sum_{\ntil=0}^{n} \sum_{q=0}^{n-\ntil}
        \raisebox{-2.5em}{\pathDecompB}
        \notag \\[1em]
         & = \
        \sum_{\ntil=0}^{n}
        \raisebox{-2.5em}{\pathDecompC}
        \ = \
        \sum_{\ntil=0}^{n}
        \raisebox{-1.2em}{\pathRedOnly}
        \hspace*{0.4em}
        \times
        \hspace*{-0.4em}
        \raisebox{-1.2em}{\pathBlueOnly}
        =
        \sum_{\ntil = 0}^{n}
        \rfunc{\ntil}{N^\prime_x}{N^\prime_y}{N^\prime_z}
        \rfunc{n-\ntil}{N_x}{N_y}{N_z}
        \,.
        \notag
    \end{align}
    \hfill $\square$

    \subsection{Proof of Eq.~\eqref{eq:fx-ito-fe}}\label{app:fx-omega}

    We use the recurrence relation~\eqref{eq:R-identity-diff} in the first equality below and the convolution identity~\eqref{eq:R-identity} in the second:
    \begin{align}
        \rfunc{n}{1}{0}{0}
         & =
        \delta_{n,0}
        +
        J_X^{2}
        \rfunc{n-1}{2}{1}{1}
        =
        \delta_{n,0}
        +
        J_X^{2}
        \sum_{\ntil = 0}^{n-1}
        \rfunc{\ntil}{1}{0}{0}
        \rfunc{n-1-\ntil}{1}{1}{1}
        \,,
    \end{align}
    where we also used $\rfunc{n}{0}{0}{0} = \delta_{n,0}$.
    Multiplying both sides by $s^{2n}$ and summing over $n$, we obtain
    \begin{align}
        f_X(s)
         & =
        1 + J_X^{2} \sum_{n = 0}^{\infty} s^{2n} \sum_{\ntil = 0}^{n-1}
        \rfunc{\ntil}{1}{0}{0}
        \rfunc{n-1-\ntil}{1}{1}{1}
        \nonumber \\
         & =
        1 + J_X^{2} \sum_{n = 0}^{\infty} \sum_{\ntil = 0}^{n-1}
        \qty(\rfunc{\ntil}{1}{0}{0} s^{2\ntil})
        \qty(\rfunc{n-1-\ntil}{1}{1}{1} s^{2(n-1-\ntil)} s^{2})
        \nonumber \\
         & =
        1 + J_X^{2} f_X(s) \omega(s)
        \,,
    \end{align}
    which is equivalent to Eq.~\eqref{eq:fx-ito-fe} for $\alpha = X$.
    The same argument applies to $f_Y(s)$ and $f_Z(s)$.
    \hfill $\square$

    \subsection{Proof of Eq.~\eqref{eq:omega-fxyz}}\label{app:omega-fxyz}
    To prove Eq.~\eqref{eq:omega-fxyz}, we first factorize $\rfunc{n}{1}{1}{1}$ by applying Eq.~\eqref{eq:R-identity} twice:
    \begin{align}
        \rfunc{n}{1}{1}{1}
        =
        \sum_{\substack{n_x + n_y + n_z = n \\ n_x, n_y, n_z \ge 0}} \rfunc{n_x}{1}{0}{0} \rfunc{n_y}{0}{1}{0} \rfunc{n_z}{0}{0}{1}
        \,.
    \end{align}
    Substituting this factorization into the definition~\eqref{eq:fx-omega} of $\omega(s)$ and using the definitions of $f_X(s)$, $f_Y(s)$, and $f_Z(s)$, we obtain
    \begin{align}
        \omega(s)
         & = \sum_{n = 0}^{\infty} \rfunc{n}{1}{1}{1} s^{2n+2}
        \nonumber                                              \\
         & =
        s^2  \qty(\sum_{n_x = 0}^{\infty} \rfunc{n_x}{1}{0}{0} s^{2n_x}) \qty(\sum_{n_y = 0}^{\infty} \rfunc{n_y}{0}{1}{0} s^{2n_y}) \qty(\sum_{n_z = 0}^{\infty} \rfunc{n_z}{0}{0}{1} s^{2n_z})
        \nonumber                                              \\
         & =
        s^2 f_X(s) f_Y(s) f_Z(s)
        \,.
    \end{align}
    \hfill $\square$

    \section{Proof of Theorem~\ref{thm:coefficient-recurrence-conservation}}
    \label{app:commutativity-Qk}

        In this appendix, we prove Theorem~\ref{thm:coefficient-recurrence-conservation}.

        \begin{coefficientrecurrencerestatement}[Restatement]
            Let $\qfamily{k}{F}$ be defined by Eq.~\eqref{eq:Qk-F}.
            If $\ffuncone{n}{\vec N}=0$ for $n<0$ and $F$ satisfies Eq.~\eqref{eq:F-identity-diff}, then $\qty[\qfamily{k}{F},H]=0$ for $2\le k\le L$.
        \end{coefficientrecurrencerestatement}

    \subsection{Setup and notation}

    Throughout this appendix, we absorb the factors $J_{\boldsymbol{A}}$ and $(-\imi)^{t-1}$ into the doubling product $\overline{\boldsymbol{A}}$.
    We redefine the component of a doubling product starting at site $i$ as
    \begin{equation}
        \overline{A_{1} A_{2} \cdots A_{l}}(i)
        \coloneqq
        J_{\overline{A_{1} \cdots A_{l}}}
        (A_1)_i(A_1A_2)_{i+1}(A_2A_3)_{i+2}\cdots (A_{l-1}A_l)_{i+l-1}(A_l)_{i+l}
        \,,
    \end{equation}
    where $l$ is the number of letters in the doubling product, which we call the \emph{length}.
    The doubling product is the sum over all translations, $\overline{\boldsymbol{A}} = \sum_{i=1}^{L} \overline{\boldsymbol{A}}(i)$.
    As in Eq.~\eqref{doubling-abbreviated}, we also denote a doubling product with length $l$ as
    \begin{align}
        \obA
        =
        \overline{A_{1} \cdots A_{l}} = \overline{B^{m_1+1}_{1} B^{m_2+1}_{2} \cdots B^{m_t+1}_{t}}
        \,,
    \end{align}
    where the flattened notation $\overline{A_{1} \cdots A_{l}}$ corresponds to the abbreviated notation~\eqref{doubling-abbreviated} with $B_r \coloneqq A_{i_r}$, $i_1 = 1$, and $i_{r+1} = i_r + m_r + 1$.
    Here $t = l - m$ is the number of distinct letter groups and $m = \sum_{r=1}^{t} m_r$ is the total number of holes.

    We let $(N_x, N_y, N_z)$ be the counts of each letter in the sequence $B_1, \ldots, B_t$, and use the shorthand notation $\vec{N} = (N_x, N_y, N_z)$, $\vec 1=(1,1,1)$, and $\vec{e}_A$ for the unit vector with $1$ in the component corresponding to $A \in \{X, Y, Z\}$.
    The coupling coefficient is then written as
    \begin{align}
        J_{\overline{A_{1} \cdots A_{l}}}
        =
        (J_X J_Y J_Z)^{m}
        \prod_{r=1}^{t}
        J_{B_r}^{1 - m_r}
        \,.
    \end{align}

    With this convention, the Hamiltonian and the operator in Theorem~\ref{thm:coefficient-recurrence-conservation} are written as
    \begin{align}
        H & = \overline{X} + \overline{Y} + \overline{Z}
        \,,
        \\
        \label{eq:xyz-Qk-3}
        \qfamily{k}{F}
          & =
        \sum_{\substack{0 \leq n+m < \floor{k/2}         \\ n,m \ge 0}}
        (-1)^{n+m}
        \sum_{\overline{\boldsymbol{A}}\in\mathcal{S}_k^{n,m}}
        \ffuncone{n}{\vec N+m}
        \,
        \overline{\boldsymbol{A}}
        \,,
    \end{align}
        where $\vec N+m\coloneqq(N_x+m,N_y+m,N_z+m)$.
    A doubling product in $\mathcal{S}_k^{n,m}$ has $m$ holes and length $l$ satisfying $l + m = k - 2n - 1$.
    The operator $\qfamily{k}{F}$ in Eq.~\eqref{eq:xyz-Qk-3} differs from that in Eq.~\eqref{eq:Qk-F} by the overall factor $(-\imi)^{k-2}$, which does not affect the commutation relation $\qty[\qfamily{k}{F}, H] = 0$.
        Throughout this appendix, we also use the halved commutator $\qty[A, B] \equiv \frac{1}{2}\qty(AB - BA)$.

    \subsection{Commutator algebra for doubling products}
    \label{app:comm-algebra-doubling}

    The commutator of a doubling product with $H$ decomposes as
    \begin{align}
        \qty[\overline{A_{1}A_{2}\cdots A_{l}}, H]
        =
        \sum_{j=0}^{l+1}
        \sum_{B \in \{X, Y, Z\}}
        \ad{\overline{B}(j)}{\overline{A_{1}A_{2}\cdots A_{l}}}
        \,,
    \end{align}
    where
    \begin{align}
        \ad{\overline{B}(j)}{\overline{A_{1}A_{2}\cdots A_{l}}}
        \coloneqq
        \sum_{i=1}^{L}
        \qty[\overline{A_{1}A_{2}\cdots A_{l}}(i), \overline{B}(i+j-1)]
        \,.
    \end{align}

    When the Hamiltonian density $\overline{B}$ acts on the letter $A_j$ inside the doubling product, we write
    \begin{align}
        \ad{\overline{B}(j)}{\overline{A_{1}A_{2} \cdots A_{j}\cdots A_{l}}}
        =
        \overline{A_{1}A_{2} \cdots \admark{B}{A_{j}}\cdots A_{l}}
        \,.
    \end{align}
    When the action is at the edges ($j=0$ or $j=l+1$), the dotted box represents a new letter:
    \begin{align}
        \ad{\overline{B}(0)}{\overline{A_{1}\cdots A_{l}}}
         & =
        \admarknew{B}\,
        \overline{A_{1} \cdots A_{l}}
        \,,
        \\
        \ad{\overline{B}(l+1)}{\overline{A_{1}\cdots A_{l}}}
         & =
        \overline{A_{1} \cdots A_{l}}\,
        \admarknew{B}
        \,.
    \end{align}

    In the rules below, $(A, B, C)$ denotes an arbitrary permutation of $(X, Y, Z)$.
    The commutator of a doubling product with the Hamiltonian generates four types of transitions: length increase/decrease with the number of holes unchanged, and hole increase/decrease with the length unchanged.
    All commutators other than those listed below vanish.

    \paragraph{Length increase.}
    When the Hamiltonian acts at the edges, it extends the length:
    \begin{align}
        \overline{\cdots A}\, \admarknew{B}
         & =
        \overline{\cdots A \, B}
        \,,
        \\
        \admarknew{B}\, \overline{A \cdots}
         & =
        \overline{B \, A \cdots} \times (-1)
        \,.
    \end{align}
    Each of the two letters $B \neq A$ gives a nonvanishing contribution to the sum over $B$.

    \paragraph{Length decrease.}
    When adjacent letters at the edges match, the length contracts:
    \begin{align}
        \overline{\cdots A \, \admark{B}{B}}
         & =
        \overline{\cdots A} \times J^2_{B}
        \,,
        \\
        \overline{\admark{B}{B} \, A \cdots}
         & =
        \overline{A \cdots} \times (-J^2_{B})
        \,.
    \end{align}

    \paragraph{Hole increase.}
    In the bulk, acting on an intermediate letter creates a hole on one side:
    \begin{align}
        \overline{\cdots A \, \admark{A}{B} \, C \cdots}
         & =
        \overline{\cdots A \, C \, C \cdots}
        \,,
        \\
        \overline{\cdots A \, \admark{C}{B} \, C \cdots}
         & =
        \overline{\cdots A \, A \, C \cdots} \times (-1)
        \,.
    \end{align}
    At the edges:
    \begin{align}
        \overline{\admark{C}{B} \, A \cdots}
         & =
        \overline{A \, A \cdots}
        \,,
        \\
        \overline{\cdots A \, \admark{C}{B}}
         & =
        \overline{\cdots A \, A} \times (-1)
        \,.
    \end{align}

    \paragraph{Hole decrease.}
    Acting on a hole (consecutive identical letters) removes it:
    \begin{align}
        \overline{\cdots A \, \admark{C}{A} \, C \cdots}
         & =
        \overline{\cdots A \, B \, C \cdots} \times (-J^2_{C})
        \,,
        \\
        \overline{\cdots A \, \admark{A}{C} \, C \cdots}
         & =
        \overline{\cdots A \, B \, C \cdots} \times J^2_{A}
        \,.
    \end{align}
    At the edges:
    \begin{align}
        \overline{\cdots A \, \admark{C}{A}}
         & =
        \overline{\cdots A \, B} \times (-J^2_{C})
        \,,
        \\
        \overline{\admark{C}{A} \, A \cdots}
         & =
        \overline{B \, A \cdots} \times J^2_{C}
        \,.
    \end{align}

    \paragraph{Disorder terms.}
    The hole-decreasing transitions also generate disorder terms whose letter sequence contains $\I$:
    \begin{align}
        \overline{\cdots A \, \admark{A}{A} \, C \cdots}
         & =
        \overline{\cdots A \, \I \, C \cdots} \times (-1)
        \,,
        \\
        \overline{\cdots A \, \admark{C}{C} \, C \cdots}
         & =
        \overline{\cdots A \, \I \, C \cdots}
        \,.
    \end{align}
    To handle these terms, we extend the notation to allow the identity operator $\I$ in the letter sequence and define $J_{\I} \coloneqq J_X J_Y J_Z$ so that the coupling factor $J_{\obA}$ remains consistent when $\I$ replaces a Pauli letter.

    Since the coefficient $\ffunc{n}{N_x+m}{N_y+m}{N_z+m}$ depends only on the letter counts $(N_x, N_y, N_z)$ and $m$, not on the positions of holes, the parent products $\overline{\cdots A \, A \, C \cdots}$ and $\overline{\cdots A \, C \, C \cdots}$ carry the same coefficient in $\qfamily{k}{F}$.
    Since the two actions above produce the same term $\overline{\cdots A \, \I \, C \cdots}$ with opposite signs, the disorder terms cancel pairwise and do not contribute to $\qty[\qfamily{k}{F}, H]$.

    \subsection{Proof of commutativity}

    From the commutator algebra above, $\qty[\qfamily{k}{F}, H]$ is a linear combination of doubling products:
    \begin{align}
        \qty[\qfamily{k}{F}, H]
         & =
        \sum_{\boldsymbol{A}} C_{\boldsymbol{A}} \, \obA
        \,,
    \end{align}
    where the sum runs over all doubling products.
    We will show that all coefficients $C_{\boldsymbol{A}}$ vanish.

    Setting $\vec N\mapsto\vec N-\vec 1$ in Eq.~\eqref{eq:F-identity-diff} and rearranging gives
    \begin{align}
        \label{eq:F-identity-diff-rewrite}
        J_{A}^2 \, \ffuncone{n-1}{\vec{N}}
         & =
        \ffuncone{n}{\vec{N}-\vec 1}
        -
        \ffuncone{n}{\vec{N}-\vec 1-\vec{e}_A}
        \,.
    \end{align}

    \begin{figure}[htbp]
        \centering
        \begin{tikzpicture}[baseline=0, scale=3]
            \drawQkStructure{12}
        \end{tikzpicture}
        \caption{
            Structure of $\qfamily{k}{F}$ for $k=12$.
            Each circle at position $(l, m)$ represents a doubling product in $\qfamily{12}{F}$ with length $l$ and $m$ holes.
            The index $n$ is determined by $l + m = k - 2n - 1$, and the shaded regions group doubling products with the same $n$.
            Arrows represent transitions induced by the commutator with the Hamiltonian: vertical arrows for length-changing and horizontal arrows for hole-changing transitions.
            Crosses represent doubling products generated by these transitions; the proof of $\qty[\qfamily{k}{F}, H] = 0$ reduces to showing that all contributions at each cross cancel.
        }
        \label{fig:Q12-structure}
    \end{figure}

    The structure of $\qfamily{k}{F}$ is illustrated in Figure~\ref{fig:Q12-structure} for $k = 12$.

    The coefficient $C_{\boldsymbol{A}}$ receives contributions from doubling products in $\qfamily{k}{F}$ that differ from $\obA$ at a single letter position.
    We decompose
    \begin{align}
        C_{\boldsymbol{A}}
         & =
        \sum_{i=0}^{l+1} C_{\boldsymbol{A}}^{(i)}
        \,,
    \end{align}
    where $C_{\boldsymbol{A}}^{(i)}$ is the contribution from doubling products that differ from $\obA$ at position $i$.

    The local cancellation structure is shown in Figure~\ref{fig:basic-structure}.
    Each circle at position $(l', m')$ represents a source doubling product in $\qfamily{k}{F}$ with index $n'$ satisfying $l' + m' = k - 2n' - 1$; this $n'$ appears as the first argument of the coefficient $\ffuncone{n'}{\cdots}$.
    Denoting the index of the $(l-1, m)$ source by $n$ (so that $l + m = k - 2n$), the four sources are as follows:
    \begin{itemize}
        \item $(l+1, m)$ with index $n-1$: length decrease (positions $i = 0$ and $i = l+1$)
        \item $(l-1, m)$ with index $n$: length increase (positions $i = 1$ and $i = l$)
        \item $(l, m+1)$ with index $n-1$: hole decrease (positions $i = 1, \ldots, l$)
        \item $(l, m-1)$ with index $n$: hole increase (positions $i = 1, \ldots, l$)
    \end{itemize}

        The contribution from an absent source circle is understood to be zero, so the same cancellation applies at the boundaries of the diagram, including $m=0$.
    We compute each $C_{\boldsymbol{A}}^{(i)}$ and show that their sum vanishes by a telescoping argument.

    \begin{figure}[tb]
        \centering
        \scalebox{1.35}{%
            \begin{tikzpicture}[baseline=0, scale=3]
                \drawBasicStructure
            \end{tikzpicture}%
        }
        \caption{
            Local structure for the cancellation of contributions at a cross.
            The cross at $(l, m)$ represents the target doubling product $\obA$ generated by the commutator.
            The four circles represent doubling products in $\qfamily{k}{F}$ that generate $\obA$ through the commutator with $H$:
            $(l \pm 1, m)$ for length-changing transitions (vertical arrows) and $(l, m \pm 1)$ for hole-changing transitions (horizontal arrows).
            The shaded regions indicate the index $n$ or $n-1$ of the source doubling products.
        }
        \label{fig:basic-structure}
    \end{figure}

    \paragraph{Contributions at position $i = l+1$ (right edge, length decrease).}

    The sources $\overline{A_1 \cdots A_{l} \, B}$ and $\overline{A_1 \cdots A_{l} \, B'}$ at $(l+1, m)$ with index $n-1$, where $\{B, B', A_l\} = \{X, Y, Z\}$, have letter counts $\vec{N} + \vec{e}_{B}$ and $\vec{N} + \vec{e}_{B'}$, respectively.
    The relevant commutator actions are
    \begin{align}
        \overline{A_1 \cdots A_{l} \, \admark{B}{B}}
         & =
        \obA \times J_{B}^2
        \,,
         &
        \overline{A_1 \cdots A_{l} \, \admark{B'}{B'}}
         & =
        \obA \times J_{B'}^2
        \,.
    \end{align}
    The contribution is
    \begin{align}
        C_{\boldsymbol{A}}^{(l+1)}
         & =
        (-1)^{(n-1)+m}
        \sum_{B \neq A_l}
        \ffuncone{n-1}{\vec{N} + m + \vec{e}_{B}} \cdot J_{B}^2
        \nonumber \\
         & =
        (-1)^{n+m}
        \sum_{B \neq A_l}
        \qty[
            \ffuncone{n}{\vec{N} + m - 1}
            -
            \ffuncone{n}{\vec{N} + m - 1 + \vec{e}_{B}}
        ]
        \nonumber \\
         & =
        (-1)^{n+m}
        \qty[
            2 \ffuncone{n}{\vec{N} + m - 1}
            -
            \sum_{B \neq A_l}
            \ffuncone{n}{\vec{N} + m - 1 + \vec{e}_{B}}
        ]
        \,,
    \end{align}
    where we used Eq.~\eqref{eq:F-identity-diff-rewrite} in the second equality.

    \paragraph{Contributions at position $i = 0$ (left edge, length decrease).}

    By left-right symmetry, the calculation parallels that of $i = l+1$.
    The commutator actions $\overline{\admark{B}{B} \, A_1 \cdots A_{l}} = \obA \times (-J_{B}^2)$ carry an opposite sign, yielding
    \begin{align}
        C_{\boldsymbol{A}}^{(0)}
         & =
        (-1)^{n+m+1}
        \qty[
            2 \ffuncone{n}{\vec{N} + m - 1}
            -
            \sum_{B \neq A_1}
            \ffuncone{n}{\vec{N} + m - 1 + \vec{e}_{B}}
        ]
        \,.
    \end{align}

    \paragraph{Contributions at position $i = l$ (right edge, length increase and hole changes).}

    \emph{Case $A_{l-1} \neq A_l$ (length increase and hole decrease).}
    The length-increase source $\overline{A_1 \cdots A_{l-1}}$ at $(l-1, m)$ with index $n$ has letter counts $\vec{N} - \vec{e}_{A_l}$.
    The relevant action is
    \begin{align}
        \overline{A_1 \cdots A_{l-1}} \, \admarknew{A_l}
         & =
        \obA
        \,,
    \end{align}
    which contributes
    \begin{align}
        (-1)^{n+m} \ffuncone{n}{\vec{N} + m - \vec{e}_{A_l}}
        \,.
    \end{align}

    The hole-decrease source $\overline{A_1 \cdots A_{l-2} \, A_{l-1} \, A_{l-1}}$ at $(l, m+1)$ with index $n-1$ has letter counts $\vec{N} - \vec{e}_{A_l}$.
    With $\{A_{l-1}, A_l, C\} = \{X, Y, Z\}$, the relevant action is
    \begin{align}
        \overline{A_1 \cdots A_{l-2} \, A_{l-1} \, \admark{C}{A_{l-1}}}
         & =
        \obA \times (-J_C^2)
        \,,
    \end{align}
    which contributes
    \begin{align}
         & (-1)^{(n-1)+(m+1)}
        \ffuncone{n-1}{\vec{N} + m + 1 - \vec{e}_{A_l}} \cdot (-J_C^2)
        \nonumber             \\
         & \quad=
        (-1)^{n+m+1}
        J_C^2 \, \ffuncone{n-1}{\vec{N} + m + \vec{e}_{A_{l-1}} + \vec{e}_C}
        \nonumber             \\
         & \quad=
        (-1)^{n+m}
        \qty[
            \ffuncone{n}{\vec{N} + m - 1 + \vec{e}_{A_{l-1}}}
            -
            \ffuncone{n}{\vec{N} + m - 1 + \vec{e}_{A_{l-1}} + \vec{e}_C}
        ]
        \nonumber             \\
         & \quad=
        (-1)^{n+m}
        \qty[
            \ffuncone{n}{\vec{N} + m - 1 + \vec{e}_{A_{l-1}}}
            -
            \ffuncone{n}{\vec{N} + m - \vec{e}_{A_l}}
        ]
        \,,
    \end{align}
    where we used Eq.~\eqref{eq:F-identity-diff-rewrite} in the third equality and $\vec{e}_{A_{l-1}} + \vec{e}_C = (1,1,1) - \vec{e}_{A_l}$ in the last.

    Combining these two contributions, we have
    \begin{align}
        C_{\boldsymbol{A}}^{(l)} \big|_{A_{l-1} \neq A_l}
         & =
        (-1)^{n+m}
        \ffuncone{n}{\vec{N} + m - 1 + \vec{e}_{A_{l-1}}}
        \,.
    \end{align}

    \emph{Case $A_{l-1} = A_l$ (hole increase).}
    The sources $\overline{A_1 \cdots A_{l-2} \, A_{l-1} \, B}$ and $\overline{A_1 \cdots A_{l-2} \, A_{l-1} \, C}$ at $(l, m-1)$ with index $n$, where $\{A_l, B, C\} = \{X, Y, Z\}$, contribute through the actions
    \begin{align}
        \overline{A_1 \cdots A_{l-2} \, A_{l-1} \, \admark{C}{B}}
         & =
        \obA \times (-1)
        \,,
         &
        \overline{A_1 \cdots A_{l-2} \, A_{l-1} \, \admark{B}{C}}
         & =
        \obA \times (-1)
        \,.
    \end{align}
    The contribution is
    \begin{align}
        C_{\boldsymbol{A}}^{(l)} \big|_{A_{l-1} = A_l}
         & =
        (-1)^{n+(m-1)}
        \qty(
        - \ffuncone{n}{\vec{N} + m - 1 + \vec{e}_{B}}
        - \ffuncone{n}{\vec{N} + m - 1 + \vec{e}_{C}}
        )
        \nonumber \\
         & =
        (-1)^{n+m}
        \sum_{B \neq A_l}
        \ffuncone{n}{\vec{N} + m - 1 + \vec{e}_{B}}
        \,.
    \end{align}

    Combining both cases, we obtain
    \begin{align}
        C_{\boldsymbol{A}}^{(l)}
         & =
        (-1)^{n+m}
        \qty[
            \overline{\delta}_{A_{l-1}, A_l}
            \ffuncone{n}{\vec{N} + m - 1 + \vec{e}_{A_{l-1}}}
            +
            \delta_{A_{l-1}, A_l}
            \sum_{B \neq A_l}
            \ffuncone{n}{\vec{N} + m - 1 + \vec{e}_{B}}
        ]
        \,,
    \end{align}
    where $\overline{\delta}_{A, B} \coloneqq 1 - \delta_{A, B}$.

    \paragraph{Contributions at position $i = 1$ (left edge, length increase and hole changes).}

    By left-right symmetry, the calculation parallels that of $i = l$.
    The commutator actions carry opposite signs: $\admarknew{A_1} \, \overline{A_2 \cdots A_{l}} = \obA \times (-1)$ for length increase, and $\overline{\admark{C}{A_2} \, A_2 \, A_3 \cdots A_{l}} = \obA \times J_C^2$ for hole decrease (where $\{A_1, A_2, C\} = \{X, Y, Z\}$).
    For hole increase with $A_1 = A_2$, we have $\overline{\admark{C}{B} \, A_2 \cdots A_{l}} = \obA$ (where $\{A_1, B, C\} = \{X, Y, Z\}$).
    Combining all cases, we obtain
    \begin{align}
        C_{\boldsymbol{A}}^{(1)}
         & =
        (-1)^{n+m+1}
        \qty[
            \overline{\delta}_{A_1, A_2}
            \ffuncone{n}{\vec{N} + m - 1 + \vec{e}_{A_2}}
            +
            \delta_{A_1, A_2}
            \sum_{B \neq A_1}
            \ffuncone{n}{\vec{N} + m - 1 + \vec{e}_{B}}
        ]
        \,.
    \end{align}

    \paragraph{Contributions at positions $i = 2, \ldots, l-1$ (bulk, hole changes).}\mbox{}\par

    \emph{Case $\{A_{i-1}, A_i, A_{i+1}\} = \{X, Y, Z\}$ (hole decrease).}
    Two sources at $(l, m+1)$ with index $n-1$ contribute through the actions
    \begin{align}
        \overline{A_1 \cdots A_{i-1} \, \admark{A_{i+1}}{A_{i-1}} \, A_{i+1} \cdots A_{l}}
         & =
        \obA \times (-J_{A_{i+1}}^2)
        \,,
        \\
        \overline{A_1 \cdots A_{i-1} \, \admark{A_{i-1}}{A_{i+1}} \, A_{i+1} \cdots A_{l}}
         & =
        \obA \times J_{A_{i-1}}^2
        \,.
    \end{align}
    The contribution is
    \begin{align}
        C_{\boldsymbol{A}}^{(i)} \big|_{\{A_{i-1}, A_i, A_{i+1}\} = \{X,Y,Z\}}
         & =
        (-1)^{(n-1)+(m+1)}
        \ffuncone{n-1}{\vec{N} + m + 1 - \vec{e}_{A_{i}}}
        \qty(
        J_{A_{i-1}}^2
        -
        J_{A_{i+1}}^2
        )
        \nonumber       \\
         & =
        (-1)^{n+m}
        \Bigl[
            \ffuncone{n}{\vec{N} + m - \vec{e}_{A_{i}}}
            -
            \ffuncone{n}{\vec{N} + m - \vec{e}_{A_{i}} - \vec{e}_{A_{i-1}}}
        \nonumber       \\
         & \qquad\qquad
            -
            \ffuncone{n}{\vec{N} + m - \vec{e}_{A_{i}}}
            +
            \ffuncone{n}{\vec{N} + m - \vec{e}_{A_{i}} - \vec{e}_{A_{i+1}}}
            \Bigr]
        \nonumber       \\
         & =
        (-1)^{n+m}
        \qty[
            \ffuncone{n}{\vec{N} + m - \vec{e}_{A_{i}} - \vec{e}_{A_{i+1}}}
            -
            \ffuncone{n}{\vec{N} + m - \vec{e}_{A_{i}} - \vec{e}_{A_{i-1}}}
        ]
        \,,
    \end{align}
    where we used Eq.~\eqref{eq:F-identity-diff-rewrite} in the second equality.

    \emph{Case $A_{i-1} = A_i \neq A_{i+1}$ (hole increase).}
    One source at $(l, m-1)$ with index $n$ contributes.
    Let $B$ be the unique letter satisfying $\{A_{i-1}, B, A_{i+1}\} = \{X, Y, Z\}$.
    The relevant action is
    \begin{align}
        \overline{A_1 \cdots A_{i-1} \, \admark{A_{i+1}}{B} \, A_{i+1} \cdots A_{l}}
         & =
        \obA \times (-1)
        \,.
    \end{align}
    The contribution is
    \begin{align}
        C_{\boldsymbol{A}}^{(i)} \big|_{A_{i-1} = A_i \neq A_{i+1}}
         & =
        (-1)^{n+(m-1)}
        \ffuncone{n}{\vec{N} + m - 1 + \vec{e}_{B}}
        \times (-1)
        \nonumber \\
         & =
        (-1)^{n+m}
        \ffuncone{n}{\vec{N} + m - \vec{e}_{A_{i}} - \vec{e}_{A_{i+1}}}
        \,.
    \end{align}

    \emph{Case $A_{i-1} \neq A_i = A_{i+1}$ (hole increase).}
    One source at $(l, m-1)$ with index $n$ contributes.
    Let $B$ be the unique letter satisfying $\{A_{i-1}, B, A_{i+1}\} = \{X, Y, Z\}$.
    The relevant action is
    \begin{align}
        \overline{A_1 \cdots A_{i-1} \, \admark{A_{i-1}}{B} \, A_{i+1} \cdots A_{l}}
         & =
        \obA
        \,.
    \end{align}
    The contribution is
    \begin{align}
        C_{\boldsymbol{A}}^{(i)} \big|_{A_{i-1} \neq A_i = A_{i+1}}
         & =
        (-1)^{n+(m-1)}
        \ffuncone{n}{\vec{N} + m - 1 + \vec{e}_{B}}
        \nonumber \\
         & =
        (-1)^{n+m+1}
        \ffuncone{n}{\vec{N} + m - \vec{e}_{A_{i-1}} - \vec{e}_{A_{i}}}
        \,.
    \end{align}

    \emph{Case $A_{i-1} = A_{i+1}$ (no contribution).}
    There is no source and $C_{\boldsymbol{A}}^{(i)} \big|_{A_{i-1} = A_{i+1}} = 0$.

    Combining all cases, we have
    \begin{align}
        C_{\boldsymbol{A}}^{(i)}
         & =
        (-1)^{n+m}
        \biggl[
            \delta_{A_{i-1}, A_{i}} \ffuncone{n}{\vec{N} + m - \vec{e}_{A_{i}} - \vec{e}_{A_{i+1}}}
            -
            \delta_{A_{i}, A_{i+1}} \ffuncone{n}{\vec{N} + m - \vec{e}_{A_{i-1}} - \vec{e}_{A_{i}}}
        \nonumber        \\
         & \hspace*{3em}
            +
            \overline{\delta}_{A_{i}, A_{i+1}}
            \overline{\delta}_{A_{i-1}, A_{i}}
            \qty(
            \ffuncone{n}{\vec{N} + m - \vec{e}_{A_{i}} - \vec{e}_{A_{i+1}}}
            -
            \ffuncone{n}{\vec{N} + m - \vec{e}_{A_{i}} - \vec{e}_{A_{i-1}}}
            )
            \biggr]
        \nonumber        \\
         & =
        (-1)^{n+m}
        \biggl[
            \overline{\delta}_{A_{i}, A_{i+1}}  \ffuncone{n}{\vec{N} + m - \vec{e}_{A_{i}} - \vec{e}_{A_{i+1}}}
            -
            \overline{\delta}_{A_{i-1}, A_{i}} \ffuncone{n}{\vec{N} + m - \vec{e}_{A_{i}} - \vec{e}_{A_{i-1}}}
            \biggr]
        \,.
    \end{align}
    Thus, the total contribution from the bulk positions is
    \begin{align}
        \sum_{i=2}^{l-1} C_{\boldsymbol{A}}^{(i)}
         & =
        (-1)^{n+m}
        \qty[
            \overline{\delta}_{A_{l-1}, A_{l}}
            \ffuncone{n}{\vec{N} + m - \vec{e}_{A_{l}} - \vec{e}_{A_{l-1}}}
            -
            \overline{\delta}_{A_{1}, A_{2}}
            \ffuncone{n}{\vec{N} + m - \vec{e}_{A_{1}} - \vec{e}_{A_{2}}}
        ]
        \,.
    \end{align}

    \paragraph{Cancellation of contributions.}

    Summing all contributions, we have
    \begin{align}
          & \sum_{i=0}^{l+1} C_{\boldsymbol{A}}^{(i)}
        =
        C_{\boldsymbol{A}}^{(0)}
        +
        C_{\boldsymbol{A}}^{(1)}
        +
        \sum_{i=2}^{l-1} C_{\boldsymbol{A}}^{(i)}
        +
        C_{\boldsymbol{A}}^{(l)}
        +
        C_{\boldsymbol{A}}^{(l+1)}
        \nonumber                                     \\
        = &
        (-1)^{n+m}
        \biggl[
            -2 \ffuncone{n}{\vec{N} + m - 1}
            +
            \sum_{B \neq A_1}
            \ffuncone{n}{\vec{N} + m - 1 + \vec{e}_{B}}
        \nonumber                                     \\
          & \hspace*{2em}
            -
            \overline{\delta}_{A_1, A_2}
            \ffuncone{n}{\vec{N} + m - 1 + \vec{e}_{A_2}}
            -
            \delta_{A_1, A_2}
            \sum_{B \neq A_1}
            \ffuncone{n}{\vec{N} + m - 1 + \vec{e}_{B}}
            -
            \overline{\delta}_{A_{1}, A_{2}}
            \ffuncone{n}{\vec{N} + m - \vec{e}_{A_{1}} - \vec{e}_{A_{2}}}
        \nonumber                                     \\
          & \hspace*{2em}
            +
            \overline{\delta}_{A_{l-1}, A_{l}}
            \ffuncone{n}{\vec{N} + m - \vec{e}_{A_{l}} - \vec{e}_{A_{l-1}}}
            +
            \overline{\delta}_{A_{l-1}, A_l}
            \ffuncone{n}{\vec{N} + m - 1 + \vec{e}_{A_{l-1}}}
            +
            \delta_{A_{l-1}, A_l}
            \sum_{B \neq A_l}
            \ffuncone{n}{\vec{N} + m - 1 + \vec{e}_{B}}
        \nonumber                                     \\
          & \hspace*{2em}
            -
            \sum_{B \neq A_l}
            \ffuncone{n}{\vec{N} + m - 1 + \vec{e}_{B}}
            +
            2 \ffuncone{n}{\vec{N} + m - 1}
            \biggr]
        \,.
    \end{align}
    The terms $\pm 2 \ffuncone{n}{\vec{N} + m - 1}$ cancel.
    For the remaining terms, we use the identity $\ffuncone{n}{\vec{N} + m - \vec{e}_{A_{1}} - \vec{e}_{A_{2}}} = \ffuncone{n}{\vec{N} + m - 1 +  \vec{e}_{A_{1, 2}}}$ when $A_1 \neq A_2$ with $\{A_1, A_2, A_{1,2}\} = \{X, Y, Z\}$, which gives
    \begin{align}
          &
        \overline{\delta}_{A_1, A_2}
        \ffuncone{n}{\vec{N} + m - 1 + \vec{e}_{A_2}}
        +
        \delta_{A_1, A_2}
        \sum_{B \neq A_1}
        \ffuncone{n}{\vec{N} + m - 1 + \vec{e}_{B}}
        +
        \overline{\delta}_{A_{1}, A_{2}}
        \ffuncone{n}{\vec{N} + m - \vec{e}_{A_{1}} - \vec{e}_{A_{2}}}
        \nonumber \\
        = &
        (\delta_{A_1, A_2} + \overline{\delta}_{A_1, A_2})
        \sum_{B \neq A_1}
        \ffuncone{n}{\vec{N} + m - 1 + \vec{e}_{B}}
        =
        \sum_{B \neq A_1}
        \ffuncone{n}{\vec{N} + m - 1 + \vec{e}_{B}}
        \,.
    \end{align}
    An analogous relation holds for the right-edge terms with $A_{l-1}$ and $A_l$.
    Substituting these back, we obtain
    \begin{align}
        \sum_{i=0}^{l+1} C_{\boldsymbol{A}}^{(i)}
         & =
        (-1)^{n+m}
        \biggl[
            \qty(
            \sum_{B \neq A_1}
            \ffuncone{n}{\vec{N} + m - 1 + \vec{e}_{B}}
            -
            \sum_{B \neq A_1}
            \ffuncone{n}{\vec{N} + m - 1 + \vec{e}_{B}}
            )
        \nonumber        \\
         & \hspace*{8em}
            +
            \qty(
            \sum_{B \neq A_l}
            \ffuncone{n}{\vec{N} + m - 1 + \vec{e}_{B}}
            -
            \sum_{B \neq A_l}
            \ffuncone{n}{\vec{N} + m - 1 + \vec{e}_{B}}
            )
            \biggr]
        \nonumber        \\
         &
        = 0
        \,.
    \end{align}

    This completes the proof of Theorem~\ref{thm:coefficient-recurrence-conservation}.

    \section{Proof of Theorem~\ref{thm:Qk-MPO}}
    \label{app:Qk-MPO}

    In this appendix, we prove that the MPO representation~\eqref{eq:MPOall} reproduces the local conserved quantities $Q_k$.

    \subsection*{Structure of the local tensor}

    The local tensor $\Gamma_k^i$ of Eq.~\eqref{eq:mpo_rep_result} is a $k \times k$ block matrix with the following properties:
    \begin{enumerate}
        \item \textbf{Almost upper triangular}: All entries below the diagonal vanish except for the bottom-left corner $(k, 1)$.
        \item \textbf{Diagonal entries}: The only nonzero diagonal entry is $(\Gamma_k^i)_{1,1} = I$.
        \item \textbf{Dual-number corner}: The bottom-left entry $(\Gamma_k^i)_{k,1} = \epsilon \sbs_i^\top$ contains the dual number $\epsilon$ satisfying $\epsilon^2 = 0$.
    \end{enumerate}

    \subsection*{Path interpretation of the trace}

    Expanding the trace over the auxiliary space, we obtain
    \begin{align}
        \Tr_a \bck{\Gamma_k^1 \Gamma_k^2 \cdots \Gamma_k^L}
        = \sum_{a_1, a_2, \ldots, a_L = 1}^{k}
        (\Gamma_k^1)_{a_1, a_2} (\Gamma_k^2)_{a_2, a_3} \cdots (\Gamma_k^L)_{a_L, a_1}.
        \label{eq:trace_expansion}
    \end{align}
    Each term in this sum can be interpreted as a closed path $(a_1 \to a_2 \to \cdots \to a_L \to a_1)$ through the auxiliary indices.

    \begin{lem}
        A product $(\Gamma_k^1)_{a_1, a_2} (\Gamma_k^2)_{a_2, a_3} \cdots (\Gamma_k^L)_{a_L, a_1}$ is nonzero only if:
        \begin{enumerate}
            \item The sequence $(a_1, a_2, \ldots, a_L)$ is constant and equal to $(1, 1, \ldots, 1)$, or
            \item The sequence contains exactly one ``descent'' from $k$ to $1$ via the element $\epsilon \sbs_i^\top$.
        \end{enumerate}
    \end{lem}

    \begin{proof}
        Since the only entry below the diagonal is the corner $(k,1)$ and the only nonzero diagonal entry is at $(1,1)$, a closed path that leaves index $1$ must increase strictly at every step until it reaches index $k$, and can return to index $1$ only through the bottom-left entry.
        The constraint $\epsilon^2 = 0$ ensures that this return occurs at most once.
    \end{proof}

    Based on the lemma, we decompose the trace as
    \begin{align}
        \Tr_a \bck{\Gamma_k^1 \Gamma_k^2 \cdots \Gamma_k^L}
         & = I + \epsilon \sum_{j=1}^{L} \sum_{s=2}^{k}
        \sum_{1 = a_1 < a_2 < \cdots < a_s = k}
        (\Gamma_k^j)_{a_1, a_2} (\Gamma_k^{j+1})_{a_2, a_3} \cdots (\Gamma_k^{j+s-2})_{a_{s-1}, a_s} \, \sbs_{j+s-1}^{\top},
        \label{eq:trace_decomposition}
    \end{align}
    where the inner sum runs over strictly ascending sequences from $1$ to $k$.
    The index $j$ labels the starting site (with periodic identification $\Gamma_k^{j+L} \equiv \Gamma_k^j$), and $s$ denotes the number of support sites.
    The ascending segment itself has $s-1$ steps, followed by the bottom-left transition $\epsilon\sbs_{j+s-1}^{\top}$.

    \subsection*{Parametrization by step sizes}

    We parametrize the ascending paths by the step sizes $b_p \equiv a_{p+1} - a_p$ for $p = 0, 1, \ldots, s-2$.
    These satisfy $b_p > 0$ and $\sum_{p=0}^{s-2} b_p = k - 1$.

    From the definition~\eqref{eq:mpo_rep_result}, the matrix element $(\Gamma_k^i)_{a_p, a_{p+1}}$ depends only on:
    \begin{itemize}
        \item whether $a_p = 1$ (first row) or $a_p > 1$ (bulk rows),
        \item the step size $b_p = a_{p+1} - a_p$,
        \item the parity of $b_p$.
    \end{itemize}

    We denote this element by $\gamma_i^{b_p}$, defined as
    \begin{equation}
        \gamma_i^{b} \equiv (\Gamma_k^i)_{a, a+b}
        =
        \begin{cases}
            \sbs_i^{(b-1)/2} & \text{if } a = 1 \text{ and } b \text{ is odd}, \\
            \Mbs_i^{(b-1)/2} & \text{if } a > 1 \text{ and } b \text{ is odd}, \\
            \ebs^{b/2 - 1}   & \text{if } b \text{ is even}.
        \end{cases}
        \label{eq:gamma_definition}
    \end{equation}

    Substituting this parametrization into Eq.~\eqref{eq:trace_decomposition}, the sum over ascending paths becomes
    \begin{align}
         & \sum_{j=1}^{L} \sum_{s=2}^{k}
        \sum_{1 = a_1 < a_2 < \cdots < a_s = k}
        (\Gamma_k^j)_{a_1, a_2} \cdots (\Gamma_k^{j+s-2})_{a_{s-1}, a_s} \, \sbs_{j+s-1}^{\top}
        \nonumber                                          \\
         & = \sum_{j=1}^{L} \sum_{s=2}^{k}
        \sum_{\substack{b_0 + b_1 + \cdots + b_{s-2} = k-1 \\ b_0: \text{odd}, \, b_p > 0}}
        \sbs_j^{(b_0-1)/2} \, \gamma_{j+1}^{b_1} \cdots \gamma_{j+s-2}^{b_{s-2}} \, \sbs_{j+s-1}^{\top}.
        \label{eq:path_sum_parametrized}
    \end{align}
    Here, the constraint that $b_0$ is odd arises because $(\Gamma_k^i)_{1, 1+b_0}$ is nonzero only when $1 + b_0$ is even.

    We classify the ascending paths according to the parity of their step sizes.
    Let $m$ denote the number of even steps and $t$ the number of odd steps, so that $t + m = s - 1$.
    We write $0 = i_0 < i_1 < \cdots < i_{t-1}$ for the odd-step positions and $\tilde{i}_1 < \tilde{i}_2 < \cdots < \tilde{i}_m$ for the even-step positions; together they partition the step indices $\{0, 1, \ldots, s-2\}$.
    Here $i_0 = 0$ because $b_0$ is odd.

    Each $\ebs$ factor inserts an identity operator in the Pauli string, corresponding to a hole; thus the number of even steps $m$ equals the number of holes in the resulting operator.
    We define the indices that label the building blocks:
    \begin{align}
        g_{i_r}         & \equiv (b_{i_r} - 1)/2 \quad \text{(index of $\sbs^{g_{i_r}}$ or $\Mbs^{g_{i_r}}$ at odd step $i_r$)},    \\
        h_{\tilde{i}_r} & \equiv b_{\tilde{i}_r}/2 - 1 \quad \text{(index of $\ebs^{h_{\tilde{i}_r}}$ at even step $\tilde{i}_r$)}.
    \end{align}
    The sum of these indices is denoted by $n \equiv \sum_{r=0}^{t-1} g_{i_r} + \sum_{r=1}^{m} h_{\tilde{i}_r}$.

    The constraint $\sum_{p=0}^{s-2} b_p = k - 1$ fixes $t$ and $s$ in terms of $n$ and $m$.
    Since $b_{i_r} = 2g_{i_r} + 1$ for odd steps and $b_{\tilde{i}_r} = 2h_{\tilde{i}_r} + 2$ for even steps, we have
    \begin{align}
        k - 1 & = \sum_{r=0}^{t-1} (2g_{i_r} + 1) + \sum_{r=1}^{m} (2h_{\tilde{i}_r} + 2) = 2n + t + 2m.
        \nonumber
    \end{align}
    Solving for $t$, we obtain $t = k - 2(n + m) - 1$.
    The support is then $s = k - 2n - m$.
    \subsection*{Expanding the building blocks}

    For fixed parameters $(n,m)$, we expand the product of building blocks using the definitions $\sbs_i^g = \sbs_i \Jbs \Rbs^g$, $\Mbs_i^g = \Mbs_i \Jbs \Rbs^g$, and $\ebs^h = \ebs \Jtilbs \, \rfunc{h}{1}{1}{1}$ from Section~\ref{sec:Building-blocks}.

    To describe the structure of the product, we define the segment lengths
    \begin{equation}
        \ell_r \equiv i_r - i_{r-1} \quad (r = 1, \ldots, t),
    \end{equation}
    where we set $i_t \equiv s$ for notational convenience.
    Each segment $r$ contains one odd step at position $i_{r-1}$ (contributing $\sbs$ or $\Mbs$) followed by $\ell_r - 1$ even steps (each contributing $\ebs$).

    After expanding and rearranging, the product of matrix elements becomes
    \begin{align}
         & \sbs_j^{(b_0-1)/2} \, \gamma_{j+1}^{b_1} \cdots \gamma_{j+s-2}^{b_{s-2}} \, \sbs_{j+s-1}^{\top}
        \nonumber                                                                                          \\
         & = \qty(\prod_{r=1}^{m} \rfunc{h_{\tilde{i}_r}}{1}{1}{1})
        \times \sbs_j \Jbs \Rbs^{g_{i_0}} \Jtilbs^{\ell_1 - 1} \Mbs_{j+i_1} \Jbs \Rbs^{g_{i_1}} \Jtilbs^{\ell_2 - 1} \cdots \Mbs_{j+i_{t-1}} \Jbs \Rbs^{g_{i_{t-1}}} \Jtilbs^{\ell_t - 1} \sbs_{j+s-1}^{\top}.
        \label{eq:product_expansion}
    \end{align}
    Here we have used the fact that $\ebs$ commutes with all other building blocks, and collected the scalar factors $\rfunc{h_{\tilde{i}_r}}{1}{1}{1}$ from the $\ebs^{h_{\tilde{i}_r}}$ terms.

    \subsection*{Evaluating the Pauli structure}

    The matrices $\Jbs$, $\Jtilbs$, and $\Rbs^n$ are all diagonal, while $\sbs$ is a row vector and $\Mbs$ encodes transitions between Pauli indices.
    Since $\Mbs$ has zeros on the diagonal, the Pauli type must change at each $\Mbs$ insertion.

    Let $\boldsymbol{A} = (A_1, A_2, \ldots, A_t)$ with $A_r \in \{X, Y, Z\}$ denote the sequence of Pauli types, where $A_r \neq A_{r+1}$ for $r = 1, \ldots, t-1$.
    The first Pauli type $A_1$ is determined by $\sbs_j$, and the final column vector $\sbs_{j+s-1}^{\top}$ carries type $A_t$.
    The contribution to the operator string is
    \begin{equation}
        \overline{A_1^{\ell_1} A_2^{\ell_2} \cdots A_t^{\ell_t}},
    \end{equation}
    where the overline denotes translation averaging (summing over starting positions $j$).

    The diagonal structure of $\Jbs$ and $\Jtilbs$ determines the coupling constant factor.
    Noting that $(\Jtilbs)_{A_r A_r} = J_X J_Y J_Z / J_{A_r}$, the product of $\Jbs$ and $\Jtilbs$ factors along the Pauli sequence evaluates to
    \begin{equation}
        J_{\boldsymbol{A}} = (J_X J_Y J_Z)^m \prod_{r=1}^{t} J_{A_r}^{2 - \ell_r}.
    \end{equation}

    Similarly, the diagonal structure of $\Rbs^g$ gives the polynomial coefficient factor.
    Using the notation $(n_r^X, n_r^Y, n_r^Z) = (\delta_{A_r, X}, \delta_{A_r, Y}, \delta_{A_r, Z})$, the sum over all ways to distribute the index $n$ among the building blocks yields
    \begin{equation}
        \sum_{\substack{\sum_{r=0}^{t-1} g_{i_r} + \sum_{r=1}^{m} h_{\tilde{i}_r} = n}}
        \qty(\prod_{r=1}^{m} \rfunc{h_{\tilde{i}_r}}{1}{1}{1})
        \qty(\prod_{r=1}^{t} \rfunc{g_{i_{r-1}}}{n_r^X}{n_r^Y}{n_r^Z}).
    \end{equation}
    By the convolution identity~\eqref{eq:R-identity}, this sum simplifies to
    \begin{equation}
        \rfunc{n}{N_x^{\boldsymbol{A}} + m}{N_y^{\boldsymbol{A}} + m}{N_z^{\boldsymbol{A}} + m},
    \end{equation}
    where $N_x^{\boldsymbol{A}}$, $N_y^{\boldsymbol{A}}$, $N_z^{\boldsymbol{A}}$ count the number of $X$, $Y$, $Z$ labels in $\boldsymbol{A}$, respectively.

    \subsection*{Final summation}

    Collecting all contributions, we obtain
    \begin{align}
         & \sum_{j=1}^{L} \sum_{1 = a_1 < a_2 < \cdots < a_s = k}
        (\Gamma_k^j)_{1, a_2} \cdots (\Gamma_k^{j+s-2})_{a_{s-1}, k} \, \sbs_{j+s-1}^{\top}
        \nonumber                                                 \\
         & = \sum_{\substack{0 \leq n+m < \lfloor k/2 \rfloor     \\ n, m \geq 0}}
        \sum_{\substack{\boldsymbol{A} = (A_1, \ldots, A_t)       \\ A_r \in \{X, Y, Z\}, \, A_r \neq A_{r+1}}}
        \sum_{0 = i_0 < i_1 < \cdots < i_{t-1} < i_t = s}
        J_{\boldsymbol{A}} \, \rfunc{n}{N_x^{\boldsymbol{A}} + m}{N_y^{\boldsymbol{A}} + m}{N_z^{\boldsymbol{A}} + m}
        \, \overline{A_1^{\ell_1} A_2^{\ell_2} \cdots A_t^{\ell_t}}
        \nonumber                                                 \\
         & = \sum_{\substack{0 \leq n+m < \lfloor k/2 \rfloor     \\ n, m \geq 0}}
        \sum_{\obA \in \mathcal{S}_k^{n,m}}
        J_{\boldsymbol{A}} \, \rfunc{n}{N_x^{\boldsymbol{A}} + m}{N_y^{\boldsymbol{A}} + m}{N_z^{\boldsymbol{A}} + m}
        \, \obA,
        \label{eq:final_sum}
    \end{align}
    where $t \equiv k - 2(n + m) -1$.

    Since $\ell_r$ counts the letters in the $r$-th group, it equals $m_r + 1$ in the abbreviated notation~\eqref{doubling-abbreviated}, so $J_{\boldsymbol{A}}$ above coincides with the definition below Eq.~\eqref{eq:xyz-Qk}.
    Comparing with the definition of $Q_k$ in Eq.~\eqref{eq:xyz-Qk}, we confirm that
    \begin{equation}
        \Tr_a \bck{\Gamma_k^1 \Gamma_k^2 \cdots \Gamma_k^L} = I + \epsilon Q_k,
    \end{equation}
    which completes the proof of Theorem~\ref{thm:Qk-MPO}.

        \section{Proof of Theorem~\ref{thm:fused-expansion}}
        \label{app:fused-expansion}

        This appendix proves Theorem~\ref{thm:fused-expansion} from a product of two eight-vertex transfer matrices near the fusion point of Baxter's R-matrix~\cite{Kulish-Reshetikhin-Sklyanin-1981,Konno2006}, and recovers the local tensor $\w_i(s; \epsilon)$ of Eq.~\eqref{eq:w-mpo}.
        To retain the $\epsilon$-dependent entries of $\w_i(s;\epsilon)$, we keep the full four-dimensional auxiliary space rather than project onto the triplet at the fusion point, and expand to first order in $\epsilon$.

        We use Baxter's R-matrix $R^{8V}_{ai}(z)$ and the transfer matrix $t^{8V}(z)$ of Eqs.~\eqref{eq:baxter-R-main}--\eqref{eq:baxter-transfer-main}, with $(\tau,\eta_{8V})$ realizing the given coupling ratios~\cite{Baxter1982,Takebe-1995}.
        The fusion procedure starts from two such R-matrices sharing the physical site $i$, with auxiliary spaces $1$ and $2$~\cite{Kulish-Reshetikhin-Sklyanin-1981,Konno2006} and spectral parameters
        \begin{equation}
            \label{eq:baxter-fusion-raw-coordinates-rev}
            z_1=z-2\eta_{8V}+\delta_0
            \,,
            \qquad
            z_2=z
            \,,
            \qquad
            \text{fusion point: }
            \delta_0=0
            \,.
        \end{equation}
        Their product defines
        \begin{equation}
            \label{eq:baxter-fusion-unprojected-L-rev}
            \mathbb L^{8V}_{A,i}(z,\delta_0)
            \coloneqq
            R^{8V}_{1i}(z_1)R^{8V}_{2i}(z_2)
            \,,
            \qquad
            A=(1,2)
            \,.
        \end{equation}

        \subsection*{Singlet--triplet form of the fused Lax operator}

        We write $\mathbb L^{8V}_{A,i}$ in the singlet--triplet basis of the two auxiliary spin-$1/2$ spaces, with phases chosen so that Eq.~\eqref{eq:baxter-fusion-singlet-triplet-L-rev} below takes the symmetric form,
        \begin{equation}
            \label{eq:baxter-fusion-singlet-triplet-basis-rev}
            \begin{aligned}
                e_0
                 & =
                \frac{\ket{\uparrow\downarrow}-\ket{\downarrow\uparrow}}{\sqrt{2}}
                \,,
                 &
                e_X
                 & =
                -\frac{\ket{\uparrow\uparrow}-\ket{\downarrow\downarrow}}{\sqrt{2}}
                \,,
                \\
                e_Y
                 & =
                \frac{\imi(\ket{\uparrow\uparrow}+\ket{\downarrow\downarrow})}{\sqrt{2}}
                \,,
                 &
                e_Z
                 & =
                \frac{\ket{\uparrow\downarrow}+\ket{\downarrow\uparrow}}{\sqrt{2}}
                \,.
            \end{aligned}
        \end{equation}

        The two weight tuples of the product are abbreviated by
        \begin{equation}
            \label{eq:baxter-fusion-weight-tuples-rev}
            x_\mu(\delta_0)
            =
            \varphi_\mu(z_1)
            \,,
            \qquad
            y_\mu
            =
            \varphi_\mu(z_2)
            \,.
        \end{equation}
        Below, $x_\mu\coloneqq x_\mu(0)$ and $F(0)$ denote fusion-point values.

        Multiplication of the two factors of Eq.~\eqref{eq:baxter-fusion-unprojected-L-rev} in the basis~\eqref{eq:baxter-fusion-singlet-triplet-basis-rev} gives
        \begin{equation}
            \label{eq:baxter-fusion-singlet-triplet-L-rev}
            \mathbb L_i^{8V}(\delta_0)
            =
            \begin{pmatrix}
                \Lambda_0I & P_X^-X_i      & P_Y^-Y_i      & P_Z^-Z_i      \\
                P_X^+X_i   & \Lambda_XI    & Q_{XY}^{Z}Z_i & Q_{XZ}^{Y}Y_i \\
                P_Y^+Y_i   & Q_{YX}^{Z}Z_i & \Lambda_YI    & Q_{YZ}^{X}X_i \\
                P_Z^+Z_i   & Q_{ZX}^{Y}Y_i & Q_{ZY}^{X}X_i & \Lambda_ZI
            \end{pmatrix}
            \,.
        \end{equation}

        The diagonal coefficients are
        \begin{align}
            \label{eq:baxter-fusion-Lambda-rev}
            \Lambda_0(\delta_0)
             & =
            x_0(\delta_0)y_0
            -x_X(\delta_0)y_X
            -x_Y(\delta_0)y_Y
            -x_Z(\delta_0)y_Z
            \,,
            \\
            \Lambda_\alpha(\delta_0)
             & =
            x_0(\delta_0)y_0
            -x_\alpha(\delta_0)y_\alpha
            +x_\beta(\delta_0)y_\beta
            +x_\gamma(\delta_0)y_\gamma
            \,,
            \qquad
            \{\alpha,\beta,\gamma\}=\{X,Y,Z\}
            \,.
            \nonumber
        \end{align}

        For a cyclic triple $(\alpha,\beta,\gamma)$ of $(X,Y,Z)$,
        \begin{equation}
            \label{eq:baxter-fusion-Ppm-rev}
            P_\alpha^\pm(\delta_0)
            =
            x_\alpha(\delta_0)y_0
            -x_0(\delta_0)y_\alpha
            \pm
            \qty(
            x_\beta(\delta_0)y_\gamma
            +x_\gamma(\delta_0)y_\beta
            )
            \,.
        \end{equation}

        For pairwise distinct $\alpha,\beta,\gamma$,
        \begin{equation}
            \label{eq:baxter-fusion-Qabg-rev}
            Q_{\alpha\beta}^{\gamma}(\delta_0)
            =
            -\imi\varepsilon_{\alpha\beta\gamma}
            \qty(
            x_0(\delta_0)y_\gamma
            +x_\gamma(\delta_0)y_0
            +x_\alpha(\delta_0)y_\beta
            -x_\beta(\delta_0)y_\alpha
            )
            \,,
            \qquad
            \varepsilon_{XYZ}=1
            \,.
        \end{equation}

        \subsection*{Scalar normalization and gauge fixing}

        Wherever $\Lambda_0P_X^+P_Y^+P_Z^+\ne0$ at the fusion point, we normalize by the scalar $\Lambda_0$ and apply the diagonal auxiliary gauge that fixes the first column to $(I,X_i,Y_i,Z_i)^\top$:
        \begin{equation}
            \label{eq:baxter-fusion-lower-left-gauge-rev}
            G
            =
            \operatorname{diag}
            \qty(
            1,
            \frac{P_X^+}{\Lambda_0},
            \frac{P_Y^+}{\Lambda_0},
            \frac{P_Z^+}{\Lambda_0}
            )
            \,,
            \qquad
            \widetilde L_i^{8V}(\delta_0)
            =
            G^{-1}\frac{\mathbb L_i^{8V}(\delta_0)}{\Lambda_0(\delta_0)}G
            \,.
        \end{equation}

        Then
        \begin{equation}
            \label{eq:baxter-fusion-gauged-L-rev}
            \widetilde L_i^{8V}(\delta_0)
            =
            \begin{pmatrix}
                I   & \mathfrak b_XX_i        & \mathfrak b_YY_i        & \mathfrak b_ZZ_i        \\
                X_i & \mathfrak c_XI          & \mathfrak d_{XY}^{Z}Z_i & \mathfrak d_{XZ}^{Y}Y_i \\
                Y_i & \mathfrak d_{YX}^{Z}Z_i & \mathfrak c_YI          & \mathfrak d_{YZ}^{X}X_i \\
                Z_i & \mathfrak d_{ZX}^{Y}Y_i & \mathfrak d_{ZY}^{X}X_i & \mathfrak c_ZI
            \end{pmatrix}
            \,.
        \end{equation}

        The coefficients are
        \begin{equation}
            \label{eq:baxter-fusion-gauge-fixed-coefficients-rev}
            \mathfrak b_\alpha
            =
            \frac{P_\alpha^+P_\alpha^-}{\Lambda_0^2}
            \,,
            \qquad
            \mathfrak c_\alpha
            =
            \frac{\Lambda_\alpha}{\Lambda_0}
            \,,
            \qquad
            \mathfrak d_{\alpha\beta}^{\gamma}
            =
            \frac{P_\beta^+}{P_\alpha^+}
            \frac{Q_{\alpha\beta}^{\gamma}}{\Lambda_0}
            \,.
        \end{equation}
        They are homogeneous of degree zero in $y$.

        \subsection*{Coordinates for the expansion at the fusion point}

        At the fusion point, write $x_\mu=x_\mu(0)=\varphi_\mu(z-2\eta_{8V})$; then there is a scalar $\nu$ such that
        \begin{equation}
            \label{eq:baxter-fusion-curve-nu-rev}
            x_\alpha^2
            =
            x_0^2-(J_\beta+J_\gamma)\nu
            \,,
            \qquad
            \{\alpha,\beta,\gamma\}=\{X,Y,Z\}
            \,,
        \end{equation}
        where $\nu=\nu(z-2\eta_{8V})$ is the factor of Eq.~\eqref{eq:baxter-invariants-main} at the argument $z-2\eta_{8V}$.
        The theta-function addition formulas~\cite[Chapter~21]{Whittaker-Watson-1927} give
        \begin{equation}
            \label{eq:baxter-fusion-nu-theta-factor-rev}
            \nu(z-2\eta_{8V})
            \propto
            \vartheta_1(z-2\eta_{8V}\mid\tau)
            \vartheta_1(z\mid\tau)
            \,.
        \end{equation}
        The nonzero proportionality factor is independent of $z$ and depends only on the overall normalization of $J_\alpha$.

        From $x_\mu$ and $\nu$ we form
        \begin{align}
            \label{eq:baxter-fusion-D-A-Pi-rev}
            D_\alpha
             & =
            x_0^2+x_\alpha^2-x_\beta^2-x_\gamma^2
            =
            2J_\alpha\nu
            \,,
            \\
            A_\alpha
             & =
            x_0x_\alpha-x_\beta x_\gamma
            \,,
            \nonumber       \\
            \Pi
             & =
            (x_0-x_X-x_Y+x_Z)(x_0-x_X+x_Y-x_Z)
            \nonumber       \\
             & \qquad\times
            (x_0+x_X-x_Y-x_Z)(x_0+x_X+x_Y+x_Z)
            \,.
            \nonumber
        \end{align}
        Pairing the four factors of $\Pi$ for any fixed $\alpha$ gives
        \begin{equation}
            \label{eq:baxter-fusion-A-identity-rev}
            \Pi
            =
            \qty(D_\alpha+2A_\alpha)\qty(D_\alpha-2A_\alpha)
            =
            D_\alpha^2-4A_\alpha^2
            \,,
            \qquad
            A_\alpha^2+\frac{\Pi}{4}
            =
            J_\alpha^2\nu^2
            \,.
        \end{equation}

        We parametrize the fusion curve by
        \begin{equation}
            \label{eq:baxter-fusion-s-branch-rev}
            s
            \coloneqq
            \frac{16\imi\nu}{\Pi^2}
            A_XA_YA_Z
            \,,
        \end{equation}
        which is rational in $x_\mu$ and $\nu$.
        Since $\vartheta_1$ has a simple zero at the origin and $\vartheta_1(-2\eta_{8V}\mid\tau)\ne0$, Eq.~\eqref{eq:baxter-fusion-nu-theta-factor-rev} shows that $\nu$ has a simple zero at $z=0$.
        At $z=0$, the identities $\vartheta_1(-u\mid\tau)=-\vartheta_1(u\mid\tau)$ and $\vartheta_j(-u\mid\tau)=\vartheta_j(u\mid\tau)$ for $j=2,3,4$ give $x_0=-1$ and $x_\alpha=1$ for $\alpha=X,Y,Z$.
        Consequently, $A_\alpha=-2$ and $\Pi=-16$.
        Thus $16\imi A_XA_YA_Z/\Pi^2=-\imi/2\ne0$, so $s$ has a simple zero at $z=0$.
        The complex inverse function theorem gives a unique local inverse $z=z(s)$ satisfying $z(0)=0$.

        With this definition, $4\nu^2/\Pi$ satisfies Eq.~\eqref{eq:eq-for-omega} at that value of $s$,
        \begin{equation}
            \label{eq:baxter-fusion-s-u-rev}
            \frac{4\nu^2}{\Pi}
            \prod_{\alpha=X,Y,Z}
            \qty(1-\frac{4\nu^2}{\Pi}J_\alpha^2)
            =
            s^2
            \,,
        \end{equation}
        because $s^2=-256\nu^2A_X^2A_Y^2A_Z^2/\Pi^4=(4\nu^2/\Pi)\prod_{\rho=X,Y,Z}(-4A_\rho^2/\Pi)$ and the identity~\eqref{eq:baxter-fusion-A-identity-rev} gives $-4A_\alpha^2/\Pi=1-4\nu^2J_\alpha^2/\Pi$.

        Since $z(s)\to0$ as $s\to0$, the combination $4\nu^2/\Pi$ is therefore the root of Eq.~\eqref{eq:eq-for-omega} that vanishes at $s=0$, namely the generating function $\omega$ of Eq.~\eqref{eq:fx-omega}:
        \begin{equation}
            \label{eq:baxter-fusion-u-coordinate-rev}
            \omega(s)
            =
            \frac{4\nu^2}{\Pi}
            \,,
            \qquad
            f_\alpha(s)
            =
            \frac{1}{1-\omega(s)J_\alpha^2}
            =
            -\frac{\Pi}{4A_\alpha^2}
            \,,
        \end{equation}
        with $f_\alpha$ defined by Eq.~\eqref{eq:fx-ito-fe}.

        The displacement of Eq.~\eqref{eq:baxter-fusion-raw-coordinates-rev} moves the argument $z_1$, so $x_\mu(\delta_0)=\varphi_\mu(z_1)$ stays on the eight-vertex curve with $\nu$ in Eq.~\eqref{eq:baxter-fusion-curve-nu-rev} evaluated at $z_1$.
        Writing $\dot x_\mu$ and $\dot\nu$ for the $\delta_0$-derivatives at $\delta_0=0$, we set $\Theta\coloneqq x_0\dot\nu-2\nu\dot x_0$.
        This combination measures the rate at which the displacement moves $x_\mu(\delta_0)$ off $P_X^-=P_Y^-=P_Z^-=0$.
        As $s\to0$ the argument $z_1$ at $\delta_0=0$ tends to a simple zero of $\nu$, so $\eval{\Theta}_{s=0}=x_0\eval{\dot\nu}_{s=0}\ne0$ and the branch $z(s)$ is regular at $s=0$.

        Substituting $\nu=s\Pi^2/(16\imi A_XA_YA_Z)$ from Eq.~\eqref{eq:baxter-fusion-s-branch-rev} writes the coordinate change as an explicit multiple of $s$,
        \begin{equation}
            \label{eq:baxter-fusion-chart-collapse-rev}
            \delta_0
            =
            s\,\kappa(s)\,\delta
            \,,
            \qquad
            s\,\kappa(s)
            =
            \nu\,
            \frac{\imi\,\Pi\,x_Xx_Yx_Z}{A_XA_YA_Z\,\Theta}
            \,,
            \qquad
            \kappa(s)
            \coloneqq
            \frac{\Pi^3\,x_Xx_Yx_Z}{16A_X^2A_Y^2A_Z^2\,\Theta}
            \,,
        \end{equation}
        with $\kappa(s)$ the same function that appears in Theorem~\ref{thm:fused-expansion}.
        Since $\kappa(0)\ne0$, one unit of $\delta$ is a displacement $\delta_0$ of order $s$.

        At $s=0$, the local inverse gives $z=0$, and Eq.~\eqref{eq:baxter-fusion-chart-collapse-rev} fixes $\delta_0=0$ for every $\delta$.
        Equation~\eqref{eq:baxter-fusion-raw-coordinates-rev} therefore gives $z_1=-2\eta_{8V}$ and $z_2=0$.
        The parity of the theta functions in Eq.~\eqref{eq:baxter-theta-weights-main} then gives
        \begin{align}
            R^{8V}_{ai}(0)
             & =
            2P_{ai}
            \,,
            \qquad
            R^{8V}_{ai}(-2\eta_{8V})
            =
            -4P^-_{ai}
            \,,
            \nonumber \\
            \mathbb L^{8V}_{A,i}(0,0)
             & =
            -8P^-_{1i}P_{2i}
            =
            -8P_{2i}P^-_{12}
            \,.
            \label{eq:baxter-fusion-s-zero-projector-rev}
        \end{align}
        Here $P_{ai}$ permutes spaces $a$ and $i$, and $P^-_{ab}$ projects spaces $a$ and $b$ onto their singlet state.
        Since $P^-_{12}$ annihilates auxiliary triplet states, Eq.~\eqref{eq:baxter-fusion-s-zero-projector-rev} shows that the triplet block vanishes at $s=0$.
        The same calculation gives $\Lambda_0=-4$ and $P_\alpha^+=4$, so the normalization and the gauge in Eq.~\eqref{eq:baxter-fusion-lower-left-gauge-rev} are regular at $s=0$.

        For any coefficient $F$ built from the tuples, $F(s,\delta)$ denotes its value at $z=z(s)$ and displacement $\delta_0=s\,\kappa(s)\,\delta$.
        A dot denotes the $\delta_0$-derivative at fixed $s$,
        \begin{equation}
            \label{eq:baxter-fusion-dot-convention-rev}
            \dot F(s)
            \coloneqq
            \eval{\partial_{\delta_0}F}_{\delta_0=0}
            =
            \frac{1}{s\,\kappa(s)}\,
            \eval{\partial_\delta F(s,\delta)}_{\delta=0}
            \,.
        \end{equation}

        \subsection*{Reduction at the fusion point}

        Equations~\eqref{eq:baxter-fusion-raw-coordinates-rev} and~\eqref{eq:baxter-fusion-chart-collapse-rev} give $z_1-z_2=-2\eta_{8V}+s\kappa(s)\delta$.
        Thus, for $s\ne0$, the fusion-point condition $z_1-z_2=-2\eta_{8V}$ is equivalent to $\delta=0$; at $s=0$, the change of variables gives $\delta_0=0$ for every $\delta$.
        At the fusion point, the singlet-to-triplet coefficients satisfy
        \begin{equation}
            \label{eq:baxter-fusion-divisor-rev}
            P_X^-(s,0)
            =
            P_Y^-(s,0)
            =
            P_Z^-(s,0)
            =
            0
            \,.
        \end{equation}

        At $z_1-z_2=-2\eta_{8V}$ Baxter's R-matrix is proportional to the singlet projector, which gives these three equations~\cite{Kulish-Reshetikhin-Sklyanin-1981,Konno2006}.
        We derive them from a theta-function addition formula~\cite[Chapter~21]{Whittaker-Watson-1927}, which also supplies the coefficients needed off the fusion point.
        With $w=z-\eta_{8V}$, $v=z+\eta_{8V}$, the common modulus $\tau$ suppressed, and $(\vartheta_X,\vartheta_Y,\vartheta_Z)=(\vartheta_2,\vartheta_3,\vartheta_4)$, that formula reads, for every cyclic triple $(\alpha,\beta,\gamma)$ of $(X,Y,Z)$,
        \begin{equation}
            \label{eq:baxter-fusion-theta-addition-specialization-rev}
            \frac{\vartheta_\alpha(w)\vartheta_1(v)}
            {\vartheta_\alpha(\eta_{8V})\vartheta_1(\eta_{8V})}
            -
            \frac{\vartheta_1(w)\vartheta_\alpha(v)}
            {\vartheta_1(\eta_{8V})\vartheta_\alpha(\eta_{8V})}
            =
            \frac{\vartheta_\beta(w)\vartheta_\gamma(v)}
            {\vartheta_\beta(\eta_{8V})\vartheta_\gamma(\eta_{8V})}
            +
            \frac{\vartheta_\gamma(w)\vartheta_\beta(v)}
            {\vartheta_\gamma(\eta_{8V})\vartheta_\beta(\eta_{8V})}
            \,.
        \end{equation}

        In the abbreviations~\eqref{eq:baxter-fusion-weight-tuples-rev}, Eq.~\eqref{eq:baxter-fusion-theta-addition-specialization-rev} gives, at the fusion point,
        \begin{equation}
            \label{eq:baxter-fusion-Pminus-addition-rev}
            x_\alpha y_0-x_0y_\alpha
            =
            x_\beta y_\gamma+x_\gamma y_\beta
        \end{equation}
        for every cyclic triple $(\alpha,\beta,\gamma)$, which is exactly $P_\alpha^-(s,0)=0$.
        The singlet-to-triplet entries in the top row of Eq.~\eqref{eq:baxter-fusion-singlet-triplet-L-rev} therefore vanish, so the product before projection maps the triplet sector into itself.
        In the gauge-fixed coefficients~\eqref{eq:baxter-fusion-gauge-fixed-coefficients-rev}, this vanishing reads $\mathfrak b_\alpha(s,0)=0$.

        With the columns ordered as $(y_0,y_X,y_Y,y_Z)$, Eq.~\eqref{eq:baxter-fusion-Pminus-addition-rev} is the homogeneous system
        \begin{equation}
            \label{eq:baxter-fusion-Pminus-system-rev}
            M(x)y
            =
            0
            \,,
            \qquad
            M(x)
            =
            \begin{pmatrix}
                x_X & -x_0 & -x_Z & -x_Y \\
                x_Y & -x_Z & -x_0 & -x_X \\
                x_Z & -x_Y & -x_X & -x_0
            \end{pmatrix}
            \,.
        \end{equation}

        The four signed maximal minors of $M(x)$ are $(\det M_{\widehat 0},\allowbreak-\det M_{\widehat X},\allowbreak\det M_{\widehat Y},\allowbreak-\det M_{\widehat Z})=-B$, with $B$ the cubic vector of Eq.~\eqref{eq:baxter-fusion-B-projective-rev}, so the cofactor identity gives $M(x)B(x)=0$.
        Since $B_0$ is not the zero polynomial, $M(x)$ generically has rank three, and its kernel is then the line spanned by $B(x)$.
        Hence $y_\mu\propto B_\mu(x)$, with
        \begin{align}
            \label{eq:baxter-fusion-B-projective-rev}
            B_0
             & =
            x_0^3-x_0x_X^2-x_0x_Y^2-x_0x_Z^2+2x_Xx_Yx_Z
            \,,
            \\
            B_\alpha
             & =
            x_0^2x_\alpha-2x_0x_\beta x_\gamma-x_\alpha^3
            +x_\alpha x_\beta^2+x_\alpha x_\gamma^2
            \,,
            \qquad
            \{\alpha,\beta,\gamma\}=\{X,Y,Z\}
            \,.
            \nonumber
        \end{align}
        Because the gauge-fixed coefficients~\eqref{eq:baxter-fusion-gauge-fixed-coefficients-rev} are of degree zero in $y$, the proportionality factor cancels, and every factorization below may be computed with $y_\mu=B_\mu$, without theta functions.

        \begin{lem}[Algebraic reduction at the fusion point]
            \label{lem:baxter-fusion-algebraic-reduction-rev}
            Substitute $y_\mu=B_\mu$ of Eq.~\eqref{eq:baxter-fusion-B-projective-rev} into Eqs.~\eqref{eq:baxter-fusion-Lambda-rev}--\eqref{eq:baxter-fusion-Qabg-rev}.
            Then, for pairwise distinct triplet labels $\alpha,\beta,\gamma$, the fusion-point values are
            \begin{align}
                \label{eq:baxter-fusion-factorization-rev}
                \Lambda_0(0)
                 & =
                \Pi
                \,,
                \\
                \Lambda_\alpha(0)
                 & =
                D_\beta D_\gamma
                \,,
                \nonumber \\
                P_\alpha^+(0)
                 & =
                4A_\beta A_\gamma
                \,,
                \nonumber \\
                Q_{\alpha\beta}^{\gamma}(0)
                 & =
                -2\imi\varepsilon_{\alpha\beta\gamma}A_\gamma D_\beta
                \,.
                \nonumber
            \end{align}

            Consequently, in the coordinates $s$, $\omega(s)$, and $f_\alpha(s)$ of Eqs.~\eqref{eq:baxter-fusion-s-branch-rev}--\eqref{eq:baxter-fusion-u-coordinate-rev}, the representative-independent triplet coefficients at the fusion point are
            \begin{equation}
                \label{eq:baxter-fusion-u-D-A-rev}
                \mathfrak c_\alpha(0)
                =
                \omega(s)J_\beta J_\gamma
                \,.
            \end{equation}

            The off-diagonal coefficients are
            \begin{equation}
                \label{eq:baxter-fusion-off-diagonal-compact-rev}
                \mathfrak d_{\alpha\beta}^{\gamma}(0)
                =
                \varepsilon_{\alpha\beta\gamma}\,
                sf_\beta(s)J_\beta
                \,.
            \end{equation}
        \end{lem}

        \begin{proof}
            For the representative cyclic triple $(\alpha,\beta,\gamma)=(X,Y,Z)$, substitution of Eq.~\eqref{eq:baxter-fusion-B-projective-rev} into Eqs.~\eqref{eq:baxter-fusion-Lambda-rev}--\eqref{eq:baxter-fusion-Qabg-rev} gives the polynomial factorizations
            \begin{align}
                \Lambda_0(0)
                 & =
                x_0B_0-x_XB_X-x_YB_Y-x_ZB_Z
                =
                \Pi
                \,,
                \nonumber \\
                \Lambda_X(0)
                 & =
                x_0B_0-x_XB_X+x_YB_Y+x_ZB_Z
                =
                D_YD_Z
                \,,
                \nonumber \\
                P_X^+(0)
                 & =
                2\qty(x_XB_0-x_0B_X)
                =
                4A_YA_Z
                \,,
                \nonumber \\
                Q_{XY}^{Z}(0)
                 & =
                -\imi\qty(x_0B_Z+x_ZB_0+x_XB_Y-x_YB_X)
                =
                -2\imi A_ZD_Y
                \,.
                \label{eq:baxter-fusion-factorization-representative-rev}
            \end{align}
            Cyclic permutation of $(X,Y,Z)$ gives all identities in Eq.~\eqref{eq:baxter-fusion-factorization-rev}.

            Since $D_\alpha=2J_\alpha\nu$ and $\omega(s)=4\nu^2/\Pi$, the diagonal ratio is $\mathfrak c_\alpha(0)=D_\beta D_\gamma/\Pi=\omega(s)J_\beta J_\gamma$.
            For the off-diagonal ratio, Eqs.~\eqref{eq:baxter-fusion-gauge-fixed-coefficients-rev} and~\eqref{eq:baxter-fusion-factorization-rev} give $\mathfrak d_{\alpha\beta}^{\gamma}(0)=-2\imi\varepsilon_{\alpha\beta\gamma}A_\alpha A_\gamma D_\beta/(A_\beta\Pi)$.
            Substituting $s=16\imi\nu A_XA_YA_Z/\Pi^2$, $f_\beta=-\Pi/(4A_\beta^2)$, and $D_\beta=2J_\beta\nu$ gives the same expression for $\varepsilon_{\alpha\beta\gamma}sf_\beta(s)J_\beta$, proving Eq.~\eqref{eq:baxter-fusion-off-diagonal-compact-rev}.
        \end{proof}

        Writing $\widetilde L_{i,\mathrm{trip}}^{8V}$ for the lower $3\times3$ triplet block of $\widetilde L_i^{8V}$, Lemma~\ref{lem:baxter-fusion-algebraic-reduction-rev} gives
        \begin{align}
            \label{eq:baxter-fusion-triplet-body-rev}
            \widetilde L_{i,\mathrm{trip}}^{8V}(0)
             & =
            \begin{pmatrix}
                \omega(s)J_YJ_Z I & sf_Y(s)J_Y Z_i    & -sf_Z(s)J_Z Y_i   \\
                -sf_X(s)J_X Z_i   & \omega(s)J_XJ_Z I & sf_Z(s)J_Z X_i    \\
                sf_X(s)J_X Y_i    & -sf_Y(s)J_Y X_i   & \omega(s)J_XJ_Y I
            \end{pmatrix}
            \nonumber \\
             & =
            \lax_i(s)
            \,,
        \end{align}
        where the last equality is the definition of $\lax_i(s)$ in Eq.~\eqref{eq:w-mpo}.

        \subsection*{First order off the fusion point}

        We move off the fusion point by displacing $\delta_0$ in Eq.~\eqref{eq:baxter-fusion-raw-coordinates-rev} at fixed $z=z(s)$, so that only the first spectral parameter $z_1$ is displaced.
        Since every coefficient in Eqs.~\eqref{eq:baxter-fusion-Lambda-rev}--\eqref{eq:baxter-fusion-Qabg-rev} is linear in $x_\mu(\delta_0)$ and $y_\mu$ is fixed, its derivative follows by substituting $\dot x_\mu$ for $x_\mu(\delta_0)$.
        The curve relation determines the derivatives $\dot x_\rho$: differentiating Eq.~\eqref{eq:baxter-fusion-curve-nu-rev} with respect to $\delta_0$ at $\delta_0=0$ and solving for $\dot x_\rho$ gives
        \begin{equation}
            \label{eq:baxter-fusion-xdot-solved-rev}
            \dot x_\rho
            =
            \frac{2x_0\dot x_0-(J_{\rho'}+J_{\rho''})\dot\nu}{2x_\rho}
            \,,
            \qquad
            \{\rho,\rho',\rho''\}=\{X,Y,Z\}
            \,,
        \end{equation}
        wherever $x_Xx_Yx_Z\ne0$.

        Where $\nu x_Xx_Yx_Z\ne0$, the curve relation gives $J_{\rho'}+J_{\rho''}=(x_0^2-x_\rho^2)/\nu$.
        Substituting this into Eq.~\eqref{eq:baxter-fusion-xdot-solved-rev} and using $\Theta=x_0\dot\nu-2\nu\dot x_0$ yields
        \begin{equation}
            \label{eq:baxter-fusion-xdot-theta-rev}
            \dot x_\rho
            =
            \frac{x_\rho\dot\nu}{2\nu}
            -
            \frac{x_0\Theta}{2\nu x_\rho}
            \,,
            \qquad
            \dot x_0
            =
            \frac{x_0\dot\nu-\Theta}{2\nu}
            \,.
        \end{equation}

        For the representative $y_\mu=B_\mu$ of Lemma~\ref{lem:baxter-fusion-algebraic-reduction-rev}, Eq.~\eqref{eq:baxter-fusion-xdot-theta-rev} separates the derivative of $P_\alpha^-$ into a tangent part and a transverse part:
        \begin{equation}
            \label{eq:baxter-fusion-Pminus-split-rev}
            \dot P_\alpha^-(s)
            =
            \frac{\dot\nu}{2\nu}P_\alpha^-(s,0)
            +
            \frac{\Theta}{2\nu x_Xx_Yx_Z}R_\alpha
            \,.
        \end{equation}
        Here $R_\alpha$ is obtained by cyclic permutation from
        \begin{align}
            R_X
             & \coloneqq
            x_Xx_Yx_ZB_X-x_0x_Yx_ZB_0
            +x_0x_Xx_ZB_Z+x_0x_Xx_YB_Y
            \nonumber    \\
             & =
            -\qty(x_0x_Y-x_Xx_Z)
            \qty(x_0x_Z-x_Xx_Y)
            \qty(x_0^2+x_X^2-x_Y^2-x_Z^2)
            \nonumber    \\
             & =
            -D_XA_YA_Z
            \,.
            \label{eq:baxter-fusion-RX-factorization-rev}
        \end{align}
        Thus $R_\alpha=-D_\alpha A_\beta A_\gamma$ for every cyclic triple $(\alpha,\beta,\gamma)$.

        Since $P_\alpha^-(s,0)=0$ and $D_\alpha=2J_\alpha\nu$, Eq.~\eqref{eq:baxter-fusion-Pminus-split-rev} becomes
        \begin{equation}
            \label{eq:baxter-fusion-Pminus-slope-rev}
            \dot P_\alpha^-(s)
            =
            -\frac{J_\alpha A_\beta A_\gamma}{x_Xx_Yx_Z}\,
            \Theta
            \,.
        \end{equation}

        Since $P_\alpha^-(s,0)=0$ by Eq.~\eqref{eq:baxter-fusion-divisor-rev}, only the factor $P_\alpha^-$ of $\mathfrak b_\alpha=P_\alpha^+P_\alpha^-/\Lambda_0^2$ varies at first order.
        The fusion-point values of Lemma~\ref{lem:baxter-fusion-algebraic-reduction-rev} then turn Eq.~\eqref{eq:baxter-fusion-Pminus-slope-rev} into the first derivative of the whole top row:
        \begin{equation}
            \label{eq:baxter-fusion-b-raw-expansion-rev}
            \dot{\mathfrak b}_\alpha(s)
            =
            \frac{4A_\beta A_\gamma}{\Pi^2}\,
            \dot P_\alpha^-(s)
            =
            \frac{f_\alpha(s)J_\alpha}{\kappa(s)}
            \,.
        \end{equation}
        The second equality uses Eq.~\eqref{eq:baxter-fusion-Pminus-slope-rev}, $A_\beta A_\gamma=A_XA_YA_Z/A_\alpha$, Eqs.~\eqref{eq:baxter-fusion-s-branch-rev} and~\eqref{eq:baxter-fusion-u-coordinate-rev}, and the definition of $\kappa(s)$.
        By the dot identity~\eqref{eq:baxter-fusion-dot-convention-rev}, the factor $1/\kappa(s)$ cancels, so that the top row at order $\delta$ is exactly the row $\sbs_i^\infty(s)$ of Eq.~\eqref{eq:w-mpo}; this is what fixes the normalization of $\delta$.
        One order further, the top row is $\delta\,\sbs_i^\infty(s)+\delta^2\va{r}_i(s)+O(\delta^3)$, which defines the second-order row $\va{r}_i(s)$.

        By Eq.~\eqref{eq:baxter-fusion-triplet-body-rev}, the triplet block at $\delta=0$ is $\lax_i(s)$.
        At $s=0$, Eq.~\eqref{eq:baxter-fusion-s-zero-projector-rev} shows that this block vanishes for every $\delta$.
        Its regular first-order correction $\delta\Delta_i(s)$ therefore satisfies $\Delta_i(0)=0$.
        Together with the top-row expansion above, this gives
        \begin{equation}
            \label{eq:baxter-fusion-pre-singular-block-rev}
            \widetilde L_i^{8V}(s,\delta)
            =
            \begin{pmatrix}
                I
                 &
                \delta\,\sbs_i^\infty(s)+\delta^2 \va{r}_i(s)+O(\delta^3)
                \\
                \sbs_i^\top
                 &
                \lax_i(s)+\delta\,\Delta_i(s)+O(\delta^2)
            \end{pmatrix}
            \,.
        \end{equation}

        In $\w_i(s; \epsilon)$ the dual number sits in the first column, whereas Eq.~\eqref{eq:baxter-fusion-pre-singular-block-rev} carries the explicit $\delta$ in the top row.
        Conjugating by $D_\delta=\operatorname{diag}(1,\delta^{-1},\delta^{-1},\delta^{-1})$, before the $O(\delta^2)$ terms are dropped, moves one power of $\delta$ across:
        \begin{equation}
            \label{eq:baxter-fusion-roadmap-rev}
            \widehat L_i^{8V}(s,\delta)
            \coloneqq
            D_\delta^{-1}\widetilde L_i^{8V}(s,\delta)D_\delta
            =
            \w_i(s; \delta)
            +
            \delta
            \begin{pmatrix}
                0 & \va{r}_i(s) \\
                0 & \Delta_i(s)
            \end{pmatrix}
            +
            O(\delta^2)
            \,.
        \end{equation}

        \subsection*{Extraction of the local conserved quantities}

        The transfer matrix of the conjugated fused tensor is
        \begin{equation}
            \label{eq:baxter-fusion-transfer-definition-rev}
            T_{\mathrm{fused}}^{8V}(s,\delta)
            =
            \Tr_a
            \qty(
            \widehat L_1^{8V}(s,\delta)
            \widehat L_2^{8V}(s,\delta)
            \cdots
            \widehat L_L^{8V}(s,\delta)
            )
            \,.
        \end{equation}

        The gauges $G$ and $D_\delta$ are site independent, so cyclicity of the auxiliary trace removes them and the trace may be evaluated on the scalar-normalized product~\eqref{eq:baxter-fusion-unprojected-L-rev}.
        Reordering that product into the two monodromy matrices and tracing over the fused auxiliary space gives
        \begin{equation}
            \label{eq:baxter-fusion-transfer-factorization-rev}
            T_{\mathrm{fused}}^{8V}(s,\delta)
            =
            \Lambda_0(s,\delta)^{-L}\,
            t^{8V}\qty(z(s)-2\eta_{8V}+\delta_0)\,
            t^{8V}\qty(z(s))
            \,,
            \qquad
            \delta_0=s\,\kappa(s)\,\delta
            \,,
        \end{equation}
        where $t^{8V}(z)$ is the eight-vertex transfer matrix~\eqref{eq:baxter-transfer-main}.

        At $\delta=0$, the local tensor~\eqref{eq:baxter-fusion-roadmap-rev} is block upper triangular, so its periodic auxiliary trace is $I+V(s)$.
        Equations~\eqref{eq:baxter-fusion-roadmap-rev} and~\eqref{eq:W-corr-mpo} give
        \begin{equation*}
            \widehat L_i^{8V}(s,\delta)-\wcorr_i(s;\delta)
            =
            \delta
            \begin{pmatrix}
                0 & \va{r}_i(s) \\
                0 & 0
            \end{pmatrix}
            +O(\delta^2)
            \,.
        \end{equation*}

        Equation~\eqref{eq:w-mpo} gives the block form
        \begin{equation*}
            \w_i^{(0)}
            \coloneqq
            \w_i(s;0)
            =
            \begin{pmatrix}
                I & \sbs_i^\infty(s) \\
                0 & \lax_i(s)
            \end{pmatrix}
            \,.
        \end{equation*}
        Note that $\widehat L_i^{8V}(s,0)=\wcorr_i(s;0)=\w_i^{(0)}$.
        Suppressing the common arguments, we write the difference as
        \begin{equation*}
            \begin{aligned}
                T_{\mathrm{fused}}^{8V}(s,\delta)-\Wcorr(s;\delta)
                 & =
                \sum_{j=1}^{L}
                \Tr_a\qty[
                    \widehat L_1^{8V}\cdots\widehat L_{j-1}^{8V}
                    \qty(\widehat L_j^{8V}-\wcorr_j)
                    \wcorr_{j+1}\cdots\wcorr_L
                ]
                \\
                 & =
                \delta\sum_{j=1}^{L}
                \Tr_a\qty[
                    \w_1^{(0)}\cdots\w_{j-1}^{(0)}
                    \left(
                    \begin{smallmatrix}
                        0 & \va{r}_j(s) \\
                        0 & 0
                    \end{smallmatrix}
                    \right)
                    \w_{j+1}^{(0)}\cdots\w_L^{(0)}
                ]
                +O(\delta^2)
                \\
                 & =
                \delta\sum_{j=1}^{L}
                \Tr_a\qty[
                    \begin{pmatrix}
                        0 & \va{r}_j(s)\lax_{j+1}(s)\cdots\lax_L(s) \\
                        0 & 0
                    \end{pmatrix}
                ]
                +O(\delta^2)
                \\
                 & =
                O(\delta^2)
                \,.
            \end{aligned}
        \end{equation*}

        After forming the expansion in $\delta$, set $\delta=\epsilon$ with $\epsilon^2=0$.
        The result above and the same identity applied to $\Wcorr(s; \epsilon)-W(s; \epsilon)$ give
        \begin{equation}
            \label{eq:baxter-fusion-transfer-first-order-rev}
            T_{\mathrm{fused}}^{8V}(s,\epsilon)
            =
            \Wcorr(s;\epsilon)
            =
            W(s; \epsilon)
            +
            \epsilon\,\mathcal D_\Delta^{(L)}(s)
            \,.
        \end{equation}
        The insertion sum is
        \begin{equation}
            \label{eq:baxter-fusion-Ddelta-rev}
            \mathcal D_\Delta^{(L)}(s)
            =
            \sum_{j=1}^{L}
            \Tr_a
            \qty(
            \lax_1(s)\cdots
            \lax_{j-1}(s)
            \Delta_j(s)
            \lax_{j+1}(s)\cdots
            \lax_L(s)
            )
            \,.
        \end{equation}
        Here the auxiliary trace is taken over the three-dimensional triplet subspace.

        Since every entry of $\lax_i(s)$ and $\Delta_i(s)$ is $O(s)$, Eq.~\eqref{eq:baxter-fusion-Ddelta-rev} gives
        \begin{equation}
            \label{eq:baxter-fusion-Cp-definition-rev}
            \mathcal D_\Delta^{(L)}(s)
            =
            \sum_{p=L}^{\infty}
            s^p\mathcal C_{p+1}^{(L)}
            \,.
        \end{equation}

        Equations~\eqref{eq:W-mpo_expand}, \eqref{eq:baxter-fusion-transfer-first-order-rev}, and~\eqref{eq:baxter-fusion-Cp-definition-rev} give
        \begin{equation}
            \label{eq:baxter-fusion-transfer-with-correction-rev}
            T_{\mathrm{fused}}^{8V}(s,\epsilon)
            =
            I+V(s)
            +
            \epsilon
            \sum_{k=2}^{L}s^{k-1}Q_k
            +
            \epsilon
            \sum_{k=L+1}^{\infty}
            s^{k-1}
            \qty(
            \widetilde Q_k^{(L)}
            +
            \mathcal C_k^{(L)}
            )
            \,.
        \end{equation}
        Without $\Delta_i(0)=0$, $\mathcal D_\Delta^{(L)}(s)$ could start at order $s^{L-1}$ and modify $Q_L$.

        The identity~\eqref{eq:baxter-fusion-transfer-factorization-rev} is Eq.~\eqref{eq:fused-expansion-factorization} after the change of variables $z=z(s)$ and the rescaling $\delta_0=s\,\kappa(s)\,\delta$ of Eq.~\eqref{eq:baxter-fusion-chart-collapse-rev}, with the scalar normalization $\Lambda_0(s,\delta)$ of Eq.~\eqref{eq:baxter-fusion-Lambda-rev}.
        After setting $\delta=\epsilon$, Eq.~\eqref{eq:baxter-fusion-transfer-with-correction-rev} agrees with Eq.~\eqref{eq:W-corr-mpo-expand}, which gives the first-order identity~\eqref{eq:fused-expansion-Wcorr}.
        This proves Theorem~\ref{thm:fused-expansion}.

        At order $\delta^0$, Eq.~\eqref{eq:baxter-fusion-transfer-factorization-rev} expresses $I+V(s)$ as a product of two commuting eight-vertex transfer matrices, so $\qty[V(s),t^{8V}(v)]=0$, $\qty[V(s),H]=0$, and $\qty[V(s),V(s')]=0$.
        Equation~\eqref{eq:V-T-relation} then proves Corollaries~\ref{cor:commutativity-Lax-H} and~\ref{cor:commutativity-Lax}; Corollary~\ref{cor:3by3-baxter} follows from the same factorization.

        \section{Solutions of Eq.~\eqref{eq:F-identity-diff}}
        \label{app:F-recurrence-boundary-solution-proof}

        The values $\ffuncone{a}{\vec 0}$ determine every solution of Eq.~\eqref{eq:F-identity-diff}:
        \begin{equation}
            \label{eq:F-recurrence-boundary-solution}
            \ffunc{n}{N_x}{N_y}{N_z}
            =
            \sum_{a=0}^{n}
            \ffuncone{a}{\vec 0}
            \rfunc{n-a}{N_x}{N_y}{N_z}
            \,.
        \end{equation}
        Let $\widetilde F_{n}^{\paren{\vec N}}$ denote the right-hand side of Eq.~\eqref{eq:F-recurrence-boundary-solution}.
        By Eq.~\eqref{eq:R-identity-diff} and its $y$- and $z$-analogues, $R_n$ satisfies Eq.~\eqref{eq:F-identity-diff}, and hence so does $\widetilde F_{n}^{\paren{\vec N}}$ by linearity.
        Moreover, $\widetilde F_{n}^{\paren{\vec 0}} = \ffuncone{n}{\vec 0}$ follows from $\rfunc{r}{0}{0}{0}=\delta_{r,0}$.
        Thus $F$ and $\widetilde F$ are solutions of Eq.~\eqref{eq:F-identity-diff} with the same origin values.

        It remains to show that a solution is determined by its origin values.
        Let $D_n(\vec N)$ be the difference of two solutions with the same origin values; then $D_n(\vec N)=0$ for $n<0$ and $D_n(\vec 0)=0$ for all $n$.
        Suppose $D_{n-1}$ vanishes identically.
        Subtracting the two recurrence relations gives $D_n(\vec N)=D_n(\vec N-\vec e_A)$ for $A\in\{X,Y,Z\}$, so $D_n$ is constant on $\mathbb Z^3$.
        Since $D_n(\vec 0)=0$, $D_n$ vanishes identically.
        By induction on $n$, $D_n\equiv0$ for all $n$; applied to $F$ and $\widetilde F$, this proves Eq.~\eqref{eq:F-recurrence-boundary-solution}.

        The choice $\ffuncone{a}{\vec 0}=\delta_{a,0}$ gives $F=R$.
        Substituting Eq.~\eqref{eq:F-recurrence-boundary-solution} into Eq.~\eqref{eq:Qk-F} gives
        \begin{equation}
            \label{eq:F-Q-triangular}
            \qfamily{k}{F}
            =
            \sum_{a=0}^{\floor{k/2}-1}
            \ffuncone{a}{\vec 0}
            \qfamily{k-2a}{R}
            \,.
        \end{equation}
        Thus, $\qfamily{k}{F}$ is a linear combination of $\qfamily{q}{R}$ with $q\le k$.

    \section{\texorpdfstring{Explicit examples of the local tensors $\Gamma^i_k$ for $5 \le k \le 8$}{Explicit examples of the local tensors for 5 <= k <= 8}}
    \label{app:gamma-examples}

        In this appendix, we give further explicit examples of the local tensors $\Gamma_k^i$ appearing in Eq.~\eqref{eq:mpo_rep_result} for $5\le k\le 8$, complementing the $k=3,4$ cases in Section~\ref{subsec:explicit_mpo}.
    For $k=5$ and $k=6$:
    \begin{align}
        \Gamma^{i}_{5}
        =
        \begin{pmatrix}
            I                    & \sbs_i^0 & O        & \sbs_i^1 & O        \\
            O                    & O        & \Mbs_i^0 & \ebs^0   & \Mbs_i^1 \\
            O                    & O        & O        & \Mbs_i^0 & \ebs^0   \\
            O                    & O        & O        & O        & \Mbs_i^0 \\
            \epsilon \sbs_i^\top & O        & O        & O        & O        \\
        \end{pmatrix}
        ,
        \hspace{2em}
        \Gamma^{i}_{6}
         & =
        \begin{pmatrix}
            I                    & \sbs_i^0 & O        & \sbs_i^1 & O        & \sbs_i^2 \\
            O                    & O        & \Mbs_i^0 & \ebs^0   & \Mbs_i^1 & \ebs^1   \\
            O                    & O        & O        & \Mbs_i^0 & \ebs^0   & \Mbs_i^1 \\
            O                    & O        & O        & O        & \Mbs_i^0 & \ebs^0   \\
            O                    & O        & O        & O        & O        & \Mbs_i^0 \\
            \epsilon \sbs_i^\top & O        & O        & O        & O        & O        \\
        \end{pmatrix}
        \,.
    \end{align}
    For $k=7$ and $k=8$:
    \begin{align}
        \Gamma^{i}_{7}
                       & =
        \begin{pmatrix}
            I                   & \sbs_i^0 & O        & \sbs_i^1 & O        & \sbs_i^2 & O        \\
            O                   & O        & \Mbs_i^0 & \ebs^0   & \Mbs_i^1 & \ebs^1   & \Mbs_i^2 \\
            O                   & O        & O        & \Mbs_i^0 & \ebs^0   & \Mbs_i^1 & \ebs^1   \\
            O                   & O        & O        & O        & \Mbs_i^0 & \ebs^0   & \Mbs_i^1 \\
            O                   & O        & O        & O        & O        & \Mbs_i^0 & \ebs^0   \\
            O                   & O        & O        & O        & O        & O        & \Mbs_i^0 \\
            \epsilon\sbs_i^\top & O        & O        & O        & O        & O        & O        \\
        \end{pmatrix}\,,                                \\
        \Gamma^{i}_{8} & = \begin{pmatrix}
                               I                    & \sbs_i^0 & O        & \sbs_i^1 & O        & \sbs_i^2 & O        & \sbs_i^3 \\
                               O                    & O        & \Mbs_i^0 & \ebs^0   & \Mbs_i^1 & \ebs^1   & \Mbs_i^2 & \ebs^2   \\
                               O                    & O        & O        & \Mbs_i^0 & \ebs^0   & \Mbs_i^1 & \ebs^1   & \Mbs_i^2 \\
                               O                    & O        & O        & O        & \Mbs_i^0 & \ebs^0   & \Mbs_i^1 & \ebs^1   \\
                               O                    & O        & O        & O        & O        & \Mbs_i^0 & \ebs^0   & \Mbs_i^1 \\
                               O                    & O        & O        & O        & O        & O        & \Mbs_i^0 & \ebs^0   \\
                               O                    & O        & O        & O        & O        & O        & O        & \Mbs_i^0 \\
                               \epsilon \sbs_i^\top & O        & O        & O        & O        & O        & O        & O        \\
                           \end{pmatrix}
        \,.
    \end{align}

        For fixed $(n,m)$, there are $3\cdot 2^{t-1}$ letter sequences $A_1\cdots A_t$ satisfying $A_\alpha\ne A_{\alpha+1}$ for $1\le\alpha<t$ and $\binom{m+t-1}{m}$ distributions of the $m$ holes.
        Summing over $p=n+m$, the number of doubling-product summands in Eq.~\eqref{eq:xyz-Qk} is
        \begin{equation*}
            N_k^{\mathrm{direct}}
            =
            3\sum_{p=0}^{\floor{(k-2)/2}}2^{k-2p-2}\binom{k-p-1}{p}
            \sim
            \frac{3(1+\sqrt{2})^k}{4\sqrt{2}}
            \,.
        \end{equation*}
        The MPO therefore compresses the exponentially many summands of the direct formula into a tensor with bond dimension $3k-2$ and only $2+\floor{k/2}+(k-1)(k-2)/2=O(k^2)$ nonzero blocks.

    \section{Derivation of the symmetric Lax operator}\label{app:simple-lax-derivation}

    Here we derive the symmetric form of the Lax operator~\eqref{eq:laxcal} from $\lax_i(s)$ by applying a gauge transformation and a change of spectral parameter.

    We introduce the diagonal gauge matrix
    \begin{align}
        G(s)
        \coloneqq
        \begin{pmatrix}
            \sqrt{s J_X f_X(s)} & 0                   & 0                   \\
            0                   & \sqrt{s J_Y f_Y(s)} & 0                   \\
            0                   & 0                   & \sqrt{s J_Z f_Z(s)}
        \end{pmatrix}
        \,.
    \end{align}
    Applying the similarity transformation $G(s) \lax_i(s) G(s)^{-1}$, we obtain
    \begin{align}
        G(s) \lax_i(s) G(s)^{-1}
        =
        \begin{pmatrix}
            \omega J_Y J_Z I                & \sqrt{s^2 J_X J_Y f_X f_Y} Z_i  & -\sqrt{s^2 J_X J_Z f_X f_Z} Y_i \\
            -\sqrt{s^2 J_X J_Y f_X f_Y} Z_i & \omega J_X J_Z I                & \sqrt{s^2 J_Y J_Z f_Y f_Z} X_i  \\
            \sqrt{s^2 J_X J_Z f_X f_Z} Y_i  & -\sqrt{s^2 J_Y J_Z f_Y f_Z} X_i & \omega J_X J_Y I
        \end{pmatrix}
        \,,
    \end{align}
    where we suppress the $s$-dependence of $f_\alpha(s)$ and $\omega(s)$.

    Equations~\eqref{eq:fx-ito-fe} and~\eqref{eq:omega-fxyz} simplify the off-diagonal entries.
    For example, the $(1,2)$ entry becomes
    \begin{align}
        s^2 J_X J_Y f_X f_Y
        = s^2 f_X f_Y f_Z \cdot \frac{J_X J_Y}{f_Z}
        = \omega J_X J_Y (1 - \omega J_Z^2)
        \,,
    \end{align}
    where the two steps use Eqs.~\eqref{eq:omega-fxyz} and~\eqref{eq:fx-ito-fe}, respectively.
    Similar calculations for the other off-diagonal entries yield
    \begingroup
    \setlength{\arraycolsep}{2pt}
    \begin{align}\label{eq:gauge-transformed-lax}
        G(s) \lax_i(s) G(s)^{-1}
        =
        \begin{pmatrix}
            \omega J_Y J_Z I                              & \sqrt{\omega J_X J_Y (1 - \omega J_Z^2)} Z_i  & -\sqrt{\omega J_X J_Z (1 - \omega J_Y^2)} Y_i \\
            -\sqrt{\omega J_X J_Y (1 - \omega J_Z^2)} Z_i & \omega J_X J_Z I                              & \sqrt{\omega J_Y J_Z (1 - \omega J_X^2)} X_i  \\
            \sqrt{\omega J_X J_Z (1 - \omega J_Y^2)} Y_i  & -\sqrt{\omega J_Y J_Z (1 - \omega J_X^2)} X_i & \omega J_X J_Y I
        \end{pmatrix}
        \,.
    \end{align}
    \endgroup

    We introduce the new spectral parameter $u \coloneqq \sqrt{\omega}$, so that $\omega = u^2$.
    We substitute $\omega = u^2$ into Eq.~\eqref{eq:gauge-transformed-lax} and factor out $\sqrt{u^3 J_X J_Y J_Z}$ from every entry.
    The resulting operator
    \begin{equation*}
        \laxcal_i(u)
        \coloneqq
        \frac{1}{\sqrt{u^3 J_X J_Y J_Z}}\,
        G(s) \lax_i(s) G(s)^{-1}
    \end{equation*}
    reproduces exactly the symmetric Lax operator of Eq.~\eqref{eq:laxcal}.
        Since the right-hand side depends only on $u$, we may take $u$ as the independent spectral parameter, thereby eliminating $s$.
    \section{Alternative proof of Corollary~\ref{cor:commutativity-Lax-H}}\label{app:commutativity-Lax-H}

        We use the telescoping method of~\cite{Fendley-xyz-2025}.
        Unlike the derivation from Theorem~\ref{thm:fused-expansion}, this proof does not use Baxter's eight-vertex transfer matrix.

    Here the Hamiltonian density is defined by $h_{i, i+1}^{\mathrm{XYZ}} \coloneqq J_XX_{i}X_{i+1}+J_YY_{i}Y_{i+1}+J_ZZ_{i}Z_{i+1}$.
    Since $H = \sum_{j=1}^{L} h_{j, j+1}^{\mathrm{XYZ}}$ is a sum of local terms, the commutativity follows if there exists an impurity tensor $E_{k,l}^{a}$, with the same index ranges as $\mathcal{L}_{k,l}^{a}$, namely $k, l \in \{X, Y, Z\}$ and $a \in \{0, X, Y, Z\}$, satisfying
    \begin{equation}\label{eq:telescope-condition}
        \qty[h_{j, j+1}^{\mathrm{XYZ}}, \mathcal{L}_{k,m}^{a} \mathcal{L}_{m,l}^{b} \sigma_j^a \sigma_{j+1}^b]
        = \qty(E_{k,m}^{a} \mathcal{L}_{m,l}^{b} - \mathcal{L}_{k,m}^{a} E_{m,l}^{b}) \sigma_j^a \sigma_{j+1}^b
        \,,
    \end{equation}
    where summation over $a, b \in \{0, X, Y, Z\}$ and $m \in \{X, Y, Z\}$ is implied, while $k$ and $l$ are free.
    For each site $j$, define the operator
    \begin{equation*}
        M_j
        \coloneqq
        \sum_{\{a_i\},\{A_i\}}
        \mathcal{L}_{A_L,A_1}^{a_1}\cdots
        \mathcal{L}_{A_{j-2},A_{j-1}}^{a_{j-1}}
        E_{A_{j-1},A_j}^{a_j}
        \mathcal{L}_{A_j,A_{j+1}}^{a_{j+1}}\cdots
        \mathcal{L}_{A_{L-1},A_L}^{a_L}
        \sigma_1^{a_1}\cdots\sigma_L^{a_L}
        \,.
    \end{equation*}
    Here the sum runs over $a_i\in\{0,X,Y,Z\}$ and $A_i\in\{X,Y,Z\}$; the factors before or after $E$ are absent for $j=1$ or $j=L$, respectively.
    Equation~\eqref{eq:telescope-condition} then gives
    \begin{equation*}
        \bck{H,T(u)}
        =
        \sum_{j=1}^{L}(M_j-M_{j+1})
        =
        0
        \,.
    \end{equation*}
    Here $M_{L+1}\equiv M_1$.

    The impurity tensor defined by
    \begin{equation}\label{eq:E-definition}
        E_{k,l}^{m} = \frac{2\imi}{u} \, \varepsilon_{klm} \, g(u J_k) \, g(u J_l) \sqrt{\frac{u J_k J_l}{J_m}}
        \,, \qquad
        E_{k,l}^{0} = 0
    \end{equation}
    satisfies Eq.~\eqref{eq:telescope-condition}.
        This identity can be verified by a direct computation.

    \section{\texorpdfstring{Expansion of $T(u)$ around $u=0$}{Expansion of T(u) around u=0}}
    \label{app:T-u-expansion}

        In this appendix, we expand the transfer matrix $T(u)$ defined in Eq.~\eqref{eq:Laxcal-mpo} about $u=0$ and derive explicit expressions for the conserved quantities appearing in Eq.~\eqref{eq:T-u-cyclic-expansion}.
        Setting $J_{XYZ}\coloneqq J_XJ_YJ_Z$, we expand $T(u)$ as
        \begin{equation}
            \label{eq:T-u-Laurent-expansion}
            T(u)
            =
            (uJ_{XYZ})^{-L/2}
            \sum_{p=0}^{\infty}
            u^p\mathcal{I}_p^{(L)}
            \,.
        \end{equation}
        By Corollaries~\ref{cor:commutativity-Lax-H},~\ref{cor:commutativity-Lax}, and~\ref{cor:Lax-Qk-commute}, the coefficients $\mathcal{I}_p^{(L)}$ commute with $H$, with one another, and with $Q_k$ for $2\le k\le L$.

        To write $\mathcal{I}_p^{(L)}$ explicitly, we introduce a periodic version of the doubling product.
        For a word $\boldsymbol{A}=(A_1,\ldots,A_L)\in\{X,Y,Z\}^L$, with $A_0\equiv A_L$, let $m(\boldsymbol{A})\coloneqq\#\{1\le j\le L\mid A_{j-1}=A_j\}$ and $t(\boldsymbol{A})\coloneqq L-m(\boldsymbol{A})$.
        Below, we write $m=m(\boldsymbol{A})$ and $t=t(\boldsymbol{A})$.
        For each such word $\boldsymbol{A}$, we define
        \begin{equation}
            \label{eq:T-u-periodic-doubling-def}
            \overline{\boldsymbol{A}}^{\mathrm{c}}
            =
            \overline{A_1 A_2 \cdots A_L}^{\mathrm{c}}
            \coloneqq
            (-\imi)^{t(\boldsymbol{A})}
            (A_LA_1)_1
            (A_1A_2)_2
            \cdots
            (A_{L-1}A_L)_L
            \,.
        \end{equation}
        The exponent is $t$ rather than the $t-1$ of Eq.~\eqref{doubling-abbreviated} because the closure wraps $A_L$ onto $A_1$.

        The words with $m$ holes form the set
        \begin{equation}
            \label{eq:T-u-doubling-set}
            \mathcal{S}_{\mathrm{c}}^{m}
            \coloneqq
            \{\boldsymbol{A}\in\{X,Y,Z\}^{L}\mid m(\boldsymbol{A})=m\}
            \,.
        \end{equation}
        Doubling products associated with different letters need not be linearly independent.
        For example, for $L=3$, $\overline{XXX}^{\mathrm{c}}=\overline{YYY}^{\mathrm{c}}=\I$.
        For $\rho\in\{X,Y,Z\}$, let
        \begin{equation}
            \label{eq:T-u-word-statistics}
            m_\rho^{\boldsymbol{A}}
            \coloneqq
            \#\{1\le j\le L\mid A_{j-1}=A_j=\rho\}
            \,,
            \qquad
            N_\rho^{\boldsymbol{A}}
            \coloneqq
            \#\{1\le j\le L\mid A_{j-1}\neq A_j=\rho\}
            \,,
        \end{equation}
        so $m=\sum_{\rho\in\{X,Y,Z\}}m_\rho^{\boldsymbol{A}}$ and $t=\sum_{\rho\in\{X,Y,Z\}}N_\rho^{\boldsymbol{A}}=L-m$.
        For a nonconstant word, $N_\rho^{\boldsymbol{A}}$ is the number of its cyclic runs labelled by $\rho$.
        The coupling factor is
        \begin{equation}
            \label{eq:T-u-closed-coupling}
            J_{\boldsymbol{A}}
            \coloneqq
            J_{XYZ}^{m}
            \prod_{\rho\in\{X,Y,Z\}}
            J_\rho^{N_\rho^{\boldsymbol{A}}-m_\rho^{\boldsymbol{A}}}
            \,.
        \end{equation}

        Each term of Eq.~\eqref{eq:T-u-pauli-string} is labelled by the word $\boldsymbol{A}=(A_1,\ldots,A_L)$ formed by its lower indices, and the entry at site $j$ is $\mathcal{L}^{a_j}_{A_{j-1},A_j}$.
        By Eq.~\eqref{eq:T-u-Lax-entries}, an entry with $A_{j-1}=A_j$ contributes the identity with scalar factor $\sqrt{uJ_{XYZ}}/J_{A_j}$, whereas an entry with $A_{j-1}\neq A_j$ contributes $\varepsilon_{A_{j-1},A_j,a_j}\sigma^{a_j}$ with scalar factor $g(uJ_{a_j})$, where $a_j$ is the letter distinct from $A_{j-1}$ and $A_j$.
        Consequently, $(\sigma^{k}\sigma^{l})_j=\imi\,\varepsilon_{k,l,a}\sigma^a_j$ gives the operator part $(-\imi)^t\prod_{j=1}^{L}(A_{j-1}A_j)_j=\overline{\boldsymbol{A}}^{\mathrm{c}}$ of Eq.~\eqref{eq:T-u-periodic-doubling-def}.
        Equation~\eqref{eq:T-u-pauli-string} therefore reads
        \begin{equation}
            \label{eq:T-u-word-sum}
            T(u)
            =
            \sum_{\boldsymbol{A}\in\{X,Y,Z\}^{L}}
            c_{\boldsymbol{A}}(u)\,
            \overline{\boldsymbol{A}}^{\mathrm{c}}
            \,,
        \end{equation}
        where the coefficient $c_{\boldsymbol{A}}(u)$ is the product of the scalar factors of the same entries,
        \begin{equation}
            \label{eq:T-u-word-coefficient}
            c_{\boldsymbol{A}}(u)
            =
            \prod_{\substack{1\le j\le L\\A_{j-1}=A_j}}
            \frac{\sqrt{uJ_{XYZ}}}{J_{A_j}}
            \prod_{\substack{1\le j\le L\\A_{j-1}\neq A_j}}
            g(uJ_{a_j})
            =
            (uJ_{XYZ})^{-\frac{L}{2}}\,
            u^{m}\,
            J_{\boldsymbol{A}}
            \prod_{\rho\in\{X,Y,Z\}}
            \qty(1-u^2J_\rho^2)^{\frac{t}{2}-N_\rho^{\boldsymbol{A}}}
            \,,
        \end{equation}
        with $J_{\boldsymbol{A}}$ given by Eq.~\eqref{eq:T-u-closed-coupling}.

        We extend $\sfunc{n}{M_X}{M_Y}{M_Z}$ to arbitrary real arguments by interpreting each binomial coefficient in Eq.~\eqref{eq:sfunc} as $\binom{n_\rho+M_\rho-1}{n_\rho}$.
        Expanding $c_{\boldsymbol{A}}(u)$ in $u$ and comparing Eq.~\eqref{eq:T-u-word-sum} with Eq.~\eqref{eq:T-u-Laurent-expansion} then gives
        \begin{equation}
            \label{eq:T-u-cyclic-expansion}
            \mathcal{I}_p^{(L)}
            =
            \sum_{n=0}^{\floor{p/2}}
            \;
            \sum_{\boldsymbol{A}\in\mathcal{S}_{\mathrm{c}}^{m}}
            \sfunc{n}{N_X^{\boldsymbol{A}}-t/2}{N_Y^{\boldsymbol{A}}-t/2}{N_Z^{\boldsymbol{A}}-t/2}\,
            J_{\boldsymbol{A}}\,
            \overline{\boldsymbol{A}}^{\mathrm{c}}
            \,.
        \end{equation}
        Here $m \equiv p-2n$ is the number of holes and $t \equiv L-m=L-p+2n$.

        As a concrete example, we give the first three conserved quantities for $L=4$.
        The first conserved quantity is
        \begin{align}
            \mathcal I_0^{(4)}
             & =
            J_Y^2J_Z^2\qty(\overline{YZYZ}^{\mathrm{c}}+\overline{ZYZY}^{\mathrm{c}})
            +
            J_Z^2J_X^2\qty(\overline{ZXZX}^{\mathrm{c}}+\overline{XZXZ}^{\mathrm{c}})
            +
            J_X^2J_Y^2\qty(\overline{XYXY}^{\mathrm{c}}+\overline{YXYX}^{\mathrm{c}})
            \nonumber \\
             & \quad+
            J_X^2J_YJ_Z\qty(\overline{XYXZ}^{\mathrm{c}}+\overline{YXZX}^{\mathrm{c}}+\overline{XZXY}^{\mathrm{c}}+\overline{ZXYX}^{\mathrm{c}})
            \nonumber \\
             & \quad+
            J_XJ_Y^2J_Z\qty(\overline{YZYX}^{\mathrm{c}}+\cdots)
            +
            J_XJ_YJ_Z^2\qty(\overline{ZXZY}^{\mathrm{c}}+\cdots)
            \,.
        \end{align}
        Here and below, each $\cdots$ denotes the remaining cyclic rotations of the words in the same parentheses.
        The next conserved quantity is
        \begin{align}
            \mathcal I_1^{(4)}
             & =
            J_XJ_Y^2J_Z^2\qty(\overline{XXYZ}^{\mathrm{c}}+\overline{XXZY}^{\mathrm{c}}+\cdots)
            +
            J_X^2J_YJ_Z^2\qty(\overline{YYZX}^{\mathrm{c}}+\overline{YYXZ}^{\mathrm{c}}+\cdots)
            \nonumber \\
             & \quad+
            J_X^2J_Y^2J_Z\qty(\overline{ZZXY}^{\mathrm{c}}+\overline{ZZYX}^{\mathrm{c}}+\cdots)
            \,.
        \end{align}
        The third one splits into its $m=2$ and $m=0$ parts, $\mathcal I_2^{(4)}=\mathcal I_{2,2}^{(4)}+\mathcal I_{2,0}^{(4)}$, where
        \begin{align}
            \mathcal I_{2,2}^{(4)}
             & =
            J_X^2J_YJ_Z^3\qty(\overline{YYYZ}^{\mathrm{c}}+\cdots)
            +
            J_X^2J_Y^3J_Z\qty(\overline{YZZZ}^{\mathrm{c}}+\cdots)
            +
            J_X^2J_Y^2J_Z^2\qty(\overline{YYZZ}^{\mathrm{c}}+\cdots)
            \nonumber \\
             & \quad+
            J_X^3J_Y^2J_Z\qty(\overline{ZZZX}^{\mathrm{c}}+\cdots)
            +
            J_XJ_Y^2J_Z^3\qty(\overline{ZXXX}^{\mathrm{c}}+\cdots)
            +
            J_X^2J_Y^2J_Z^2\qty(\overline{ZZXX}^{\mathrm{c}}+\cdots)
            \nonumber \\
             & \quad+
            J_XJ_Y^3J_Z^2\qty(\overline{XXXY}^{\mathrm{c}}+\cdots)
            +
            J_X^3J_YJ_Z^2\qty(\overline{XYYY}^{\mathrm{c}}+\cdots)
            +
            J_X^2J_Y^2J_Z^2\qty(\overline{XXYY}^{\mathrm{c}}+\cdots)
            \,,
            \\
            \mathcal I_{2,0}^{(4)}
             & =
            -2J_X^2J_Y^2J_Z^2
            \qty(
            \overline{YZYZ}^{\mathrm{c}}
            +\overline{ZYZY}^{\mathrm{c}}
            +\overline{ZXZX}^{\mathrm{c}}
            +\overline{XZXZ}^{\mathrm{c}}
            +\overline{XYXY}^{\mathrm{c}}
            +\overline{YXYX}^{\mathrm{c}}
            )
            \nonumber \\
             & \quad-
            J_X^2J_YJ_Z\qty(J_Y^2+J_Z^2)
            \qty(\overline{XYXZ}^{\mathrm{c}}+\cdots)
            -
            J_XJ_Y^2J_Z\qty(J_Z^2+J_X^2)
            \qty(\overline{YZYX}^{\mathrm{c}}+\cdots)
            \nonumber \\
             & \quad-
            J_XJ_YJ_Z^2\qty(J_X^2+J_Y^2)
            \qty(\overline{ZXZY}^{\mathrm{c}}+\cdots)
            \,.
        \end{align}

        The conserved quantities obtained from the $u$-expansion of $T(u)$ are nonlocal: at fixed $p$, their Pauli strings generally extend around the whole chain, in contrast to the local conserved quantities $Q_k$ of Eq.~\eqref{eq:MPOall}.

        \section{Conservation and mutual commutativity in the XXX limit}
        \label{app:W-XXX-commutativity}
        In the XXX case $J_X=J_Y=J_Z=1$, the operators $W(s; \epsilon)$ at different spectral parameters mutually commute.
        The same argument gives $\qty[W(s; \epsilon),H_{\mathrm{XXX}}]=0$, so the wrap-around contributions $\widetilde{Q}_k^{(L)}$ with $k>L$ appearing in Eq.~\eqref{eq:W-mpo_expand} are also conserved on a finite chain.

        \subsection*{Isotropic tensor and Catalan reparametrization}

        Writing the isotropic limit with the superscript $\mathrm{XXX}$, the expansion~\eqref{eq:W-mpo_expand} reads
        \begin{equation}
            \label{eq:W-XXX-generating-MPO-expansion}
            W^{\mathrm{XXX}}(s; \epsilon)
            =
            I+V^{\mathrm{XXX}}(s)
            +
            \epsilon
            \sum_{k=2}^{L}
            s^{k-1} Q_k^{\mathrm{XXX}}
            +
            \epsilon
            \sum_{k=L+1}^{\infty}
            s^{k-1} \widetilde{Q}_k^{\mathrm{XXX},(L)}
            \,,
        \end{equation}
        Here $V^{\mathrm{XXX}}(s)$ is the transfer matrix formed from the lower-right $3\times3$ block of the isotropic local tensor.
        For $2\le k\le L$, the operators $Q_k^{\mathrm{XXX}}$ are the local conserved quantities of the XXX chain~\cite{grabowski-xxx-1994,GRABOWSKI1995299,yamada2023matrix}.

        The three functions $f_X(s)$, $f_Y(s)$, and $f_Z(s)$ reduce to a common function $f(s)$.
        With $t\coloneqq sf(s)$, used in place of $s$ as the argument of $W^{\mathrm{XXX}}$ and $V^{\mathrm{XXX}}$ below, Eqs.~\eqref{eq:fx-ito-fe} and~\eqref{eq:eq-for-omega} give
        \begin{equation}
            \begin{aligned}
                f(s)
                 & =
                \frac{1}{1-\omega(s)}
                \,,
                 &
                \omega(s)\qty(1-\omega(s))^3
                 & =
                s^2
                \,,
                \\
                t
                 & =
                \frac{s}{1-\omega(s)}
                =
                s+O(s^3)
                \,,
                 &
                t^2
                 & =
                \omega(s)\qty(1-\omega(s))
                \,.
            \end{aligned}
        \end{equation}
        The branch $\omega=t^2+O(t^4)$ is
        \begin{equation}
            \omega(s)=C(t^2)
            \,,
            \qquad
            C(x)\coloneqq\frac{1-\sqrt{1-4x}}{2}
            =
            \sum_{n=0}^{\infty}C_nx^{n+1}
            \,,
        \end{equation}
        where $C_n=\frac{1}{n+1}\binom{2n}{n}$ is the $n$th Catalan number.

        In this limit the first row of Eq.~\eqref{eq:w-mpo} is $t\,\sbs_i$ and its lower-right block is $C(t^2)\ebs+t\Mbs_i$.
        Conjugating the auxiliary space by $\operatorname{diag}\qty(1/t,1,1,1)$, which leaves the transfer matrix unchanged, moves the factor $t$ from the first row to the first column:
        \begin{equation}
            \label{eq:W-XXX-local-tensor}
            \w_i^{\mathrm{XXX}}(t; \epsilon)
            =
            \begin{pmatrix}
                I              & X_i      & Y_i      & Z_i      \\
                \epsilon t X_i & C(t^2) I & t Z_i    & -t Y_i   \\
                \epsilon t Y_i & -t Z_i   & C(t^2) I & t X_i    \\
                \epsilon t Z_i & t Y_i    & -t X_i   & C(t^2) I
            \end{pmatrix}
            \,.
        \end{equation}

        \subsection*{Commuting XXX transfer-matrix families}

        Promoting $p=\epsilon t$ and $c=C(t^2)$ in Eq.~\eqref{eq:W-XXX-local-tensor} to parameters independent of $t$ gives
        \begin{equation}
            \label{eq:XXX-4d-general-L}
            \mathcal{L}^{(4)}_i(p,c,t)
            =
            \begin{pmatrix}
                I     & X_i    & Y_i    & Z_i    \\
                p X_i & c I    & t Z_i  & -t Y_i \\
                p Y_i & -t Z_i & c I    & t X_i  \\
                p Z_i & t Y_i  & -t X_i & c I
            \end{pmatrix}
            \,.
        \end{equation}
        Its transfer matrix is
        \begin{equation}
            \label{eq:XXX-4d-transfer-family}
            W(p,c,t)
            \coloneqq
            \Tr_a
            \bck{
                \mathcal{L}^{(4)}_1(p,c,t)
                \cdots
                \mathcal{L}^{(4)}_L(p,c,t)
            }
            \,,
        \end{equation}
        so that
        \begin{equation}
            \label{eq:W-XXX-tensor-as-L4}
            W^{\mathrm{XXX}}(t; \epsilon)
            =
            W\qty(\epsilon t, C(t^2), t)
            \,.
        \end{equation}

        The lower-right $3\times3$ block of Eq.~\eqref{eq:XXX-4d-general-L} is the local tensor
        \begin{equation}
            \label{eq:XXX-3d-local-tensor}
            \mathcal{L}_i^{(3)}(c,t)
            \coloneqq
            c\,\ebs
            +
            t\,\Mbs_i
            \,,
        \end{equation}
        with $\ebs$ and $\Mbs_i$ from Eqs.~\eqref{eq:building_blocl_e} and~\eqref{eq:building_blocl_M}.
        Its transfer matrix is
        \begin{equation}
            \label{eq:XXX-3d-transfer-family}
            V(c,t)
            \coloneqq
            \Tr_a
            \bck{
                \mathcal{L}_1^{(3)}(c,t)
                \cdots
                \mathcal{L}_L^{(3)}(c,t)
            }
            \,.
        \end{equation}
        At $p=0$ the tensor~\eqref{eq:XXX-4d-general-L} is block triangular, so
        \begin{equation}
            \label{eq:XXX-p-zero-splitting}
            W(0,c,t)
            =
            I+V(c,t)
            \,.
        \end{equation}

        Let
        \begin{equation}
            \label{eq:XXX-fundamental-lax}
            L_{a,i}(x)
            \coloneqq
            x I
            +
            \sum_{\alpha\in\{X,Y,Z\}}
            \sigma^{\alpha}_{a}\sigma^{\alpha}_{i}
        \end{equation}
        denote the fundamental XXX Lax operator on a two-dimensional auxiliary space $a$.
        Its transfer matrix is $\tau(x)\coloneqq\Tr_a\qty[L_{a,1}(x)\cdots L_{a,L}(x)]$.

        \begin{lem}
            \label{lem:XXX-4d-family}
            Let the ordinary parameters $(p,c,t)$ satisfy
            \begin{equation}
                \label{eq:XXX-4d-constraint}
                p=c-c^2-t^2
            \end{equation}
            with $c\ne1$, and let $(x,y)$ be the pair determined by $(c,t)$, up to exchange, by
            \begin{equation}
                \label{eq:XXX-4d-cd-inversion}
                xy=\frac{3c+1}{c-1}
                \,,
                \qquad
                x+y=\frac{4\imi t}{c-1}
                \,.
            \end{equation}
            Then
            \begin{equation}
                \label{eq:XXX-4d-transfer-factorization}
                W(p,c,t)
                =
                (xy-3)^{-L}\,
                \tau(x)\,\tau(y)
                \,,
            \end{equation}
            so $W(p,c,t)$ commutes with $H_{\mathrm{XXX}}$, and any two such transfer matrices commute.
        \end{lem}

        \begin{proof}
            Multiplying two copies of Eq.~\eqref{eq:XXX-fundamental-lax} on auxiliary spaces $1$ and $2$ and using the Pauli algebra gives
            \begin{equation}
                \label{eq:XXX-fusion-product}
                L_{1,i}(x)L_{2,i}(y)
                =
                \qty(xy+K_{12})I
                +
                \sum_{\gamma}
                \qty[
                    x\sigma^{\gamma}_{2}
                    +
                    y\sigma^{\gamma}_{1}
                    +
                    \imi\qty(\boldsymbol{\sigma}_1\times\boldsymbol{\sigma}_2)^{\gamma}
                ]
                \sigma^{\gamma}_{i}
                \,,
                \qquad
                K_{12}
                \coloneqq
                \sum_{\alpha}
                \sigma^{\alpha}_{1}\sigma^{\alpha}_{2}
                \,.
            \end{equation}

            In the singlet--triplet basis~\eqref{eq:baxter-fusion-singlet-triplet-basis-rev}, each term of Eq.~\eqref{eq:XXX-fusion-product} occupies the block fixed by its parity under exchanging the two auxiliary spaces.
            With $P^{(0)}=\ket{e_0}\bra{e_0}$, the scalar term is $xy+K_{12}=\qty(xy-3)P^{(0)}+\qty(xy+1)\qty(I-P^{(0)})$, since $K_{12}=I-4P^{(0)}$.
            The even combination of Pauli matrices is block diagonal and the two odd ones are purely singlet--triplet:
            \begin{equation}
                \label{eq:XXX-fusion-blocks}
                \begin{aligned}
                    \sigma_1^\gamma+\sigma_2^\gamma
                     & =
                    -2\imi
                    \sum_{\alpha,\beta\in\{X,Y,Z\}}
                    \qty(E_\gamma)_{\alpha\beta}
                    \ket{e_\alpha}\bra{e_\beta}
                    \,,
                    \qquad
                    \qty(E_\gamma)_{\alpha\beta}
                    \coloneqq
                    \varepsilon_{\gamma\alpha\beta}
                    \,,
                    \\
                    \sigma_1^\gamma-\sigma_2^\gamma
                     & =
                    2\qty(\ket{e_0}\bra{e_\gamma}+\ket{e_\gamma}\bra{e_0})
                    \,,
                    \\
                    \imi\qty(\boldsymbol{\sigma}_1\times\boldsymbol{\sigma}_2)^\gamma
                     & =
                    \qty[\sigma_1^\gamma-\sigma_2^\gamma,P^{(0)}]
                    =
                    2\qty(\ket{e_\gamma}\bra{e_0}-\ket{e_0}\bra{e_\gamma})
                    \,,
                \end{aligned}
            \end{equation}
            where the cross product is rewritten using $\qty[\sigma_1^\gamma,K_{12}]=-2\imi\qty(\boldsymbol{\sigma}_1\times\boldsymbol{\sigma}_2)^\gamma$.

            Substituting Eq.~\eqref{eq:XXX-fusion-blocks} into Eq.~\eqref{eq:XXX-fusion-product} and contracting with $\sigma_i^\gamma$ yields the triplet block $-\imi(x+y)\Mbs_i$, using $\Mbs_i=\sum_\gamma E_\gamma\sigma_i^\gamma$.
            Dividing by the singlet entry $xy-3$ leaves a tensor of the shape~\eqref{eq:XXX-4d-general-L} except that its first row carries the coefficient $(y-x-2)/(xy-3)$ and its first column the coefficient $(y-x+2)/(xy-3)$.
            Label $(x,y)$ so that $y-x\ne2$, which is always possible because exchanging $x$ and $y$ reverses the sign of $y-x$.
            Conjugating by $D\coloneqq\operatorname{diag}\qty(\frac{xy-3}{y-x-2},1,1,1)$ then normalizes the first row:
            \begin{equation}
                \label{eq:XXX-fused-normalized}
                D\,
                \frac{L_{1,i}(x)L_{2,i}(y)}{xy-3}
                \,D^{-1}
                =
                \mathcal{L}^{(4)}_i
                \qty(
                \frac{\qty(y-x)^2-4}{\qty(xy-3)^2},
                \frac{xy+1}{xy-3},
                \frac{-\imi\qty(x+y)}{xy-3}
                )
                \,.
            \end{equation}
            These three arguments satisfy the constraint~\eqref{eq:XXX-4d-constraint}, and inverting the last two gives Eq.~\eqref{eq:XXX-4d-cd-inversion} with $xy-3=4/(c-1)$.
            The two Lax factors act on distinct auxiliary spaces, so that the trace is $(xy-3)^{-L}\tau(x)\tau(y)$, which gives the factorization~\eqref{eq:XXX-4d-transfer-factorization}.

            The fundamental XXX transfer matrices mutually commute and commute with $H_{\mathrm{XXX}}$~\cite{grabowski-xxx-1994,GRABOWSKI1995299}.
            The factorization~\eqref{eq:XXX-4d-transfer-factorization} therefore gives both commutativity statements.
        \end{proof}

        Writing $p(c,t)\coloneqq c-c^2-t^2$ for the constrained value of the first argument, Lemma~\ref{lem:XXX-4d-family} states that
        \begin{equation}
            \label{eq:XXX-family-commutativity}
            \qty[W\qty(p(c,t),c,t),H_{\mathrm{XXX}}]
            =
            0
            \,,
            \qquad
            \qty[W\qty(p(c,t),c,t),W\qty(p(c^\prime,t^\prime),c^\prime,t^\prime)]
            =
            0
            \,.
        \end{equation}
        Both left-hand sides are polynomial in $(c,t)$ and $(c^\prime,t^\prime)$ and vanish for $c,c^\prime\ne1$, hence identically, so these identities hold coefficient by coefficient and remain valid for dual-number arguments.

        Since $C(t^2)-C(t^2)^2=t^2$, the curve $c=C(t^2)$ lies in the family with $p=0$.
        Equation~\eqref{eq:XXX-p-zero-splitting} therefore gives
        \begin{equation}
            \label{eq:XXX-curve-zero-order}
            W\qty(0,C(t^2),t)
            =
            I+V^{\mathrm{XXX}}(t)
            \,,
            \qquad
            V^{\mathrm{XXX}}(t)\coloneqq V\qty(C(t^2),t)
            \,,
        \end{equation}
        so the relations~\eqref{eq:XXX-family-commutativity} on this curve are those of $V^{\mathrm{XXX}}(t)$.

        \subsection*{\texorpdfstring{Conservation of $W^{\mathrm{XXX}}(t; \epsilon)$}{Conservation of WXXX(t; epsilon)}}

        Write $C=C(t^2)$ and consider the transfer matrix
        \begin{equation}
            \label{eq:XXX-dual-family-member}
            \widehat W(t;\epsilon)
            \coloneqq
            W\qty(
                \epsilon t,
                C-\epsilon t,
                t-\epsilon\qty(1-C)
            )
            \,.
        \end{equation}
        Its arguments satisfy the constraint~\eqref{eq:XXX-4d-constraint}:
        \begin{equation*}
            \begin{aligned}
                (C-\epsilon t)-(C-\epsilon t)^2-\qty[t-\epsilon\qty(1-C)]^2
                 & =
                C-C^2-t^2-\epsilon t+2\epsilon tC+2\epsilon t\qty(1-C)
                \\
                 & =
                \epsilon t
                \,.
            \end{aligned}
        \end{equation*}
        Hence $\widehat W(t;\epsilon)$ belongs to the mutually commuting family~\eqref{eq:XXX-family-commutativity}.

        Using the local tensor $\w_i^{\mathrm{XXX}}(t;\epsilon)$ from Eq.~\eqref{eq:W-XXX-local-tensor}, the two transfer matrices have the block forms
        \begin{equation*}
            \begin{aligned}
                W^{\mathrm{XXX}}(t;\epsilon)
                 & =
                \Tr_a\qty[
                    \w_1^{\mathrm{XXX}}(t;\epsilon)
                    \cdots
                    \w_L^{\mathrm{XXX}}(t;\epsilon)
                ]
                \,,
                &
                \w_i^{\mathrm{XXX}}(t;\epsilon)
                 & =
                \begin{pmatrix}
                    I                      & \sbs_i              \\
                    \epsilon t\sbs_i^\top & \mathcal A_i
                \end{pmatrix}
                \,,
                \\
                \widehat W(t;\epsilon)
                 & =
                \Tr_a\qty[
                    \widehat{\w}_1(t;\epsilon)
                    \cdots
                    \widehat{\w}_L(t;\epsilon)
                ]
                \,,
                &
                \widehat{\w}_i(t;\epsilon)
                 & \coloneqq
                \begin{pmatrix}
                    I                      & \sbs_i                    \\
                    \epsilon t\sbs_i^\top & \widehat{\mathcal A}_i
                \end{pmatrix}
                \,,
            \end{aligned}
        \end{equation*}
        where
        \begin{equation*}
            \mathcal A_i
            \coloneqq
            C\ebs+t\Mbs_i
            \,,
            \qquad
            \widehat{\mathcal A}_i
            \coloneqq
            \qty(C-\epsilon t)\ebs
            +
            \qty[t-\epsilon\qty(1-C)]\Mbs_i
            \,.
        \end{equation*}

        The Catalan relation $C(1-C)=t^2$ gives
        \begin{equation*}
            \begin{aligned}
                C\qty(\widehat{\mathcal A}_i-\mathcal A_i)
                 & =
                -\epsilon tC\ebs-\epsilon C\qty(1-C)\Mbs_i
                \\
                 & =
                -\epsilon t\qty(C\ebs+t\Mbs_i)
                =
                -\epsilon t\mathcal A_i
                \,.
            \end{aligned}
        \end{equation*}

        Set $\Delta_i=\widehat{\w}_i(t;\epsilon)-\w_i^{\mathrm{XXX}}(t;\epsilon)$.
        The ordered products satisfy
        \begin{equation*}
            \begin{aligned}
                \widehat W(t;\epsilon)-W^{\mathrm{XXX}}(t;\epsilon)
                 & =
                \sum_{j=1}^{L}
                \Tr_a\qty[
                    \widehat{\w}_1(t;\epsilon)\cdots\widehat{\w}_{j-1}(t;\epsilon)
                    \Delta_j
                    \w_{j+1}^{\mathrm{XXX}}(t;\epsilon)\cdots\w_L^{\mathrm{XXX}}(t;\epsilon)
                ]
                \\
                 & =
                \sum_{j=1}^{L}
                \Tr_a\qty[
                    \w_1^{\mathrm{XXX}}(t;0)\cdots\w_{j-1}^{\mathrm{XXX}}(t;0)
                    \Delta_j
                    \w_{j+1}^{\mathrm{XXX}}(t;0)\cdots\w_L^{\mathrm{XXX}}(t;0)
                ]
                \,.
            \end{aligned}
        \end{equation*}

        At $\epsilon=0$, the two local tensors coincide: $\widehat{\w}_i(t;0)=\w_i^{\mathrm{XXX}}(t;0)$.
        Since $\Delta_j=O(\epsilon)$, any $\epsilon$-dependent entry in another tensor would contribute $O(\epsilon^2)=0$, which proves the second equality.
        By Eq.~\eqref{eq:W-XXX-local-tensor}, $\w_i^{\mathrm{XXX}}(t;0)$ is block upper triangular, while the only nonzero block of $\Delta_j$ is $\widehat{\mathcal A}_j-\mathcal A_j$.
        Hence the auxiliary trace of each telescoping term reduces to the trace of its triplet block, and
        \begin{equation*}
            \begin{aligned}
                C\qty[\widehat W(t;\epsilon)-W^{\mathrm{XXX}}(t;\epsilon)]
                 & =
                \sum_{j=1}^{L}
                \Tr_a\qty[
                    \mathcal A_1\cdots \mathcal A_{j-1}
                    C\qty(\widehat{\mathcal A}_j-\mathcal A_j)
                    \mathcal A_{j+1}\cdots \mathcal A_L
                ]
                \\
                 & =
                -\epsilon Lt\,V^{\mathrm{XXX}}(t)
                \,.
            \end{aligned}
        \end{equation*}

        Moreover, $\widehat W(t;0)=I+V^{\mathrm{XXX}}(t)$, so $\epsilon V^{\mathrm{XXX}}(t)=\epsilon\qty[\widehat W(t;\epsilon)-I]$.
        Substituting $C=C(t^2)$ and eliminating $V^{\mathrm{XXX}}(t)$ gives
        \begin{equation}
            \label{eq:W-XXX-normalized-specialization}
            C(t^2)W^{\mathrm{XXX}}(t;\epsilon)
            =
            \qty[C(t^2)+\epsilon Lt]\widehat W(t;\epsilon)
            -
            \epsilon Lt I
            \,.
        \end{equation}

        The right-hand side of Eq.~\eqref{eq:W-XXX-normalized-specialization} contains only $\widehat W(t;\epsilon)$ from Eq.~\eqref{eq:XXX-dual-family-member} and the identity operator.
        The factor $C(t^2)$ is a nonzero scalar series.
        Therefore,
        \begin{equation}
            \qty[W^{\mathrm{XXX}}(t; \epsilon),W^{\mathrm{XXX}}(t^\prime; \epsilon^\prime)]=0
            \,,
        \end{equation}
        including the term proportional to $\epsilon\epsilon^\prime$, and
        \begin{equation}
            \qty[W^{\mathrm{XXX}}(t; \epsilon),H_{\mathrm{XXX}}]=0
            \,.
        \end{equation}

        Extracting coefficients and using $t=s+O(s^3)$ then shows that the coefficients in the $s$-expansion commute with $H_{\mathrm{XXX}}$.
        In particular, by Eq.~\eqref{eq:W-XXX-generating-MPO-expansion}, the wrap-around terms $\widetilde{Q}_k^{\mathrm{XXX},(L)}$ with $k>L$ are conserved in the XXX limit.

\end{appendix}

\bibliography{ref.bib}

\nolinenumbers

\end{document}